\documentclass[referee]{jfm}
\usepackage{amsmath}
\usepackage{graphicx}
\usepackage[debug]{psfrag}
\usepackage[sort&compress]{natbib}
\newcommand{\te}[1]{\mbox{\boldmath $#1$}}
 \newcommand{\Frac}[2]{\frac{\displaystyle #1}{\displaystyle #2}}
\newcommand{\bea}[1]{\begin{array}{#1}}
\newcommand{\eea}{\end{array}}
\newcommand{\Div}{\te{\nabla \cdot}}
\newcommand{\lambdadot}{\dot{\lambda}}
\newcommand{\airyai}{\ensuremath\mathrm{Ai}}
\newcommand{\airybi}{\ensuremath\mathrm{Bi}}

\newcommand{\mysqrt}[1]{{#1}^{1/2}}
\newcommand{\der}[2]{\Frac{\mathrm{d}#1}{\mathrm{d}#2}}
\newcommand{\pder}[2]{\frac{\displaystyle\partial#1}{\displaystyle\partial#2}}
\newcommand{\textfrac}[2]{\ensuremath{#1/#2}}
\newcommand{\textfracp}[2]{\ensuremath{(#1/#2)}}

\newcommand{\ol}{\overline}
\newcommand{\be}{\begin{equation}}
\newcommand{\ee}{\end{equation}}
\newcommand{\ba}{\begin{eqnarray}}
\newcommand{\ea}{\end{eqnarray}}

\begin{document}

\title[Frictional Cosserat model for slow granular flows]{A frictional
Cosserat model for the slow shearing of granular materials}
\author[L. S. Mohan, K. Kesava Rao and P. R. Nott]{L.\ns S\ls R\ls I\ls N\ls I\ls V\ls
A\ls S\ls A\ns M\ls O\ls H\ls A\ls N\footnotemark,\ns K.\ns K\ls
E\ls S\ls A\ls V\ls A\ns R\ls A\ls O$^{1}$,\ns \and \\ P\ls R\ls
A\ls B\ls H\ls U\ns R.\ns N\ls O\ls T\ls T$^{2}$\ns}
\affiliation{Department of Chemical Engineering, Indian Institute
of Science\\ Bangalore 560$\,$012, INDIA\\ [\affilskip]
 e-mail: $^1$kesava@chemeng.iisc.ernet.in,
 $^2$prnott@chemeng.iisc.ernet.in}

\maketitle

\begin{abstract}
\footnotetext[1]{Currently at Fluent India Pvt.\ Ltd., South
Koregaon park, Pune 411$\,$001, India \rule{0pt}{3em}}

    A rigid-plastic Cosserat model for slow frictional
flow of granular materials, proposed by us in an earlier paper,
has been used to analyze plane and cylindrical Couette flow.  In
this model, the hydrodynamic fields of a classical continuum are
supplemented by the couple stress and the intrinsic angular
velocity fields. The balance of angular momentum, which is
satisfied implicitly in a classical continuum, must be enforced in
a Cosserat continuum. As a result, the stress tensor could be
asymmetric, and the angular velocity of a material point may
differ from half the local vorticity.  An important consequence of
treating the granular medium as a Cosserat continuum is that it
incorporates a material length scale in the model, which is absent
in frictional models based on a classical continuum.  Further, the
Cosserat model allows determination of the velocity fields
uniquely in viscometric flows, in contrast to classical frictional
models. Experiments on viscometric flows of dense, slowly
deforming granular materials indicate that shear is confined to a
narrow region, usually a few grain diameters thick, while the
remaining material is largely undeformed.  This feature is
captured by the present model, and the velocity profile predicted
for cylindrical Couette flow is in good agreement with reported
data.  When the walls of the Couette cell are smoother than the
granular material, the model predicts that the shear layer
thickness is independent of the Couette gap $H$ when the latter is
large compared to the grain diameter $d_p$.  When the walls are of
the same roughness as the granular material, the model predicts
that the shear layer thickness varies as $(H/d_p)^{1/3}$ in the
limit $(H/d_p) \gg 1$, for plane shear under gravity and
cylindrical Couette flow.

\end{abstract}

\section{Introduction}
\label{sec-intro}

        The slow flow of densely packed granular materials is a
subject of considerable importance, primarily due to the obvious
commercial and technological implications resulting from a better
understanding of granular flows. Granular materials are
transported and processed in a variety of industrial operations,
and furthering our knowledge of their rheology opens up the
possibility of improving the design of these processes. The phrase
`slow flow' denotes the regime of where the density is high and
the deformation rate low, so that momentum is mainly transferred
during sustained contact between the grains, and the stress
generated by grain collisions is insignificant.  If we define the
ratio of the stress arising from grain collisions to the total
stress, ${\cal R} \equiv \rho_p \, d_p^2 \, \dot{\gamma}^2/N$,
where $\rho_p$ and $d_p$ are the density and mean diameter of the
grains, $\dot{\gamma}$ is the nominal shear rate and $N$ is the
total (normal or shear) stress, the slow flow regime corresponds
to ${\cal R} \ll 1$. This is the case in many terrestrial flows,
where gravity consolidates the medium to a state where sustained
frictional contact occurs between the particles.

        Experimental observations suggest that the drag force exerted
by a granular medium when it flows slowly past a solid surface is
not steady, but fluctuates in time (\citealp{budny79};
\citealp{tuzun_nedderman85}; \citealp{munchandersen93};
\citealp{nasuno98}; \citealp{albert_etal00}).  The oscillations
diminish in magnitude as the flow velocity increases.  In some
cases, the amplitude of the oscillations is of the same order of
magnitude as the mean value.  This phenomenon is referred to as
{\em stick-slip}.  The frequency of the oscillations depends on
the relative velocity between the granular material and the
surface, and also on various other factors.  In some of the
studies cited above, the frequency was in the range of 0.05 - 100
Hz, and the relative velocity was in the range of 5 $\mu$m/s - 10
mm/s.

    In this work we do not address the issue of stick-slip; rather
we propose a hydrodynamic model for slow granular flows that can
predict only the time averaged values of the stress and velocity
fields. It is assumed that the time-averaging is over a large
number of stick-slip events.  A prominent feature of the time
averaged stress in slow flow is its rate independence. For
instance, \citet{albert_etal99} found the drag force $\ol{F}$ on a
vertical rod dipped into a rotating bed of glass beads to be
independent of the velocity, for velocities in the range of 0.1 -
1.5 mm/s.  \citet{tardos98} sheared granular materials in the
annular gap of a vertical Couette cell, and measured the torque
required to rotate the inner cylinder at a constant speed. For
speeds in the range of 2.5 - 50 cm/s, which correspond to nominal
shear rates in the range of 2 - 40 s$^{-1}$, $\ol{F}$ decreased
very slightly as the shear rate increased; at higher shear rates,
$\ol{F}$ increased with the shear rate.

        As a first approximation in modelling the above experiments, we
may assume that the drag force does not vary with the velocity.
Within the framework of continuum mechanics, this is equivalent to
the assumption that the constitutive equation relating the
stress tensor to $\te{D}$ is unaffected if all components of
$\te{D}$ are changed by a common factor. This is termed {\em rate
independent} behaviour. Rate independent models for slow flow, which are
usually based on concepts in metal plasticity and soil mechanics,
have used with some success for flow in hoppers and
bunkers~\citep{brennen_pearce78, michalowski87, cleaver93}.

        However, these models fail for viscometric flows,
in which the direction of variation of the velocity is orthogonal
to the direction of flow. Experimental observations of viscometric
flows, such as flow through vertical
channels~\citep{nedderman_laohakul80, natarajan_etal95} or shear
in a cylindrical Couette cell~\citep{tardos98} show large portions
of the material not suffering sustained deformation, and shear
occurring only in thin
layers~\citep{nedderman_laohakul80,gudehus_tejchman91}.  Moreover,
the thickness of the shear layer is influenced by the nature of
the boundaries; it is less when the flowing medium is confined by
smooth walls than that in the case of rough
walls~\citep{nedderman_laohakul80}. In contrast, conventional
plasticity-based models predict no shear
layers~\citep{tejchman_wu93, mohan_etal97}, with the entire region
behaving like a rigid block and slipping relative to the walls.

        Owing to the occurrence of narrow shear layers, the nominal
shear rate differs significantly from the actual shear rate,
making it difficult to interpret rheological data. A theory that
predicts the extent of the locked, or ``plug'', region and the
velocity field within the shear layer would therefore be useful.

        The inability of frictional models to determine the velocity
field in viscometric flows has been attributed to the absence of a
material length scale in their constitutive equations~\citep[cited
in~\citet{muhlhaus_vardoulakis87}]{muhlhaus86}. An attempt to
correct this deficiency of the classical frictional models was
recently made by~\citet{mohan_etal99}. They modeled the granular
medium as a Cosserat continuum, which includes a material length
scale in the constitutive equations. Cosserat plasticity models
have been applied to problems in granular flow earlier
\citep{muhlhaus_vardoulakis87,muhlhaus89,tejchman_wu93,
tejchman_gudehus93,tejchman_wu94}, but the models in these studies
are posed in terms of strain increments as they only address
unsteady flows, and no results are reported for steady flow.

        In this work, we use the Cosserat model of
\citet{mohan_etal99} to examine steady shear between parallel
plates and concentric cylinders. Unlike a classical continuum, a
Cosserat continuum involves two additional field variables,
namely, the {\em intrinsic angular velocity} vector $\te{\omega}$,
and the {\em couple stress} tensor ${\bf M}$. The former
represents the local rate of spin of material elements, and the
latter represents the couple per unit area exerted by the medium.
Hence the mass and linear momentum balances must be supplemented
by the angular momentum balance relating the angular velocity
vector, the stress tensor and the couple stress tensor. (In a
classical continuum, the angular momentum balance is identically
satisfied, and the angular velocity is equal to half the
vorticity.) For steady, fully developed flow, spatial gradients of
$\te{M}$ cause $\te{\sigma}$ to be asymmetric. To our knowledge,
these effects have not been directly measured in the laboratory so
far, and careful experiments in this direction would be of value.
However, the consequences of our model are easily measurable even
in simple viscometric flows, as we shall indicate later.

        While there is  no direct experimental evidence of stress
asymmetry or deviation of the angular velocity from half of the
vorticity, there is, however, sufficient motivation for treating a
granular medium as a Cosserat continuum. In his analysis of the
stress in a fluid composed of diatomic molecules, \citet{dahler59}
shows that an asymmetric Cauchy stress can result from the
presence of non-central interactions forces between grains. Grain
interactions in slow flows are dominated by frictional forces,
which are inherently non-central. \citet{jenkins_etal89}
constructed a micro-mechanical model for an assembly of identical
spheres. They found that an asymmetric stress resulted when the
distribution of contact normals was anisotropic. However, they
secured symmetry of the stress tensor by suitably enforcing the
rotation of particles. In his simulations of rapid shear of rough
circular discs between parallel plates (the disks interacted
through instantaneous collisions), \citet{campbell93a} observed
that the stress tensor was asymmetric and that non-zero couple
stresses existed near the walls. Therefore, a frictional Cosserat
model may be an appropriate, and perhaps even the correct
description, for slow granular flow.

\section{Governing equations}
\label{sec-eqns_motion}

        As in a classical continuum, the mass and linear momentum balances
are given by
\begin{eqnarray}
       \Frac{D \rho}{D t} + \rho \Div \te{u}& = & 0, \label{eqn-contin}\\
\rho \Frac{D \te{u}}{D t} + \Div \te{\sigma} - \rho \,\te{b} & = &
0. \label{eqn-mom_bal}
\end{eqnarray}
Here $D/Dt$ represents the material derivative, \te{u} the
velocity, $\rho \equiv \rho_p \, \nu$ is the bulk density of the
medium, $\rho_p$ is the intrinsic density of the grains, $\nu$ is
the solids fraction, \te{\sigma} the Cauchy stress tensor (defined
in the compressive sense), and \te{b} the body force per unit mass.
As discussed in the preceding section, for a Cosserat continuum
these must be supplemented by the angular momentum
balance~\citep[p.~233]{jaunzemis67}, which takes the form
\begin{equation}
    \rho\,  \Frac{D ({\cal I} \,\te{\cdot \omega})}{D t} + \Div \te{M} -
        \te{\varepsilon \!:\! \sigma} - \rho \, \te{\zeta} = 0.
\label{eqn-ang_mom_bal}
\end{equation}
 Here ${\cal I}$ is the intrinsic inertia tensor, $\te{\omega}$ is
the intrinsic angular velocity, \te{M} is couple stress,
\te{\varepsilon} is the alternating tensor and \te{\zeta} is the
body couple. On a plane with unit normal $\te{n}$, the couple per
unit area transmitted to the side towards which $\te{n}$ points
(by the material on the other side) is $\te{n \cdot M}$ . We
follow the commonly used right-hand convention for the signs of
$\te{\omega}$ and $\te{n \cdot M}$. In general, the distribution
of size, shape, and orientation of the particles  is required to
determine the inertia tensor ${\cal I}$. As the present work is
confined to steady, fully developed flow, the term involving
${\cal I}$ in \eqref{eqn-ang_mom_bal} vanishes.

        Constitutive relations for \te{\sigma} and \te{M} for slow
granular flow were proposed by \citet{muhlhaus_vardoulakis87} and
\citet{tejchman_wu93} as extensions of the yield conditions and
flow rules used in classical plasticity. Their relations are posed
in terms of strains and strain increments, which they employed to
study the temporal development of shear bands. Here we rewrite
these relations in terms of the rate of deformation tensor, to
render them in a form that is appropriate for sustained flow.

\subsubsection{Yield condition}
\label{subsec-YC}

        A commonly used model for a block sliding on a plane is that
the shear force $S_f$ is proportional to the normal force $N_f$,
the constant of proportionality being the coefficient of friction
$\mu$.  When the block does not slide, $S_f < \mu N_f$.  The yield
condition is a continuum analog of this relation. For a classical
continuum, it is of the form $F(\te{\sigma},\nu) = 0$, where $F$
is a scalar function of the stress tensor \te{\sigma} and the
solids fraction $\nu$. As elastic effects are neglected, the
material is rigid if $F < 0$. If $F = 0$, either plastic or
irrecoverable deformation occurs, or the material is in a state of
incipient yielding.  The extended von Mises yield condition, used
in some studies, has $F \equiv \tau(J_2) - Y(J_1,\nu)$
where $J_1$ is the first invariant (i.e.\ the trace) of
\te{\sigma} and $J_2$ is the second invariant of the deviatoric
stress tensor $\te{\sigma}'$ (see below). To generalise this for a
Cosserat continuum, we write $\tau$ as a function of $J_2$ and the
couple stress $\te{M}$, as described below.

    Following \citet{besdo74} (cited in~\citet{lippmann95}),
\citet{deBorst93}, and \citet{tejchman_wu93}, we take the yield
condition to be given by
\begin{subequations}
\begin{equation}
  \label{eqn-yc1}
  F \equiv \tau  -  Y  = 0,
%\end{equation}
\end{equation}
where
\begin{equation}
  \label{eqn-yc2}
  \tau \equiv \mysqrt{\left(a_{1} \sigma_{ij}' \sigma_{ij}' + a_{2}
      \sigma_{ij}' \sigma_{ji}' + \Frac{1}{(L d_{p})^{2}} M_{ij}
      M_{ij} \right)},
\end{equation}
\end{subequations}
 $\sigma_{ij}' \equiv \sigma_{ij} - \textfracp{1}{3}\,
\sigma_{kk}\, \delta_{ij}$ is the deviatoric stress, $\delta_{ij}$
is the Kronecker delta, $a_{1}$, $a_{2}$, and $L$ are material
constants, and $d_{p}$ is the grain diameter. \citet{deBorst93}
assumed that the yield function $Y$  depends on the mean stress
$\sigma \equiv \textfrac{\sigma_{kk}}{3}$ and a hardening
parameter $h$.  Here $h$ is taken as the solids fraction $\nu$,
and hence $Y = Y(\sigma,\nu)$.  The parameter $L\, d_p$ determines
the characteristic material length scale, and is perhaps related
to the length of force-chains \citep{howell_etal99}.

        Only two of the three parameters $a_{1}$, $a_{2}$ and $L$
in~\eqref{eqn-yc2} are independent, because the third may be
absorbed in the definition of
$Y(\sigma,\nu)$~(see~\eqref{eqn-yc1}). Following
\citet{muhlhaus_vardoulakis87} we set $a_{1} + a_{2} =
\textfrac{1}{2}$, without loss of generality.  When $\te{M}$ vanishes,
the quantity $\tau$ in (\ref{eqn-yc2}) reduces $\sqrt{J_2}$, and hence
the extended von Mises yield condition is recovered.

While it is desirable to determine the parameters independently, neither
experiments nor micro-mechanical models are currently available to guide
their choice. However, the condition $\tau^2 \geq 0$ imposes the bound
$\vert a_2/a_1 \vert \leq 1$, which follows from the result that the
expression for $\tau^2$ in \eqref{eqn-yc2} is a quadratic form. We retain
the value of 10  used by
\citet{mohan_etal99} for $L$, as it nicely fits experimental data for flow
down vertical channels, and also their choice of $A \equiv
\textfrac{a_2}{a_1} = 1/3$. Nevertheless, we explore the
sensitivity of the model predictions to these parameters in
\S\ref{subsec-par_sens}.

    The yield function $Y(\sigma,\nu)$ is usually written in the form
$Y = Y_1(\alpha) \sigma_c(\nu) \sin\phi$, where $\alpha =
\sigma/\sigma_c$, and $\sigma_{c}(\nu)$ is the mean stress at
\emph{critical state}, which is a state of deformation at constant density
(see, for example, \citealp {jackson83}). The parameter $\phi$ is a
material constant called the angle of internal friction.
For the problems considered in this paper, it is shown later that
the material is everywhere at critical state, and hence $\alpha = 1$. The
function $Y_1(\alpha)$ may then be set equal to 1 without loss of generality,
and the yield condition reduces to
\begin{equation}
F = \tau - \sigma_c(\nu) \sin \phi = 0 \,;\,\, \sigma = \sigma_c(\nu)
\label{ycrit}
\end{equation}
at a critical state.  We expect $\sigma_{c}$ to vanish when the
solids fraction is below that of loose random packing
$\nu_{\min}$, when the grains are no longer in sustained contact,
and to increase rapidly as $\nu$ approaches the solids fraction at
dense random packing, $\nu_{\max}$.

    However, we simplify the analysis in this work by assuming that
the material is incompressible. Hence $\sigma_c$ is treated as an
primitive variable, and an explicit expression for $\sigma_c(\nu)$
is not required.

\subsubsection{Flow rule}
\label{subsec-flowrule}

    The flow rule relates the rate of deformation tensor to the
stress tensor. A commonly used flow rule in classical frictional
models is the plastic potential flow rule, which is expressed as
\begin{equation}
D_{ij} \equiv\frac{1}{2}\,\left( \pder{v_{i}}{x_{j}} +
\pder{v_{j}}{x_{i}} \right) = \lambdadot \pder{G}{\sigma_{ji}}.
\label{eqn-flowrule_classical}
\end{equation}
Here $D_{ij}$ is the rate of deformation tensor, $G(\te{\sigma},
\nu)$ is scalar function called the \emph{plastic potential}, and
$\lambdadot$ is a scalar factor which must be determined as a part
of the solution.  As detailed information on the plastic potential
$G$ is not usually available, we use the associated flow rule
\citep[p.~43]{schofield_wroth68}, for which
\begin{equation}
  G \equiv F = \tau - Y.
\label{eqn-fr_associated}
\end{equation}
This form for the flow rule, in conjunction with a yield condition
defined by (\ref{eqn-yc1})-(\ref{eqn-yc2}), accounts for density
changes accompanying deformation.  Together, they constitute a
rate-independent constitutive relation, which is a desirable
feature for slow granular flows.

    \citet{tejchman_wu93} used the approach of \citet{muhlhaus89} to
modify \eqref{eqn-flowrule_classical} for a Cosserat continuum,
but posed it in terms of  strain increments as they were
addressing unsteady deformation. Because we are interested in
sustained flow, we replace the plastic strain increments used
by~\citet{tejchman_wu93} by the appropriate deformation rates
~\citep{muhlhaus89}, and thereby write the flow rule as
\begin{subeqnarray}
% define subequation numbers
\gdef\thesubequation{\theequation \textit{a,b}}
 \te{E} \equiv \nabla \, \te{v} + \te{ \varepsilon} \cdot \te{\omega} =
\lambdadot \, \pder{F}{{\te{\sigma}}^t} \, , \quad \te{H} \equiv
\nabla \te{\omega} = \lambdadot \, \pder{ F }{ {\te{M}}^t}.
\label{eqn-flowrule}
\end{subeqnarray}
% reinstate the original definition of \thesubequation
\returnthesubequation
 Here $\te{ \varepsilon}$ is the alternating
tensor, and the superscript $t$ indicates the transpose. Equations
\eqref{eqn-flowrule} may be written in Cartesian tensor notation
as
\begin{subeqnarray}
% define subequation numbers
\gdef\thesubequation{\theequation \textit{a,b}}
  E_{ij} = \pder{v_{i}}{x_{j}} + \varepsilon_{ijk}\omega_{k} =
  \lambdadot \pder{F}{\sigma_{ji}} \, , \quad H_{ij} =
  \pder{\omega_{i}}{x_{j}} = \lambdadot \pder{F}{M_{ji}}.
\label{eqn-fr0}
\end{subeqnarray}
% reinstate the original definition of \thesubequation
\returnthesubequation
 We note here that $E_{ij}$ is the sum of the
rate of deformation tensor $D_{ij}$ and an objective antisymmetric
tensor representing the difference between the spin tensor  and the
particle spin $- \varepsilon_{ijk}\omega_{k}$. The quantities $E_{ij}$ and
$H_{ij}$ are conjugate to the stress $\sigma_{ji}$ and the couple stress
$M_{ji}$, respectively, in the sense that the dissipation rate per unit
volume by the contact forces and couples is given by $-(\sigma_{ji}E_{ij}
+ M_{ji}H_{ij})$
\citep{muhlhaus89}.

    If $\te{M}$ vanishes and $\te{\sigma}$ is symmetric,
(\ref{eqn-fr0}) reduces to
\begin{equation}
D_{ij} = \lambdadot \pder{F}{\sigma_{ji}}.
\label{assoc-classical}
\end{equation}
and
\begin{equation}
\varepsilon_{ijk}\omega_{k} =  \frac{1}{2} (\pder{v_j}{x_i} -
\pder{v_i}{x_j}) \,; \quad \pder{\omega_i}{x_j} = 0.
\label{omegavort}
\end{equation}
Equation (\ref{assoc-classical}) is identical to the associated
flow rule for the classical frictional model (defined
by~\eqref{eqn-flowrule_classical} with $G = F$). For steady flow,
equation (\ref{omegavort}) implies that $\te{\omega}$ is a
constant, and is  equal to half the vorticity. On the other hand,
the vorticity need not be a constant in a classical frictional
model.  Thus, only some of the features of the latter are
recovered in this special case.

    As explained in \S\ref{sec-summary}, the most important
aspects of our results, i.e.~the ability of the model to predict
the velocity fields in viscometric flows and the dependence of
the shear layer thickness on the properties of the material and
the flow cell, are not specific to the particular forms of the
constitutive relation chosen above, but have greater validity. The
forms of the yield condition and flow rule we have chosen above
serve to illustrate the importance of Cosserat effects in slow
granular flows.

        The application of the above Cosserat model to flow in vertical
channels has been described in \citet{mohan_etal99}. Here we
consider two other viscometric flows: (i) plane shear, and (ii)
cylindrical Couette flow. As rheological measurements are usually
made in these geometries, it is useful to develop models for these
cases.

\section{Application to plane shear}
\label{sec-plane_shear}

\begin{figure}
  \begin{center}
    \vspace{1in}
    \psfrag{V}{$V$}
    \psfrag{H}[]{$H\,\,$}
    \psfrag{N}{$N$}
    \psfrag{g}[t][]{$g$}
    \psfrag{x}[b][]{$x$}
    \psfrag{y}{$y$}
    \includegraphics[width=0.45\textwidth]{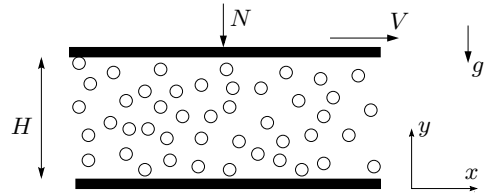}
\caption{Plane shear between parallel walls.  The lower wall is
stationary and the upper wall moves at a constant speed $V$. A
constant normal stress $N$ is applied on the upper wall. The
angles of wall friction of the lower and upper walls are
$\delta_L$ and $\delta$, respectively.
\label{fig-schem_plane_shear}}
  \end{center}
\end{figure}

       Consider steady, fully developed flow between horizontal walls
which are separated by a gap $H$ (see
figure~\ref{fig-schem_plane_shear}). The upper wall moves in the
positive $x$ direction with a constant speed $V$, the lower wall
is stationary, and a constant compressive normal stress $N$ is
applied to the upper wall. (For the case of zero gravity,
considered in section \ref{subsec-zerog}, the bottom wall is also
subjected to a compressive normal stress of magnitude $N$.)
Typically in an experiment, the initial gap, before the walls are
set in motion, is fixed,  at say $H_0$. Owing to
dilation and compaction during shearing, this can change to a
different value $H$ once steady flow is established.

    The velocity fields are of the form
\begin{equation}
\bea{c}
    v_x = v_x (y) \, , \quad v_y = 0 \, , \quad v_z = 0 \, ,\\
    \omega_x = 0  \, , \quad \omega_y = 0 \, , \quad{\omega}_z =
{\omega}_z(y) \, ,
\eea
\label{eqn-vel}
\end{equation}
and all the stresses are assumed to depend only on the $y-$
coordinate. Equations (\ref{eqn-vel}) and (\ref{eqn-contin}) imply
that $\textfrac{D \rho}{D t} = 0$. Hence the material is at a
critical state, and the yield condition is given by (\ref{ycrit}).

    The mass balance~\eqref{eqn-contin} is identically satisfied, and
the $x-$ and $y-$ components of the linear momentum balance
\eqref{eqn-mom_bal} reduce to
\begin{subeqnarray}
% define subequation numbers
\gdef\thesubequation{\theequation \textit{a,b}}
\der{\sigma_{yx}}{y}   =  0\,, \quad \der{\sigma_{yy}}{y} = - \rho
\, g. \label{eqn-xymom}
\end{subeqnarray}
% reinstate the original definition of \thesubequation
\returnthesubequation
 As there is no externally imposed body
couple, the $z-$ component of the angular momentum balance reduces
to
\begin{equation}
\der{M_{yz}}{y} + \sigma_{xy} - \sigma_{yx}  = 0.
\label{eqn-zangmom}
\end{equation}
The diagonal components of $E_{ij}$ are zero, and hence the flow
rule~\eqref{eqn-fr0} implies that
\begin{subeqnarray}
  0 & = & \Frac{\lambdadot}{6\tau}
  (2 \sigma_{xx}' - \sigma_{yy}' - \sigma_{zz}'),\\
  0 & = & \Frac{\lambdadot}{6\tau}
  (2 \sigma_{yy}' - \sigma_{xx}' - \sigma_{zz}'),\\
  0 & = & \Frac{\lambdadot}{6\tau}
  (2 \sigma_{zz}' - \sigma_{xx}' - \sigma_{yy}'). \label{eqn-fr_diag}
\end{subeqnarray}
As $\sigma_{ij}' = \sigma_{ij} - \sigma \delta_{ij}$, we have
$\sigma_{xx}' + \sigma_{yy}' + \sigma_{zz}' = 0$. Hence,
\eqref{eqn-fr_diag}  imply equality of the normal stresses,
\begin{equation}
  \label{eqn-ns}
  \sigma_{xx} = \sigma_{yy} = \sigma_{zz} = \sigma = \sigma_{c}(\nu).
\end{equation}
>From \eqref{eqn-vel} and \eqref{eqn-fr0}, we have
\begin{eqnarray}
E_{xz} = E_{zx} = E_{yz} = E_{zy} = 0, \nonumber
\end{eqnarray}
which imply that all the shear stresses except $\sigma_{xy}$ and
$\sigma_{yx}$ vanish. Similarly, all the couple stresses except
$M_{yz}$ vanish.  Hence the yield condition \eqref{eqn-yc1}
reduces to
\begin{equation}
{\tau}^2 =  a_{1}(\sigma_{xy}^{2} + \sigma_{yx}^{2}) + 2 a_{2}
\sigma_{xy}
  \sigma_{yx} + \Frac{M_{yz}^2}{(L\, d_p)^2} =  (\sigma_{c}\sin\phi)^{2}.
  \label{eqn-yc3}
\end{equation}
The remaining equations of the flow rule~\eqref{eqn-fr0} are
\begin{eqnarray}
  E_{xy} & = & \der{v_x}{y} + \omega_z = \Frac{\lambdadot}{\tau}
  \left(a_{1} \sigma_{yx} + a_{2} \sigma_{xy}
  \right),\label{eqn-fr4}\\
  E_{yx} & = & - \omega_z = \Frac{\lambdadot}{\tau}
  \left(a_{1} \sigma_{xy} + a_{2} \sigma_{yx}
  \right),\label{eqn-fr5} \\
  H_{zy} & = & \der{\omega_z}{y} = \Frac{\lambdadot}{\tau}
  \Frac{M_{yz}}{(L d_{p})^{2}}.
\label{eqn-fr6}
\end{eqnarray}
Eliminating $\lambdadot$ from \eqref{eqn-fr4} - \eqref{eqn-fr6}, we get
\begin{eqnarray}
  \der{v_x}{y}& = & - \Frac{(A + 1)(\sigma_{xy} +
    \sigma_{yx})\,\omega_z}{\sigma_{xy} + A
    \sigma_{yx}},\label{eqn-fr1}\\
 \der{\omega_z}{y} & = & - \Frac{\omega_z}{ (L d_p)^2}
    \Frac{2(A + 1) \,m}{(\sigma_{xy} + A
    \sigma_{yx})}. \label{eqn-fr2}
\end{eqnarray}

    Equations \eqref{eqn-xymom}, \eqref{eqn-zangmom}, \eqref{eqn-yc3},
and \eqref{eqn-fr1} -- \eqref{eqn-fr2} constitute the governing
equations for plane shear. They may be cast in dimensionless form
by introducing the dimensionless variables
\begin{equation}
\bea{c}
    \xi = \Frac{y}{H} \, , \quad u = \Frac{v_x}{V}\, , \quad \ol{\sigma}_{ij}
    = \Frac{\sigma_{ij}}{N}\, ,\\
    m = \Frac{1}{L\sqrt{2(A+1)}} \Frac{M_{yz}}{(N d_p)}\, , \quad
\mbox{and} \quad \omega = \Frac{\omega_z\, H}{V}.
\eea
\label{dimlessvar}
\end{equation}

    The governing equations now take the form
\begin{eqnarray}
\der{\ol{\sigma}_{yx}}{\xi}   =  0,\label{eqn-nd_xmom} \\
\der{\ol{\sigma}_{yy}}{\xi} + B \nu = 0, \label{eqn-nd_ymom} \\
\varepsilon \, \alpha \der{m}{\xi} +
\Frac{1}{2(A+1)}(\ol{\sigma}_{xy} - \ol{\sigma}_{yx}) = 0,
\label{eqn-nd_zangmom} \\
 (\ol{\sigma}_{xy}^{2} + \ol{\sigma}_{yx}^{2}) + 2 A
\ol{\sigma}_{xy} \ol{\sigma}_{yx} + 4(A+1)^2m^2
- 2(A+1)(\ol{\sigma}_{c}\sin\phi)^{2} = 0, \label{eqn-nd_yc} \\
\der{u}{\xi} + \Frac{(A + 1)(\ol{\sigma}_{xy} +
    \ol{\sigma}_{yx})\,\omega}{\ol{\sigma}_{xy} +
A \ol{\sigma}_{yx}} = 0, \label{eqn-nd_fr1}\\
\varepsilon \alpha \der{\omega}{\xi} +  \Frac{2(A +
1) \,m \, \omega}{(\ol{\sigma}_{xy} + A \ol{\sigma}_{yx})} = 0.
\label{eqn-nd_fr2}
\end{eqnarray}
where $\varepsilon \equiv d_p/H$, $B\equiv\textfrac{\rho_p g
H}{N}$ is ratio of the gravitational head at the base to the
applied normal stress on the top plate, and $\alpha \equiv
\textfrac{L}{\sqrt{2(A+1)}}$.

The yield condition \eqref{eqn-nd_yc} may be solved for
$\ol{\sigma}_{xy}$ to get
\begin{equation}
\ol{\sigma}_{xy} = - A \ol{\sigma}_{yx} \underline{+} \sqrt{ (A^2
-1 ) \overline{\sigma}^2_{yx} + 2 (A + 1) \ol{\sigma}^2_c {\sin}^2
\phi - 4 (A + 1)^2 m^2} \label{eqn-nd_sigxy}
\end{equation}
where $\ol{\sigma}_c = \sigma_c/N$.

\section{Boundary Conditions}
\label{sec-BC}

        Equations \eqref{eqn-nd_xmom} - \eqref{eqn-nd_sigxy} require five
boundary conditions. We first consider boundary conditions for the
linear and angular momentum balances
\eqref{eqn-nd_xmom}-\eqref{eqn-nd_zangmom}. The first is the
specification of the normal stress $N$ acting on the upper wall,
i.e.
\begin{equation}
\ol{\sigma}_{yy}(1) = 1. \label{eqn-nd_bc1}
\end{equation}
If the material slips relative to a wall, we use the usual
friction boundary condition (\citealp{brennen_pearce78},
\citealp[p.~41]{nedderman92}),
\begin{equation}
  - \ol{\sigma}_{yx}/\ol{\sigma}_{yy} = \tan \delta,
\label{eqn-nd_bc2}
\end{equation}
where $\delta$, the angle of wall friction, is a property of the
wall and the granular material. This is an approximate
time-averaged boundary condition and is not expected to capture
rapid events such as stick-slip. For the special case of shear in
the absence of gravity between identical walls, \eqref{eqn-nd_bc2}
applies on both walls, which is equivalent to specifying
\eqref{eqn-nd_bc2} at one wall and $m(\xi=1/2)=0$, as elaborated
in \S\ref{subsec-zerog}. We therefore have the requisite number of
boundary conditions for
\eqref{eqn-nd_xmom}-\eqref{eqn-nd_zangmom}.

    As discussed in \citet{kaza82}and \citet[p.~161]{nedderman92}, the
classical frictional model suggests that $\tan \delta \leq \sin
\phi$.  A ``fully rough'' wall is defined as one for which the
angle of wall friction  satisfies
\begin{equation}
\tan \delta = \sin \phi.
\label{eqn-bc_fullyrough}
\end{equation}
Physically we may try to realize this by coating the wall with a
monolayer of the granular material. In some experiments involving
walls coated with sand, polystyrene beads, and glass beads, the
measured values of $\delta$ were found to be within 1$^{\circ}$ of
the values predicted by \eqref{eqn-bc_fullyrough} \citep{kaza82}.
However, \eqref{eqn-bc_fullyrough} should not be taken too
seriously, as it rests on assumptions which are difficult to
verify.  Moreover, it is well known that grains do not pack as
densely near a solid boundary as in the bulk; hence it is
reasonable to suppose that a fully rough wall is an idealization
that is difficult to achieve practically.

        When the material does not slip relative to a wall, (as is the
case in most of the problems considered in
\S\ref{sec-results_planecouette}), friction is not fully mobilized
and an alternative to \eqref{eqn-nd_bc2} must be specified. One
choice is to specify the value of $m$ ($= m_w$) at the wall.  In
this work, we use the condition
\begin{equation}
        \ol{\sigma}_{xy} = \ol{\sigma}_{yx}
\label{eqn-nd_bc4}
\end{equation}
at a wall where the material does not slip relative to it. This is
equivalent to specifying $m_w$, as it can then be expressed in
terms of $\sigma_{yx}$ using (\ref{eqn-nd_bc4}) and the yield
condition \eqref{eqn-nd_yc}.  Equation (\ref{eqn-nd_bc4}) was
motivated by an expectation that Cosserat effects such as the
stress asymmetry would vanish outside the shear layer. However,
the results presented later do not support this conjecture.
Fortunately, the velocity profile is relatively insensitive to the
value of $m_w$, as discussed in section
\ref{sec-results_planecouette}.

        We now consider boundary conditions for the flow rule
\eqref{eqn-nd_fr1}-\eqref{eqn-nd_fr2}.  If the material slips
relative to a boundary, we assume as in our earlier work
\citep{mohan_etal99} that
\begin{equation}
  \te{v}-\te{v_w} = -K d_{p}\,\te{n}\! \times\! \te{\omega},
  \label{eqn-bc3}
\end{equation} where $K$ is a dimensionless constant which reflects the
roughness of the wall, \te{n} is the unit normal at the wall
(pointing into the granular material) and \te{v_w} is the linear
velocity of the wall. This boundary condition relates the angular
velocity of the material adjacent to the wall to the slip
velocity.  It was introduced by~\citet{tejchman_gudehus93} and
\citet{tejchman_wu93}, who formulated it in terms of displacement
and rotation. Here we use an equivalent form in terms of velocity
and angular velocity.

    An explanation of this condition was provided by
\citet{mohan_etal99}, which we repeat here for the sake of
completeness. If we consider a single spherical particle moving
past a stationary wall, the linear velocity $\te{v}'$ of its
center of mass and the angular velocity $\te{\omega}'$ about an
axis passing through the center of mass are related by $\te{v}' =
\textfracp{d_{p}}{2}\, \te{n}\! \times\! \te{\omega}'$ if the
particle rolls without slipping. Conversely, if the particle
slides without rolling, $\te{\omega}' = 0$ but $\te{v}'$ is
arbitrary. For the boundary condition~\eqref{eqn-bc3}, these
limits correspond to $K \rightarrow \textfrac{1}{2}$ and $K
\rightarrow \infty$, respectively.  If the particles are rough or
angular, rolling and slipping may be altogether absent, reducing
the lower limit of $K$ to zero.  In a continuum description, we
expect that $K$ will decrease as the wall roughness increases. It
is reasonable to expect that $K$ and the angle of wall friction
$\delta$ are related, as they both characterize the roughness of
the wall.  However, as data bearing on this is lacking, both $K$
and $\delta$ will be treated as independent parameters.

        For the present problem, \eqref{eqn-bc3} may be written in
dimensionless form as
\begin{equation}
u(1) = 1 + \varepsilon K \omega(1) \label{eqn-nd_bc3u}
\end{equation}
if \eqref{eqn-nd_bc2} holds at the upper wall, and
\begin{equation}
u(0) = - \varepsilon K \omega(0) \label{eqn-nd_bc3l}
\end{equation}
if \eqref{eqn-nd_bc2} holds at the lower wall.

If the material does not slip relative to the upper wall, we have
\begin{equation}
               u(1) = 1.
\label{eqn-nd_bc5u}
\end{equation}
Similarly, if the material does not slip relative to the lower
wall, we have
\begin{equation}
        u(0) = 0
\label{eqn-nd_bc5l}
\end{equation}

\section{Results for plane shear}
\label{sec-results_planecouette}

    We first consider the case of zero gravity, for
which the parameter $B$ in \eqref{eqn-nd_ymom} is set to zero.  We
then consider finite gravity, for which $B$ determines the
gravitational overburden at the base in comparison with the
applied normal load.

\subsection{Shear in the absence of gravity}
\label{subsec-zerog}

We first consider the case where the two walls are identical,
i.e.\ $\delta_L = \delta$, where $\delta_L$ and $\delta$ are the
angles of wall friction of the lower and upper walls,
respectively. The domain is symmetric about the mid-plane
$\xi=1/2$, and \eqref{eqn-nd_xmom}-\eqref{eqn-nd_fr2} admit a
solution wherein the stresses are symmetric and $m$ is
antisymmetric about $\xi=1/2$. Similarly $u - 1/2$ is
antisymmetric while the angular velocity $\omega$ is symmetric
about $\xi=1/2$. It then suffices to solve the problem in the
half-domain $1/2 \leq \xi \leq 1$, with boundary conditions
\begin{equation}
   m(1/2) = 0,\;\;\; u(1/2) = 1/2. \label{eqn-nd_bc6}
\end{equation}
and \eqref{eqn-nd_bc1}, \eqref{eqn-nd_bc2}( at $\xi = 1$), and
\eqref{eqn-nd_bc3u}. The momentum balances \eqref{eqn-nd_xmom} and
\eqref{eqn-nd_ymom} and the  boundary conditions \eqref{eqn-nd_bc1} and
\eqref{eqn-nd_bc2} imply that
\begin{equation}
\ol{\sigma}_{yy} = 1\;\;\ \mbox{and}\;\;  \ol{\sigma}_{yx} =
-\tan\delta. \label{sigxy_g0}
\end{equation}
Hence \eqref{eqn-ns} implies that the solids fraction $\nu$ is
also a constant, and its value is given by $\ol{\sigma}_c (\nu) =
1$.  Substituting for $\ol{\sigma}_{xy}$ from \eqref{eqn-nd_sigxy}
into the angular momentum balance \eqref{eqn-nd_zangmom} yields
the following  equation for the couple stress:
\begin{equation}
\alpha \, \varepsilon \der{m}{\xi} = -a \mp \sqrt{b^2 - m^2}
%=\frac{\ol{\sigma}_{yx} - \ol{\sigma}_{xy}}{2 (A + 1)}
\label{eqn-eta_g0}
\end{equation}
where
\begin{equation}
a=\Frac{\tan\delta}{2} \quad \mbox{and} \quad b = (a^2 +
\Frac{1}{2(A+1)}(\sin^2\phi - \tan^2\delta))^{1/2}.
\label{eqn-a_b_def}
\end{equation}
This must be solved for $m$ in the region $1/2 \leq \xi \leq 1$
with boundary condition \eqref{eqn-nd_bc6}. The shear stress
$\ol{\sigma}_{xy}$ then follows from \eqref{eqn-nd_sigxy}.

    Let us consider the choice of signs in the right-hand side of
\eqref{eqn-eta_g0}.  If the negative sign is used in front of the
square root, $\textfrac{dm}{d\xi} < 0$.  Equation
(\ref{eqn-nd_bc6}) implies that $|m|$ increases as $\xi$ increases
from 1/2. Therefore, for small enough values of $\varepsilon$, $m = -b$ at
some value of $\xi = \xi_r < 1$, and hence a real valued solution cannot
be constructed for $\xi > \xi_r$.  We therefore choose the positive sign
in front of the square root.   This leads to an acceptable solution
because, though $\textfrac{dm}{\xi} \geq 0$, the structure of
\eqref{eqn-eta_g0} ensures that $|m| \leq \sqrt{b^2 - a^2}$.  This
choice of roots is used throughout the paper.  It corresponds to
the use of the negative sign in front of the square root in
\eqref{eqn-nd_sigxy}.

    The differential equation \eqref{eqn-eta_g0} may be
integrated by making the substitution $m = b \sin \psi$, to get
\begin{equation}
\frac{\xi - 1/2}{\alpha \, \varepsilon} = \psi +
\frac{a}{\sqrt{b^2 - a^2}} \ln \left(\frac{c + \tan (\psi/2)}{c -
\tan (\psi/2)} \right) \label{eqn-eta_soln_g0}
\end{equation}
with $c = \textfrac{\sqrt{b-a}}{\sqrt{b+a}}$.  The velocity fields
must be determined by integrating \eqref{eqn-nd_fr1} and
\eqref{eqn-nd_fr2} numerically subject to the boundary conditions
\eqref{eqn-nd_bc3u} and the second of \eqref{eqn-nd_bc6}.

       Before discussing results for arbitrary values of the angle of wall
friction $\delta$, it is instructive to consider  the special case
of a fully rough wall, defined by \eqref{eqn-bc_fullyrough}, for
which $b = a$. In this case, \eqref{eqn-eta_g0} and the first of
\eqref{eqn-nd_bc6} imply that $m$ and all its derivatives with
respect to $\xi$ vanish at $\xi = 1/2$, resulting in $m(\xi) = 0$.
It then follows that
\begin{subequations}
\ba
    \ol{\sigma}_{xy} & = & \ol{\sigma}_{yx} = -\tan \delta, \\
    \omega & = & \omega_1 \, \mbox{(constant)}, \;\; \mbox{and}\\
    u & = & 1/2 - 2 \omega_1 (\xi - 1/2).
    \label{vel_g0}
\ea
Using the boundary condition \eqref{eqn-nd_bc3u}, we get
\begin{equation}
\omega_1 = -1/(2 + 2 \varepsilon K) \label{omega1_g0}.
\end{equation}
\end{subequations}
As expected from the direction of motion of the
upper plate (figure \ref{fig-schem_plane_shear}), $\omega_1$ is
negative, implying that the particles rotate about the $z-$ axis
in the clockwise direction.  If the gap width is large compared to
the particle diameter, $\varepsilon \rightarrow 0$ and therefore
$u(1) \rightarrow 1$. Thus the velocity slip at the walls vanishes
when thick layers are sheared.

    Thus the couple stress vanishes, the  stress tensor is
symmetric,  and the angular velocity equal to half the  vorticity,
all of which are features of a classical continuum. However, the
classical frictional model does not lead to a unique velocity
profile, unlike the Cosserat model. As noted earlier, this is
because the two models are not identical for this special case.

    For $\tan\delta < \sin\phi$, rather than determining $m$ by
solving \eqref{eqn-eta_soln_g0} and then the velocity fields by
integrating \eqref{eqn-nd_fr1}-\eqref{eqn-nd_fr2}, all the fields
are obtained by integrating  \eqref{eqn-nd_zangmom} -
\eqref{eqn-nd_fr2}, after substituting for $\ol{\sigma}_{xy}$ from
\eqref{eqn-nd_sigxy}. This is done because such a numerical
solution procedure is anyway required for plane shear under
gravity and cylindrical Couette flow, as analytical solutions
could not be constructed for these problems. The system of
equations were solved numerically using the \textsc{lsoda}
routine~\citep{petzold83} from the \textsc{odepack} library in
\textsc{netlib}.  Our numerical procedure was verified by
favourable comparison with the analytical solution given by
\eqref{eqn-eta_soln_g0}, and the asymptotic solutions in
\S\ref{appendix} (see figures \ref{fig-asymp_g0} and
\ref{fig-asymp_g_c}).

\begin{figure}
  \begin{center}
    \vspace{1in}
    \psfrag{xi}[br][][1][-90]{$\xi$}
    \psfrag{vel}[t]{$u$}
    \psfrag{omega}[t]{$\omega + 1/2\der{u}{\xi}$}
    \psfrag{sigma}[t]{$\ol{\sigma}_{yx} - \ol{\sigma}_{xy}$}
    \psfrag{m}[t]{$m$}
    \hspace*{0.1in} (a) \hspace{2.2in} (b) \\
    \includegraphics[width=0.45\textwidth]{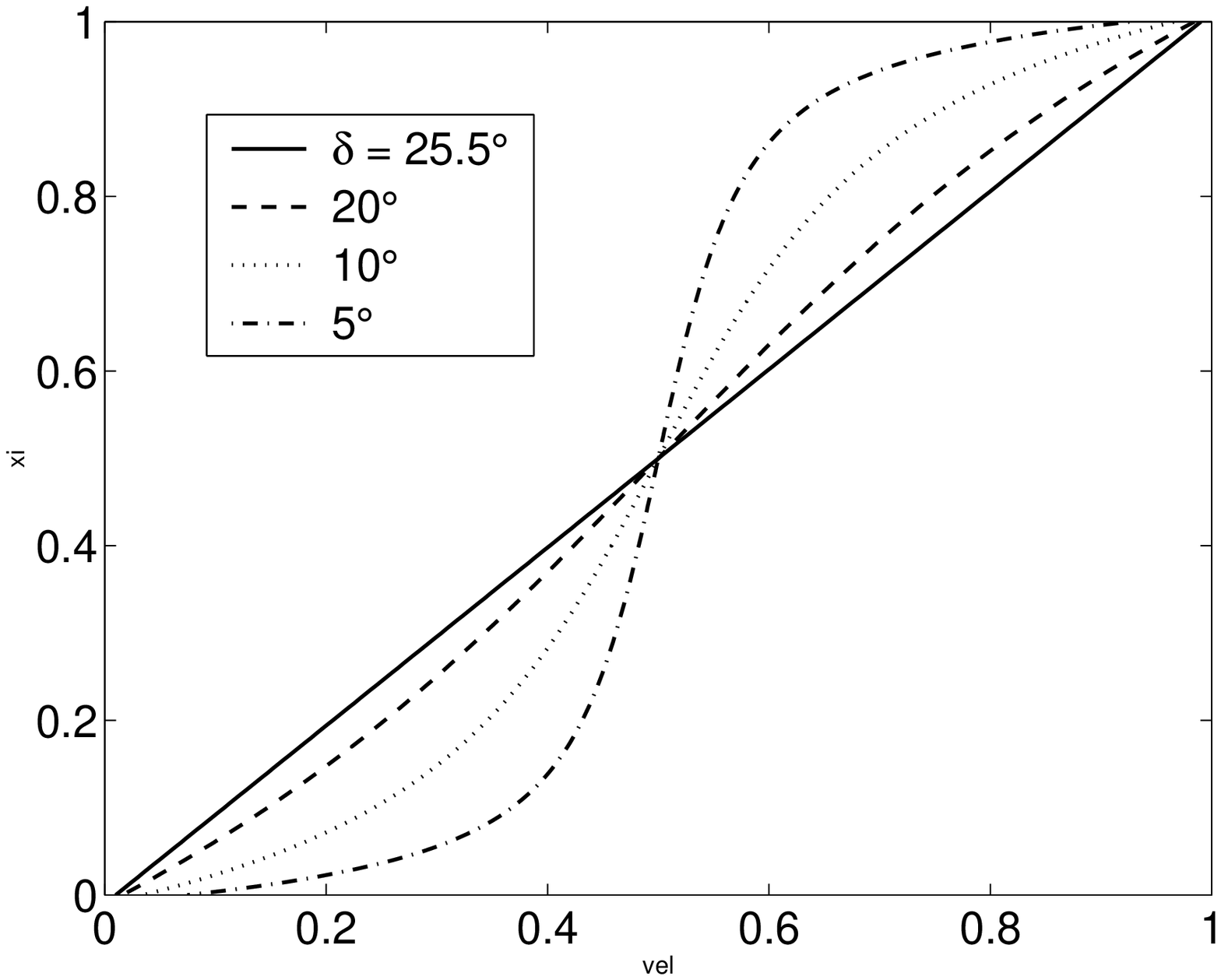}
    \hspace*{0.4in} \includegraphics[width= 0.45\textwidth]{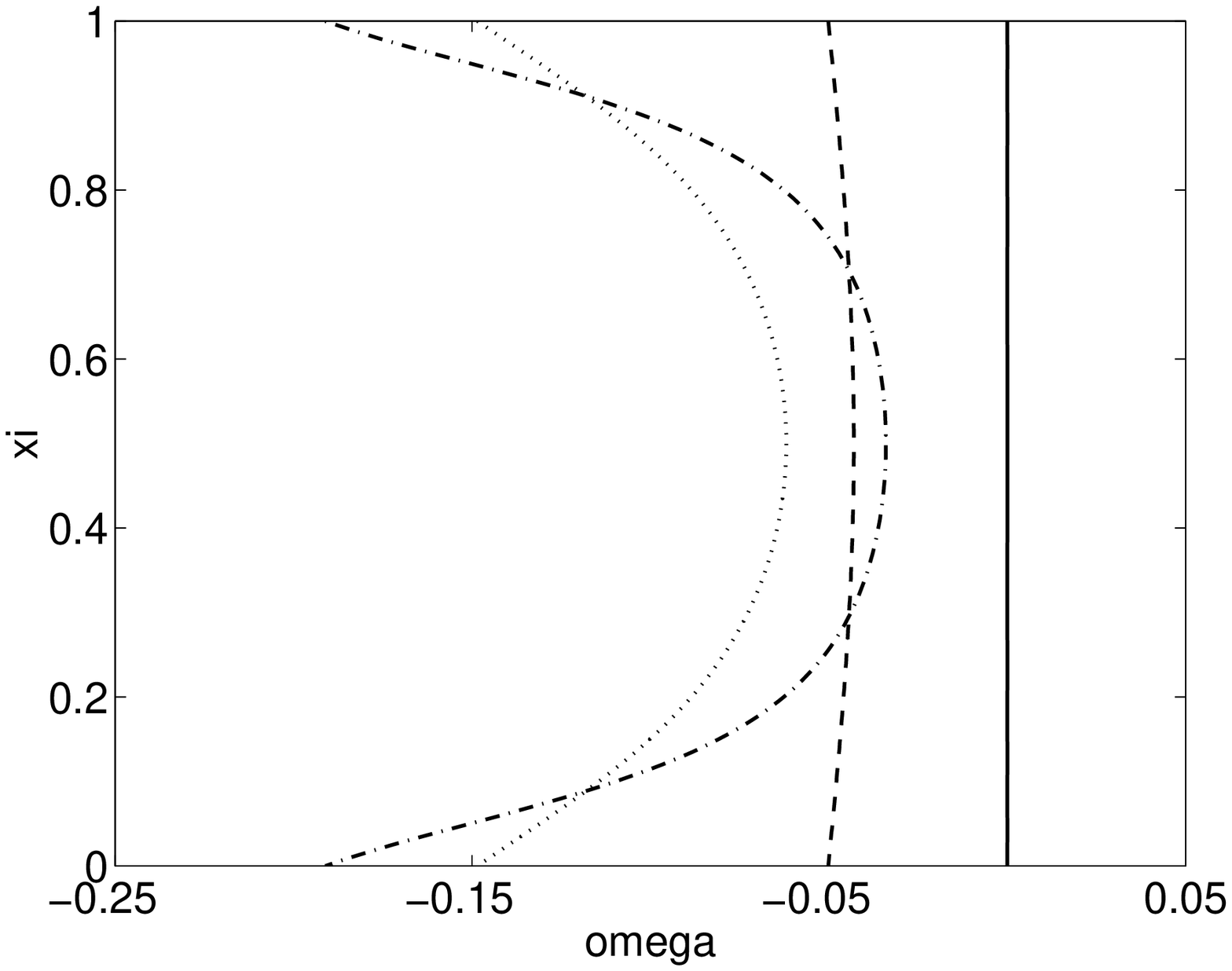}\\
    \vspace{0.5in}
    \hspace*{0.1in} (c) \hspace{2.2in} (d) \\
    \includegraphics[width= 0.45\textwidth]{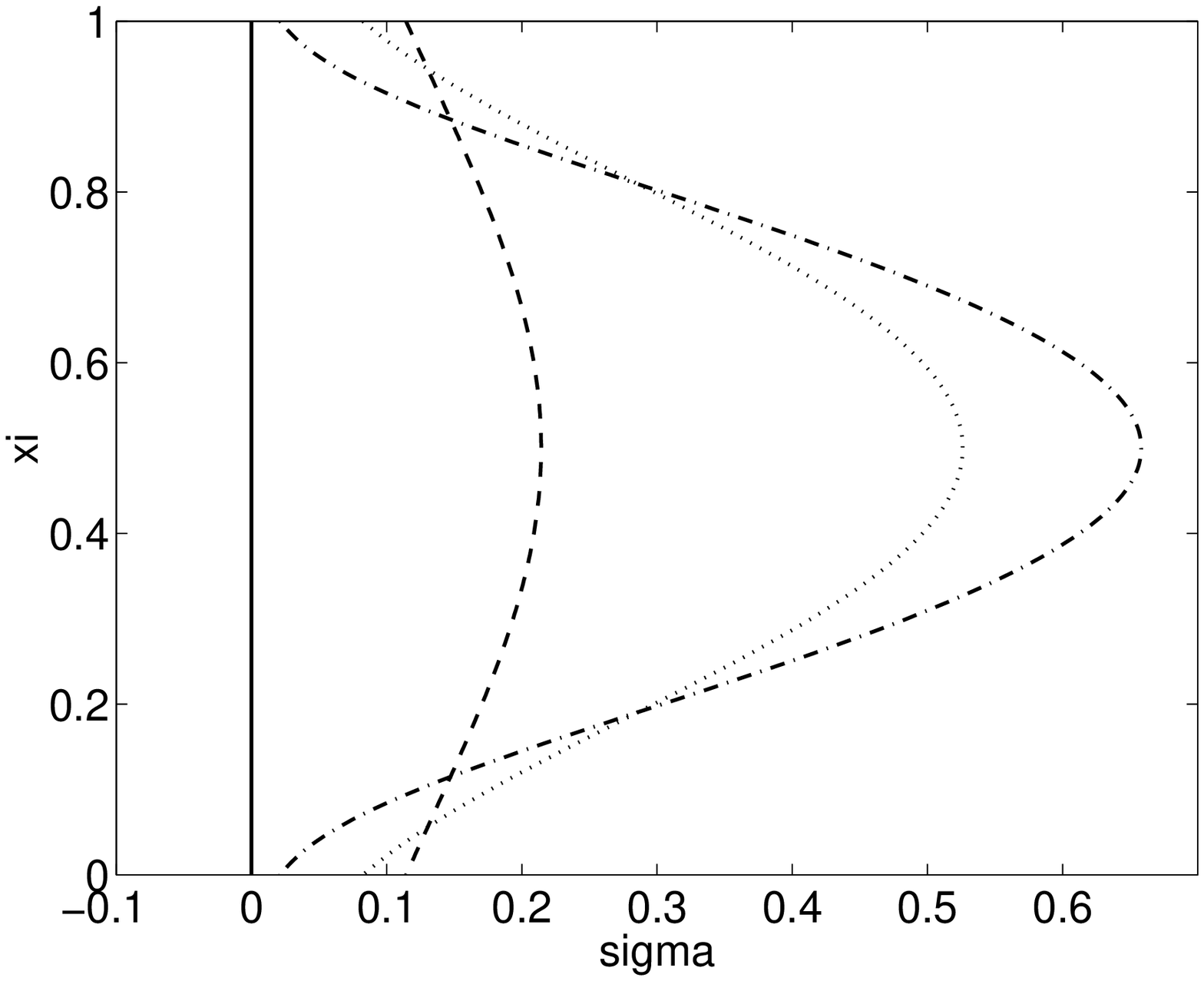}
    \hspace*{0.4in} \includegraphics[width=0.45\textwidth]{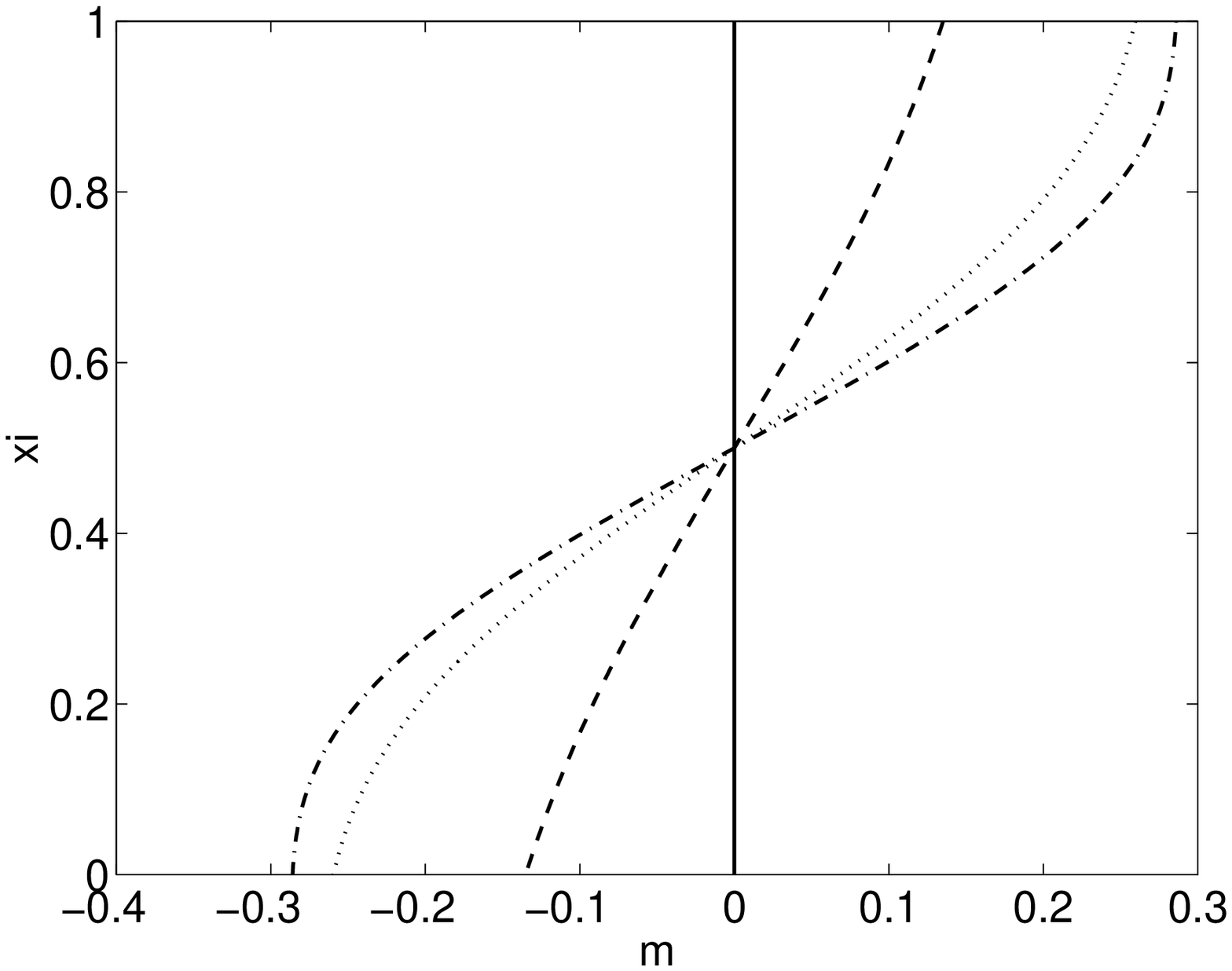}
    \caption{Results of the Cosserat model for shear between
identical walls in the absence of gravity, for different values of
the angle of wall friction $\delta$.  Panel (b) shows the
difference between $\omega$ and half the vorticity
($-1/2\,\textfrac{du}{d\xi}$), and panel (c) shows the asymmetry
in the shear stress.  The material shears uniformly when the walls
are fully rough (solid line). The thickness of the shear layer
decreases as the difference between $\sin\phi$ and $\tan\delta$
increases. Parameter values are: $\phi=28.5^\circ$,
$\varepsilon=0.04$, $L=10$, $A=1/3$ and
$K=0.5$.\label{fig-zero_g}}
  \end{center}
\end{figure}

       Figure~\ref{fig-zero_g} shows the velocity and stress profiles
for different values of the angle of wall friction $\delta$. The
velocity profile is linear for the case of fully rough walls
($\tan \delta = \sin \phi$), and  becomes curved as the difference
between $\sin\phi$ and $\tan\delta$ increases. Correspondingly,
the thickness of the shearing layer decreases. Panels (b) and (c)
illustrate the deviation of $\omega$ from half the vorticity and
the stress asymmetry, respectively. The choice of signs in
(\ref{eqn-nd_sigxy}) forces $\ol{\sigma}_{yx} - \ol{\sigma}_{xy}$
to be non-negative.

    When the angle of friction of the lower wall ($\delta_L$) differs
from that of the upper wall ($\delta$), the solution is no longer
symmetric about $\xi = 1/2$.  We assume without loss of generality
that $\delta_L > \delta$.  If the  friction boundary condition
\eqref{eqn-nd_bc2} is satisfied at the upper wall, $-
\ol{\sigma}_{yx}/\ol{\sigma}_{yy} = \tan \delta < \tan\delta_L$.
Hence \eqref{eqn-nd_bc2} is not satisfied at the lower wall.
Conversely, if \eqref{eqn-nd_bc2} is satisfied at the lower wall,
$- \ol{\sigma}_{yx}/\ol{\sigma}_{yy} = \tan \delta_L > \tan
\delta$ at the upper wall.  As this is not permissible within the
present framework, \eqref{eqn-nd_bc2} is used at the upper wall,
along with the velocity boundary condition \eqref{eqn-nd_bc3u}. At
the lower wall we use boundary condition \eqref{eqn-nd_bc4} for
the couple stress, and the no-slip condition \eqref{eqn-nd_bc5l}.

\begin{figure}
  \begin{center}
    \vspace{1in}
    \psfrag{xi}[br][][1][-90]{$\xi$}
    \psfrag{vel}[t][]{$u$}
    \psfrag{omega}[t][]{$\omega$}
    \psfrag{sigma}[t][]{$\ol{\sigma}_{xy}, \; \ol{\sigma}_{xy}$}
    \psfrag{m}[t][]{$m$}
    \hspace*{0.1in} (a) \hspace{2.2in} (b) \\
    \includegraphics[width= 0.45\textwidth]{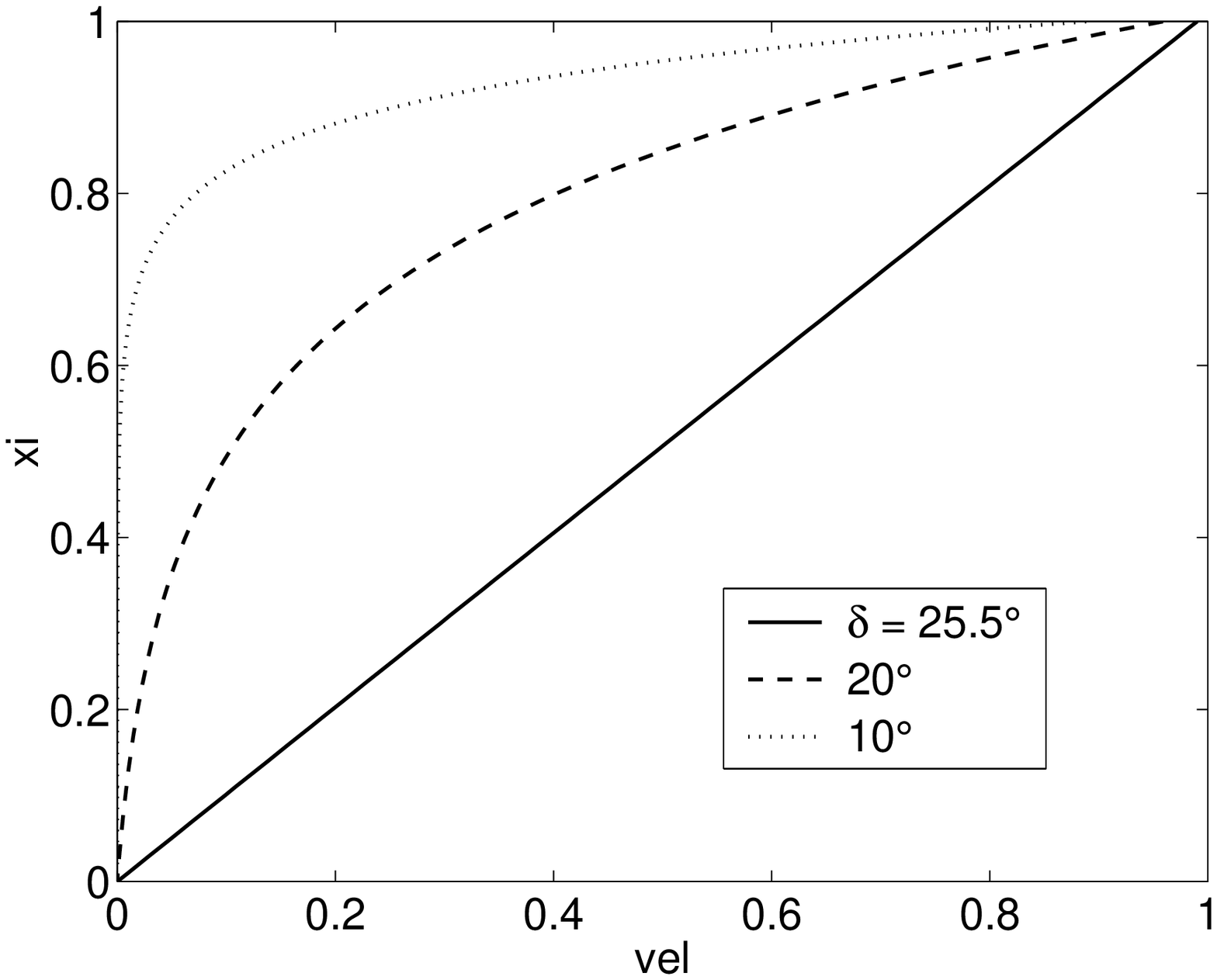}
    \hspace{0.4in} \includegraphics[width= 0.45\textwidth]{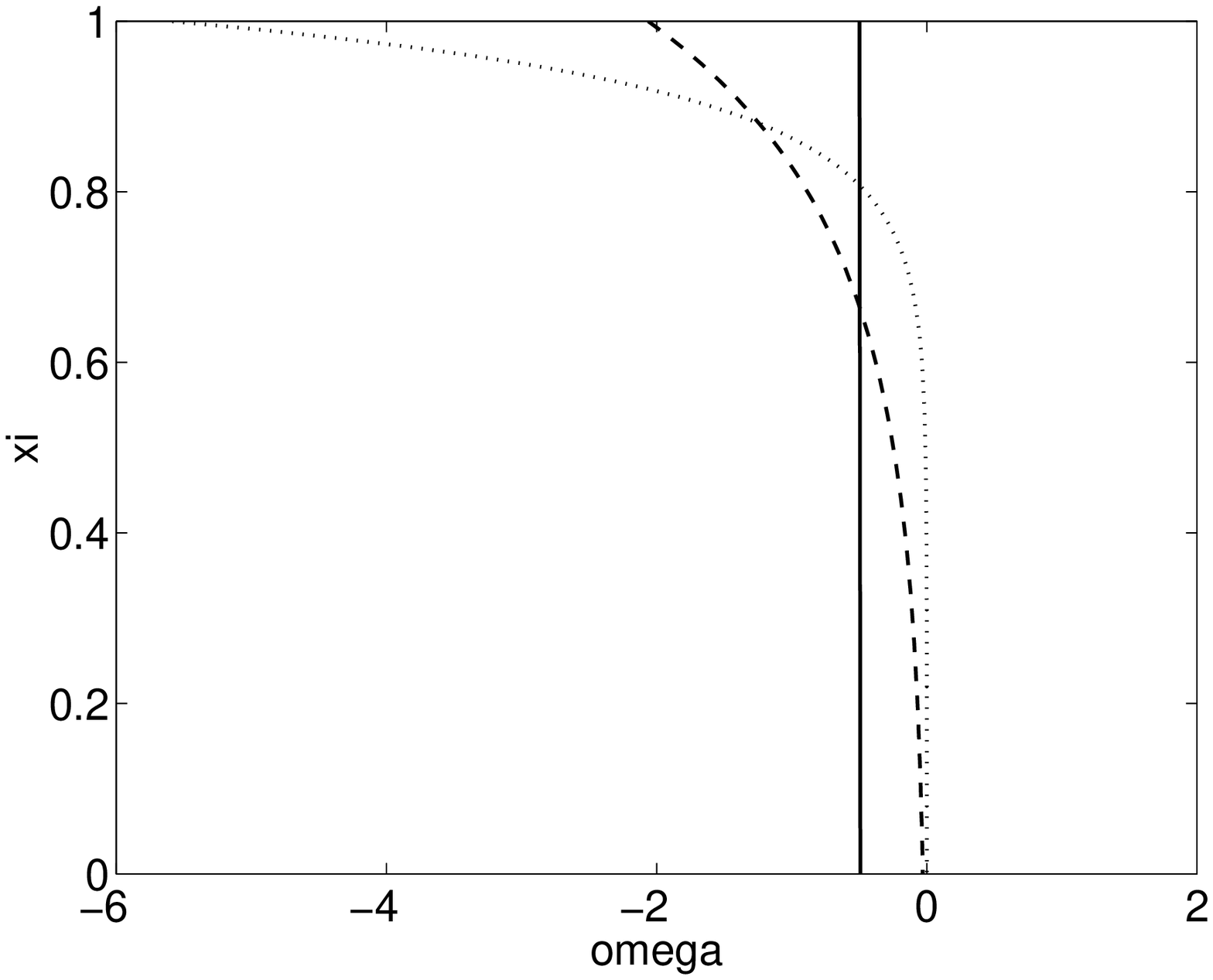} \\
%    \vspace{0.5in}
%    \hspace{1.8in} (c) \hspace{2.6in} (d)
%    \includegraphics[width= 0.45\textwidth]{sigma_aw.eps}
%    \hspace{0.2in} \includegraphics[width=0.45\textwidth]{m_aw.eps}
%    \caption{Profiles of the stress and kinematic fields for shear between
%    dissimilar walls in the absence of gravity.}
    \caption{The velocity and angular velocity profiles for plane shear in
the absence of gravity when the lower wall is rougher than the
upper wall, i.e.~$\delta_L > \delta$.  For this problem, the
stress is symmetric, the couple stress is constant (see eqn.\
\ref{eqn-eta_soln_aw_g0}), and the angular velocity is equal to
half the vorticity.  The material shears uniformly when the upper
wall is fully rough (solid line).  Other parameters are given in
the caption of figure~\ref{fig-zero_g}.\label{fig-zero_g_aw}}
\end{center}
\end{figure}

 Equations (\ref{eqn-nd_zangmom}) and \eqref{eqn-nd_bc4} imply that
$\textfrac{dm}{d\xi} = 0$ at $\xi=0$. It then follows from
\eqref{eqn-eta_g0} that all higher derivatives of $m$ vanish at
$\xi=0$. Hence, the couple stress is constant and the shear
stresses are equal,
\begin{subeqnarray}
    m & = & m_0 = (b^2 - a^2)^{1/2},\\
    \ol{\sigma}_{xy} & = & \ol{\sigma}_{yx} = -\tan\delta.
\label{eqn-eta_soln_aw_g0}
\end{subeqnarray}
Substitution of these
in \eqref{eqn-nd_fr1} yields
\[  \der{u}{\xi} = - 2 \omega.
\]
Thus, the stress tensor is symmetric and the angular velocity is
equal to half the vorticity, as in a classical continuum. However,
the couple stress is finite and the velocity field is uniquely
determined, in contrast to the predictions of a classical
frictional model. The velocity fields are readily obtained by
integrating \eqref{eqn-nd_fr1}-\eqref{eqn-nd_fr2}, resulting in
\begin{subequations}
\ba
    \omega & = & \omega_1 \, \exp(-k(1 - \xi)/\varepsilon),\\
    u & = & \Frac{-2 \omega_1 \varepsilon}{k} (\exp(-k(1 -
    \xi)/\varepsilon) - \exp(-k/\varepsilon)),
\label{eqn-soln_aw_vels}
\ea
where
\begin{equation}
    \omega_1 = -\,\varepsilon^{-1} (2/k + K - 2 e^{-k/\varepsilon}/k)^{-1}
\slabel{eqn-omega1_g0aw}
\end{equation}
is the angular velocity at the upper wall and
\begin{equation}
    k = m_0/(\alpha \, a).
\label{eqn-k_def}
\end{equation}
\end{subequations}
We see from equation \eqref{eqn-soln_aw_vels} that the shear layer
thickness increases with the roughness of the upper wall.  The
velocity profile becomes progressively linear as $\tan \delta
\rightarrow \sin \phi$ (or $k \rightarrow 0$). However, the
velocity profile in this limit still differs from the uniformly
shearing solution given in \eqref{vel_g0}, owing to the difference
in velocity boundary conditions at the lower wall in the two
cases. Profiles of the velocity and angular velocity (which equals
half the vorticity) fields are shown in figure~\ref{fig-zero_g_aw}
for three values of $\delta$.  The increasing localization of
shear near the upper wall as $\delta$ decreases is apparent.  In
this case, the velocity slip at the upper wall is nonzero in the
limit $\varepsilon \rightarrow 0$. This is in contrast to the case
of shear between fully rough walls, where the velocity slip
vanishes in this limit.

    When the Couette gap is large compared to the grain size,
i.e.~$\varepsilon \ll 1$, an approximate solution   may be
obtained by an asymptotic analysis, described in Appendix
\S\ref{appendix}. For shear between identical walls, the couple stress is
constant and the Cauchy stress is symmetric, i.e.
\begin{subeqnarray}
% define subequation numbers
  \gdef\thesubequation{\theequation \textit{a,b}}
  m = m_0 = (b^2 - a^2)^{1/2}, \quad \ol{\sigma}_{xy} = \ol{\sigma}_{yx} =
-\tan\delta,
\end{subeqnarray}
\returnthesubequation
 except in a boundary layer of thickness $
\sim \mathrm{O}(\varepsilon)$ near $\xi = 1/2$. In the boundary
layer, $m$ decreases from $m_0$ to zero at $\xi = 1/2$, but this
variation has no effect on the velocity fields (see
\S\ref{subsubsec-asymp_g0}). There is no boundary layer when one
of the walls is rougher.  The leading order velocity fields for
the two cases, given in \eqref{eqn-asymp_soln_vels_g0} and
\eqref{eqn-asymp_soln_vels_g0_aw}, may be combined and written as,
\addtocounter{equation}{-1}
\begin{subeqnarray}
% increment subequation counter by 2 to take it from a to c
\addtocounter{subequation}{2}
    \omega = \Frac{-(1 - u_0)}{\varepsilon (K + 2/k)}
\exp \left( -k(1 -     \xi)/\varepsilon  \right),\\
    u - u_0 = \Frac{ (1 - u_0) 2/k}{(K + 2/k)}\exp \left( -k(1 -
\xi)/\varepsilon  \right).
\label{eqn-asymp_vels}
\end{subeqnarray}
Here $u_0$ is the reference velocity, equal to $1/2$ (the velocity
at the symmetry axis) for shear between identical walls, and zero
(the velocity of the lower wall) for a rougher lower wall.  The
rapid decay of $\omega$ and $u$ with distance from the wall is
apparent.  Equation \eqref{eqn-asymp_vels} is not valid when the
walls are fully rough, as we have assumed in its derivation that
$k \gg \varepsilon$ (see \S\ref{subsubsec-asymp_g0}); as stated
earlier, all Cosserat effects vanish and the material shears
uniformly in the gap when the walls are fully rough.  The
asymptotic solution for the velocity fields
(\ref{eqn-asymp_vels}c-d) agrees well with the numerical solution
for shear between identical walls and for the case of a rougher
lower wall (figure~\ref{fig-asymp_g0}). The dependence of the
shear layer thickness on $\varepsilon$ is discussed in
\S\ref{sec-sl_thickness}.

\begin{figure}
  \begin{center}
    \vspace{1in}
    \psfrag{xi}[br][][1][-90]{$\xi$}
    \psfrag{vel}[t][]{$u$}
    \psfrag{m}[t][]{$m$}
    \hspace*{0.1in} (a) \hspace{2.2in} (b) \\
    \includegraphics[width= 0.45\textwidth]{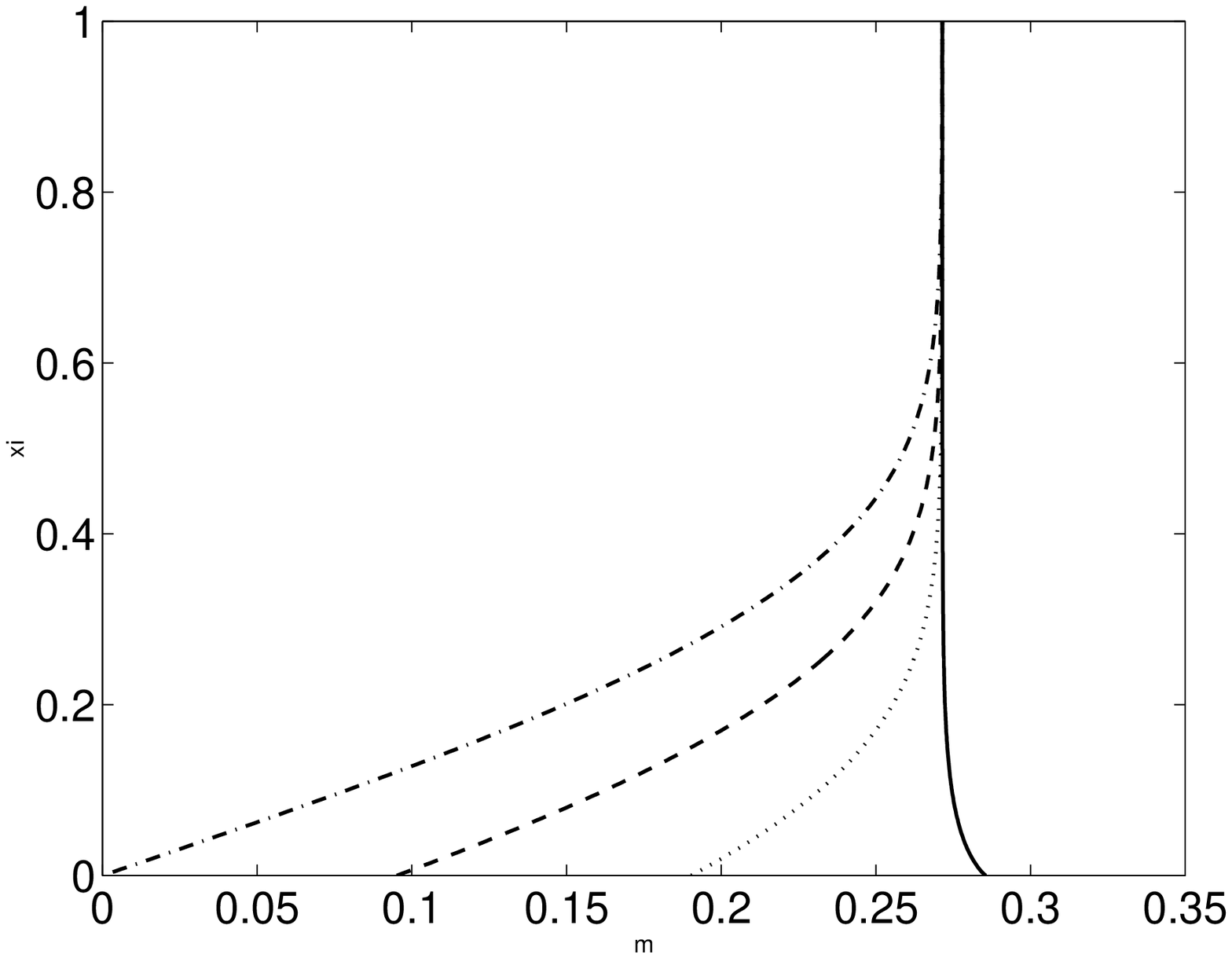}
    \hspace{0.2in} \includegraphics[width= 0.45\textwidth]{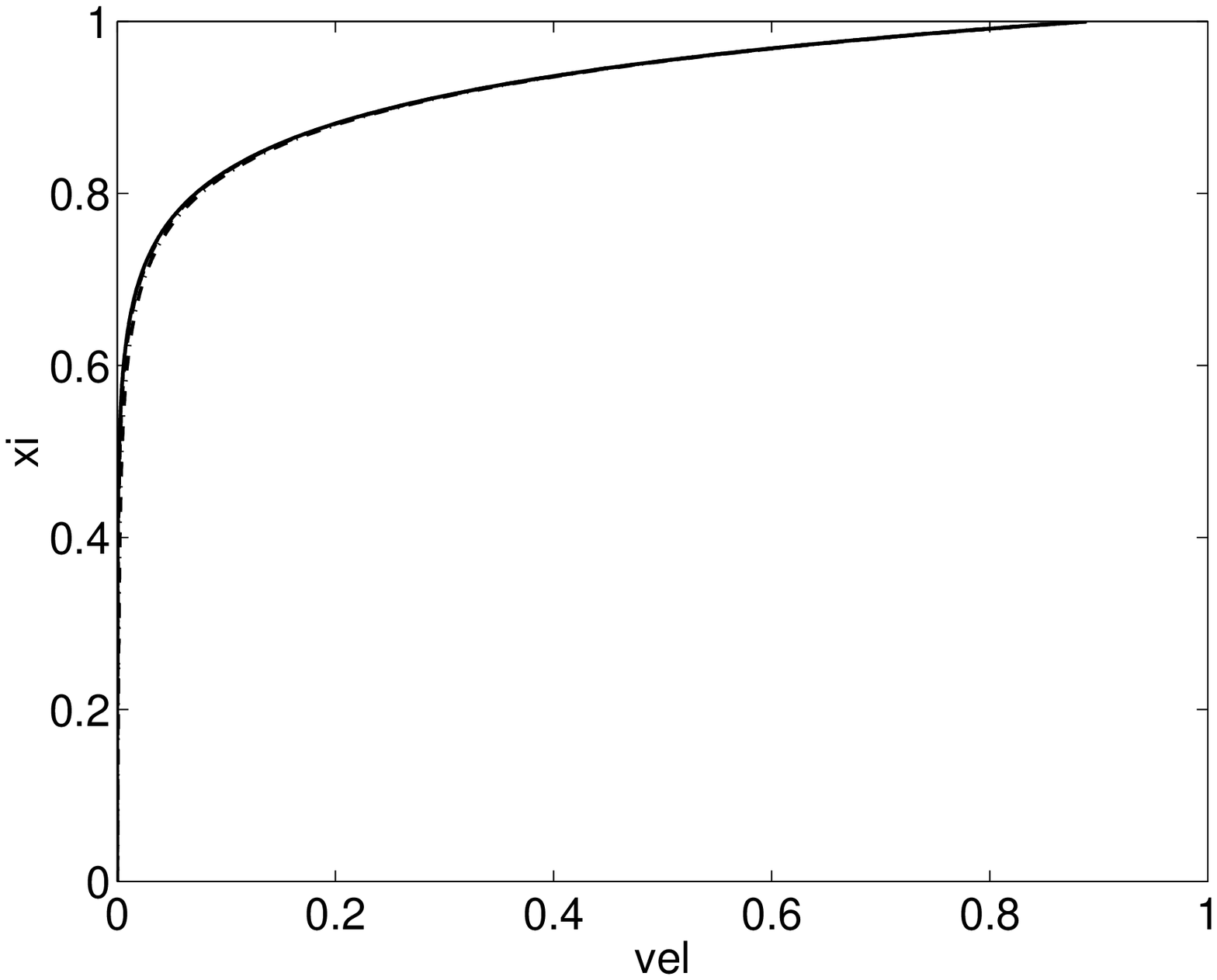}\\
    \caption{Effect of $m(0)$, the boundary condition for the couple
stress, on the (a) couple stress and (b) linear velocity fields,
for shear in the absence of gravity.  The curves are for four
values of $m(0)$ in the allowed range $0 < m(0) < b$; the solid
line is for $m(0) = b$. The velocity profiles for all the cases
are indistinguishable. Here $\delta = 10^\circ$, $\delta_L >
\delta$, and other parameters are given in the caption of
figure~\ref{fig-zero_g}.\label{fig-bc}}
\end{center}
\end{figure}

    As discussed in \S\ref{sec-BC}, we do not have rigourous basis
for the boundary condition \eqref{eqn-nd_bc4} for the case of a
rougher lower wall.  A study of the sensitivity of the model
predictions to varying the boundary condition is therefore in
order. The imposition of any condition for the angular momentum at
the lower wall is equivalent to specifying the value of the couple
stress there.  In particular, the boundary condition
\eqref{eqn-nd_bc4} is equivalent to $m(0) = m_0$, where $m_0$ is
defined by (\ref{eqn-eta_soln_aw_g0}a).  For the shear stress
$\ol{\sigma}_{xy}$ to be real, \eqref{eqn-eta_g0} implies that the
maximum value $m(0)$ can take is $b$, defined in
\eqref{eqn-a_b_def}; for $\omega$ and the velocity gradient to
increase in magnitude with distance from the lower wall (as is
physically reasonable) the flow rule
\eqref{eqn-nd_fr1}-\eqref{eqn-nd_fr2} dictates that the minimum
value $m(0)$ can take is zero. As shown in figure~\ref{fig-bc},
the value of $m(0)$ has a negligible effect on the velocity
profile. The reason for this result is that the couple stress
approaches a common asymptote as $\xi \to 1$ for all values of
$m(0)$ (see also \S\ref{subsubsec-asymp_g}). As the velocity and
angular velocity fields decay exponentially with distance from the
upper wall, with a decay length set by the couple stress near the
upper wall, the boundary condition for $m(0)$ is largely
unimportant in determining the velocity fields.

\subsection{Shear in a gravitational field}
\label{subsec_gravity}

    Let us first consider the case where both walls have the same
roughness, so that $\delta_L = \delta$.  Equations
\eqref{eqn-nd_xmom} and \eqref{eqn-nd_ymom} imply
\begin{equation}
\ol{\sigma}_{yx} = \mbox{constant}, \;\;
\der{\ol{\sigma}_{yy}}{\xi} = - B \, \nu. \label{sigxyg}
\end{equation}
Hence the ratio $\textfrac{|\ol{\sigma}_{yx}|}{\ol{\sigma}_{yy}}$
decreases with distance from the upper wall, and
\eqref{eqn-nd_bc2} cannot be satisfied at the lower wall if it
holds at the upper wall. Therefore, for the same reasons given in
\S\ref{subsec-zerog} (for the case $\delta < \delta_L$), the
boundary conditions are \eqref{eqn-nd_bc1}, \eqref{eqn-nd_bc2} and
\eqref{eqn-nd_bc3u} at the upper wall ($\xi = 1$), and
\eqref{eqn-nd_bc4} and \eqref{eqn-nd_bc5l} at the lower wall ($\xi
= 0$).

       As the normal stress $\ol{\sigma}_{yy}$ increases with distance from
the upper wall, \eqref{eqn-ns} implies that the solids fraction
$\nu$ also increases. However, in the regime of high solids
fraction, a moderate change in $\sigma_{c}$ causes only a small
change in $\nu$. As gravitational compaction alone is usually
sufficient to bring the solids fraction to a level close to
maximum packing, we treat the material as incompressible, and the
mean stress at critical state $\sigma_c$ as an independent field.
This approximation is used only to simplify the analysis; the
 variation in the body force can be easily be accounted for.

\begin{figure}
  \begin{center}
    \psfrag{xi}[br][][1][-90]{$\xi$}
    \psfrag{vel}[t][]{$u$}
    \psfrag{omega}[t][]{$\omega + 1/2\der{u}{\xi}$}
    \psfrag{sigma}[t][]{$\ol{\sigma}_{yx} - \ol{\sigma}_{xy}$}
    \psfrag{m}[t][]{$m$}
    \vspace{1in}
    \hspace*{0.1in} (a) \hspace{2.2in} (b) \\
    \includegraphics[width= 0.45\textwidth]{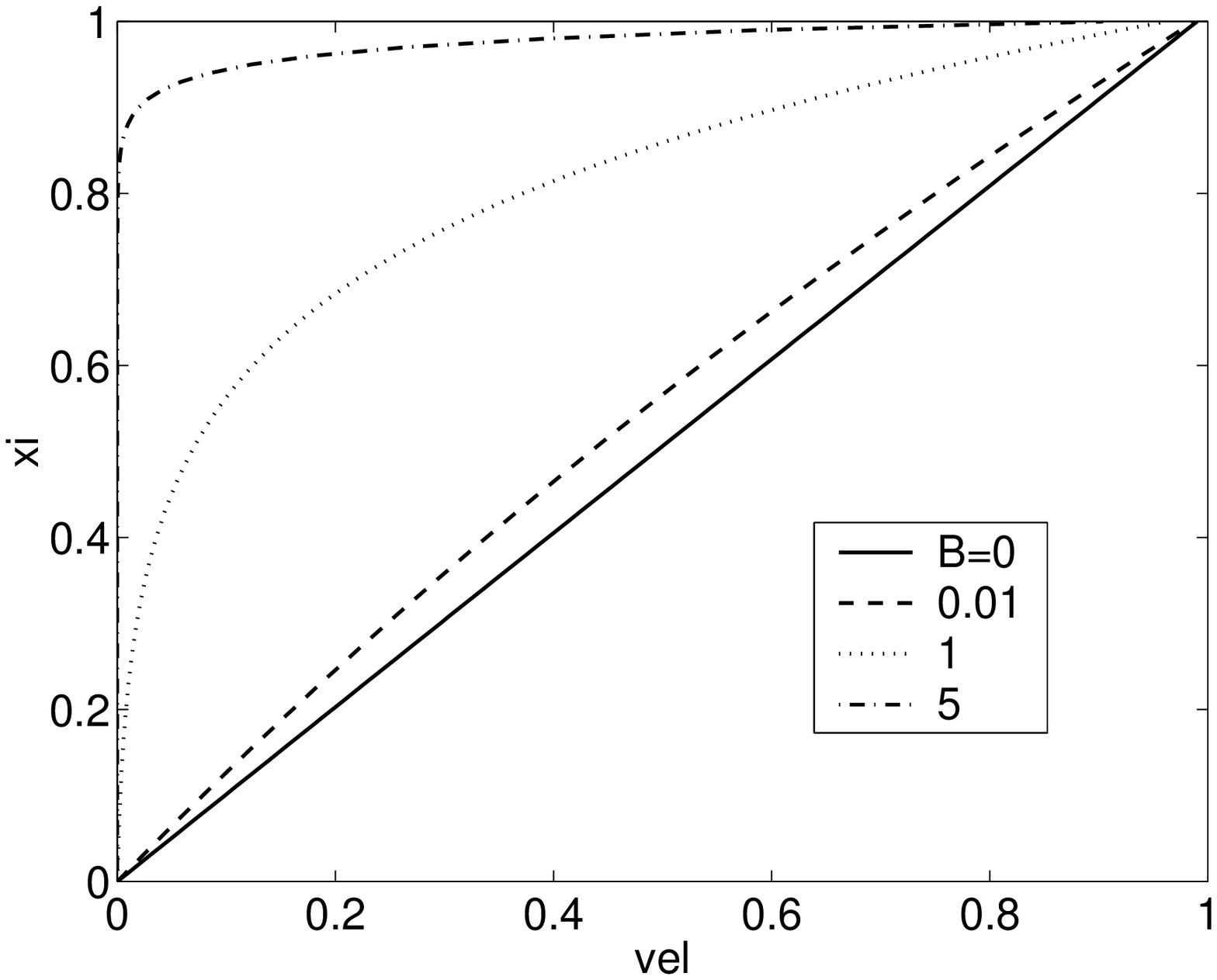}
    \hspace{0.2in} \includegraphics[width= 0.45\textwidth]{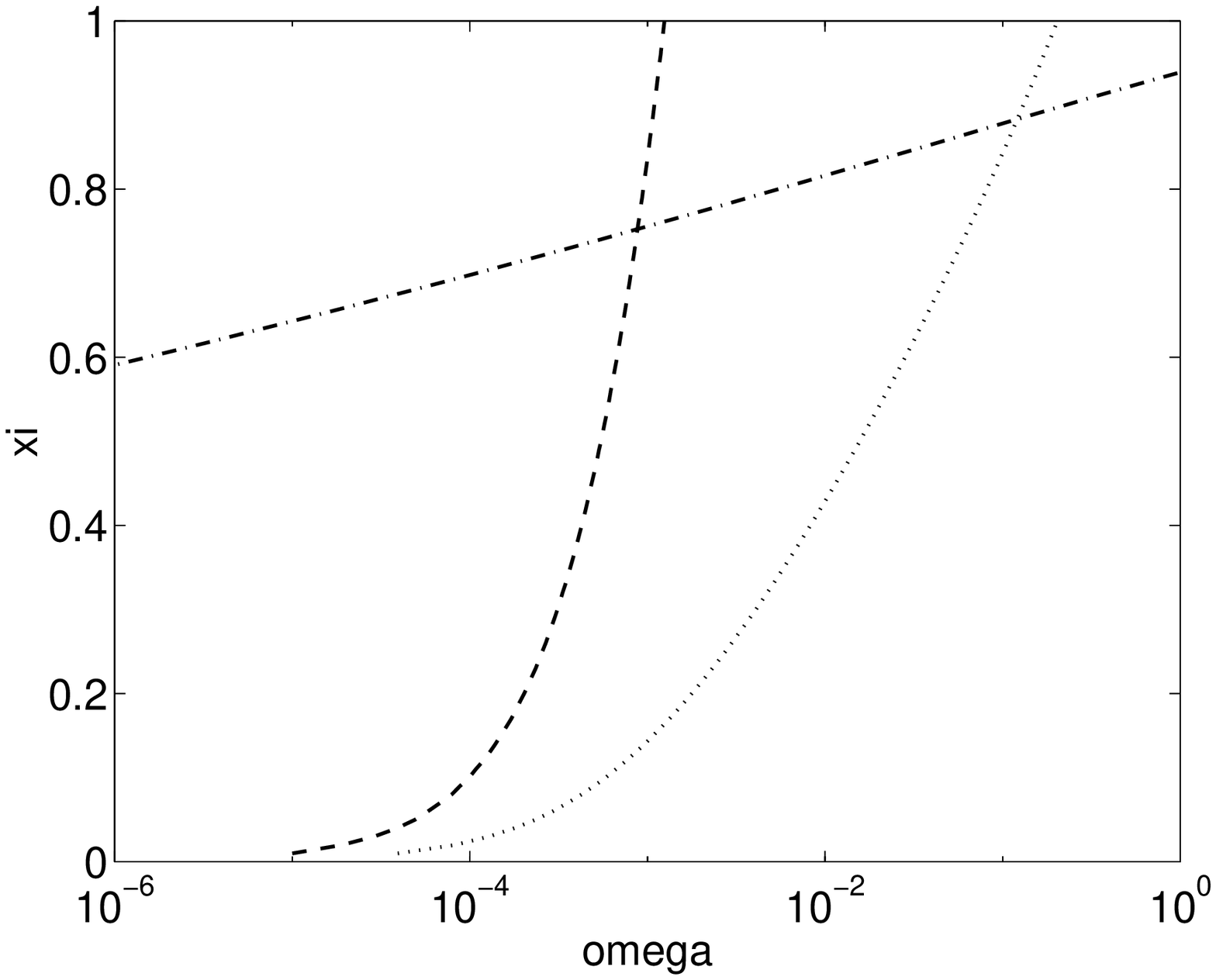} \\
    \vspace{0.5in}
    \hspace*{0.1in} (c) \hspace{2.2in} (d) \\
    \includegraphics[width= 0.45\textwidth]{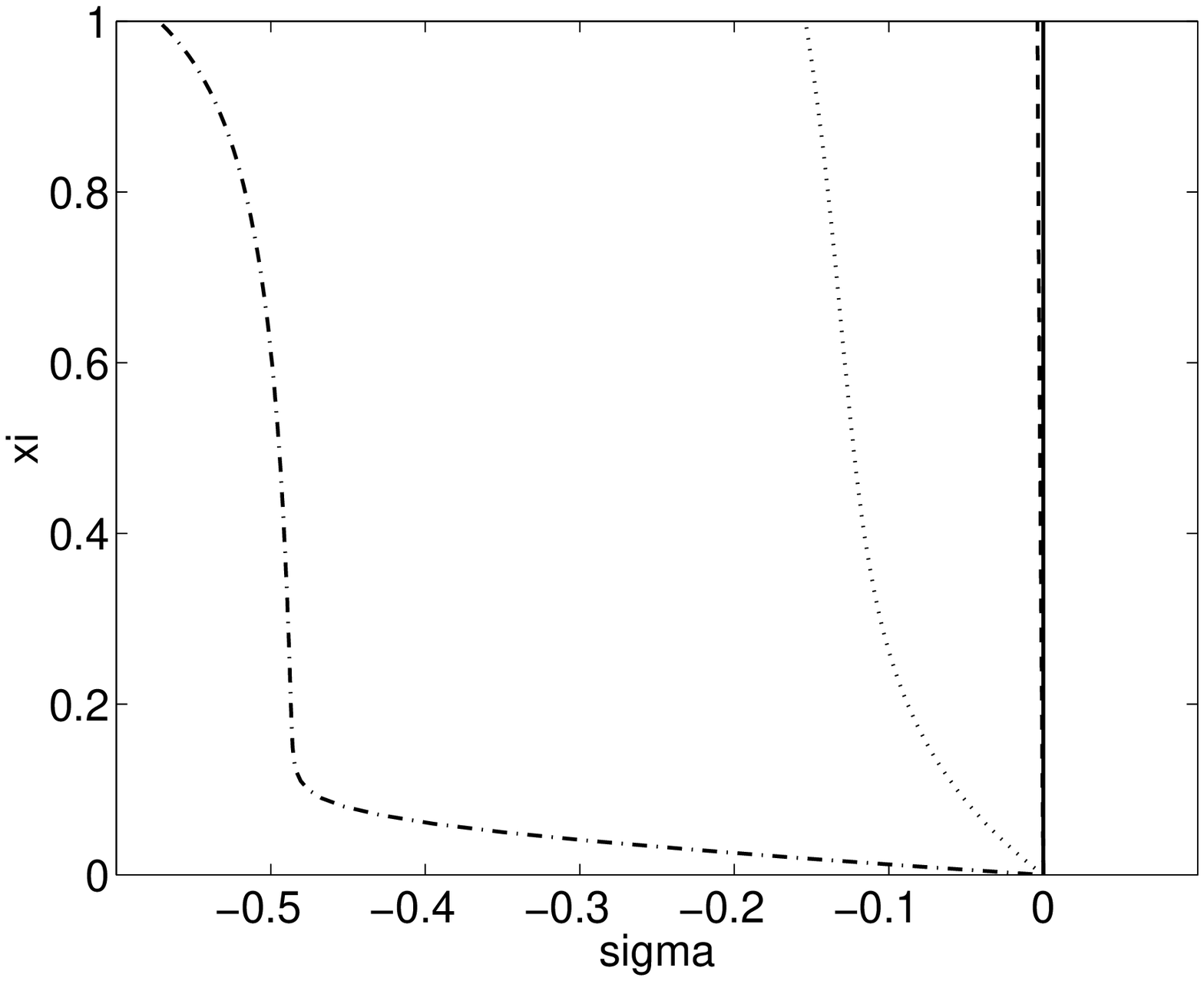}
    \hspace{0.2in} \includegraphics[width=0.45\textwidth]{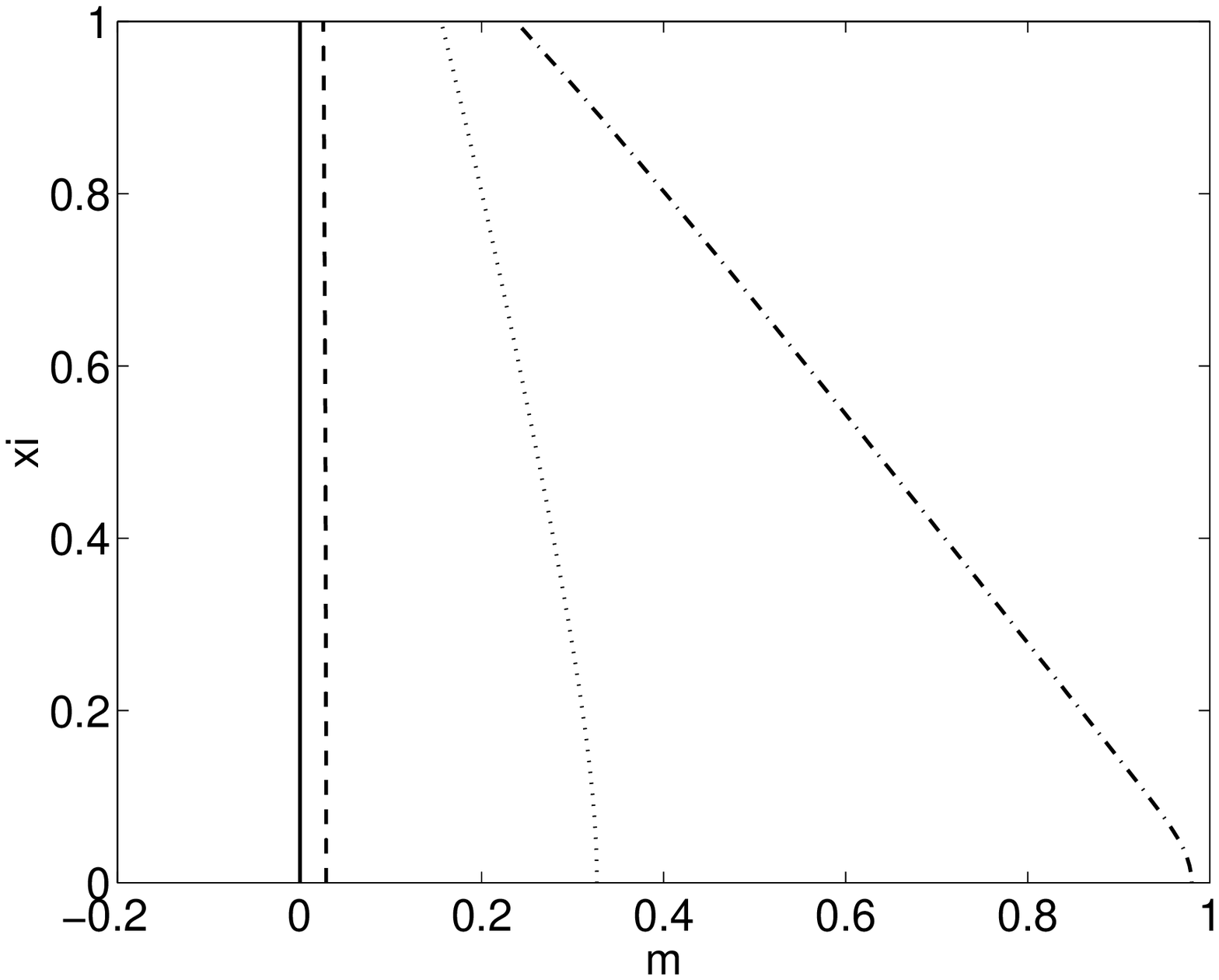}
    \caption{Predictions of the model for plane shear between
fully rough walls under gravity.  Panel (b) shows the difference
between $\omega$ and half the vorticity, and panel (c) shows the
asymmetry in the shear stress.  The solid line is absent in panel
(b) because $\omega + 1/2\, du/d\xi = 0$ for this case.  The
parameter $B \equiv (\rho_p g H)/N$, indicates the strength of the
gravitational body force. Other parameters are as in the caption
of figure~\ref{fig-zero_g}.\label{fig-finite_g}}
  \end{center}
\end{figure}

    Figure~\ref{fig-finite_g} shows results for shear between fully
rough walls under gravity, for four values of the parameter $B$.
The solution for $B=0$ corresponds to the uniform shear solution
in the absence of gravity.  As $B$ increases, the velocity profile
becomes curved, and shear is localized in a thin layer adjacent to
the upper wall for large $B$.  While the couple stress increases
with distance from the upper wall, other Cosserat effects, such as
the difference between $\omega$ and half the vorticity and the
asymmetry in the Cauchy stress are maximum at the upper wall.

    For large Couette gap, the asymptotic analysis in
\S\ref{subsubsec-asymp_g} provides an approximate solution.  If
the walls are not fully rough, the leading order solution  for the
velocity fields is identical to that for gravity-free shear, given by
\eqref{eqn-asymp_vels} with $u_0 = 0$.  If the walls are fully
rough, there is a boundary layer of thickness $ \sim
\varepsilon^{2/3}$ near the upper wall within which $m$ deviates
from $m_0$.  Consequently, the shear layer is thinner than in the zero
gravity case(see \S\ref{sec-sl_thickness}).  The leading order
asymptotic solution for the couple stress and velocity fields are
compared with the numerical solution in
figure~\ref{fig-asymp_g_c}, showing good agreement. As a small value of
$\varepsilon$ has been used, the slip velocity at the upper wall is
negligible.

    If $\delta_L$ is sufficiently smaller than $\delta$, the friction
boundary condition \eqref{eqn-nd_bc2} can be satisfied at the
lower wall. Assuming that the density of the material remains
constant across the layer (see above), it is seen that
\eqref{eqn-nd_bc2} holds at the lower wall if $\tan\delta_L \le
\tan\delta/(1+B\nu)$, and at the upper wall if $\tan\delta_L \ge
\tan\delta/(1+B\nu)$.  The shear layer is adjacent to the lower
wall in the former case, and adjacent to the upper wall (as in
figure \ref{fig-finite_g}.

\section{Cylindrical Couette flow}
\label{sec-cyl_couette}

    The cylindrical Couette cell is a common device for making rheological
measurements.  The material is confined between two vertical
coaxial cylinders of radii $R_i$ and $R_i + H$.  Typically, the
inner cylinder is rotated at constant speed, while the outer
cylinder is kept stationary.

\begin{figure}
  \begin{center}
    \vspace{1in}
    \psfrag{o}[t][]{\hspace*{-1ex}$\omega$}
    \psfrag{H}{$H$}
    \psfrag{g}[t][]{$\,\,g$}
    \psfrag{t}{$\theta$}
    \psfrag{R}[t][r]{$\;\;R_i$}
    \psfrag{r}{$r$}
    \psfrag{z}{$z\,\,$}
\includegraphics[width=0.45\textwidth]{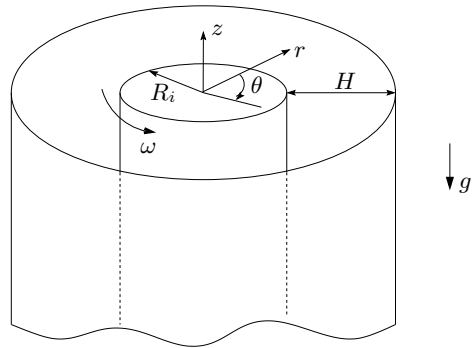}
\caption{Schematic figure of cylindrical Couette flow.  The outer
cylinder is  stationary and the inner cylinder rotates at
constant angular velocity $\omega$.\label{fig-schem_cyl_couette}}
\end{center}
\end{figure}

    Using the cylindrical coordinates indicated in figure
\ref{fig-schem_cyl_couette}, the velocity field for steady
axisymmetric flow is assumed to be of the form
\begin{equation}
v_r = 0, \quad v_z = 0, \quad v_{\theta} = v_{\theta}(r).
\label{eqn-vel_cyl}
\end{equation}
The only non-vanishing component of the vorticity is in the $z$-
direction, and we expect the same for the intrinsic angular
velocity,
\begin{equation}
\omega_{r} = \omega_{\theta} = 0, \quad \omega_{z} = \omega_z(r).
\label{eqn-omega_cyl}
\end{equation}
As in the case of plane shear (see \S\ref{sec-plane_shear}), the
diagonal components of \eqref{eqn-flowrule} imply equality of
normal stresses,
\begin{equation}
\sigma_{rr} = \sigma_{\theta \theta}= \sigma_{zz} = \sigma_{c}
(\nu) \label{eqn-ns_cyl}
\end{equation}
It also follows from the flow rule \eqref{eqn-flowrule} that all
shear stresses except $\sigma_{r \theta}$ and $\sigma_{\theta r}$
vanish, and all couple stresses except $M_{rz}$ vanish.

    We note as an interesting aside that, as $\sigma_{rz} = 0$, there is
no shear stress in the vertical direction on the walls of the
Couette cell, regardless of either the wall roughness or the
height of the material in the cell. In contrast, it is well known
that the shear stress at the walls supports a part of the weight of a
\emph{static} granular column, leading to the so-called Janssen
saturation of the stress with distance from the upper surface
\citep[p.~84]{janssen95,nedderman92}. While clear evidence as to
whether or not this feature is preserved during flow is lacking,
the data of \citet{tardos98} appear to be inconsistent with the
Janssen solution. Nevertheless, measurements of $\sigma_{r z}$
in cylindrical Couette flow have not been reported in the
literature, and such measurements would provide a valuable check
on the model predictions.

    We introduce the dimensionless variables
\begin{equation}
\bea{c}
    u = \Frac{v_{\theta}}{V_i} \, , \quad \omega =
    \Frac{\omega_z H}{V_i} \, , \quad \xi = \Frac{r - R_i}{H}
    \, , \quad \zeta = \Frac{z}{H} \, , \\
    \ol{\sigma}_{ij} = \Frac{\sigma_{ij}}{\rho_p \, g \, H} \, , \quad
    \ol{\sigma}_c = \Frac{\sigma_c}{\rho_p \, g H} \, , \quad m =
    \Frac{1}{L\sqrt{2(A+1)}} \, \Frac{M_{rz}}{\rho_p \, g H d_p},
\label{eqn-dimless_vars_cyl}
\eea
\end{equation}
 where $R_i$ and $V_i$ are the radius and the velocity of the inner
cylinder, respectively.  The yield condition \eqref{eqn-yc1} then
reduces to
\begin{equation}
(\ol{\sigma}_{r \theta}^{2} + \ol{\sigma}_{\theta r}^{2}) + 2 A
\ol{\sigma}_{\theta r} \ol{\sigma}_{r \theta} + 4(A+1)^2m^2 =
2(A+1)(\ol{\sigma}_{c}\sin\phi)^{2}.
\label{eqn-nd_yc_cyl}
\end{equation}

    As in \S\ref{sec-plane_shear}, the remaining components of the flow
rule, after elimination of the factor $\lambdadot/\tau$,  yield
\begin{eqnarray}
  \der{u}{\xi} -\Frac{u}{(\ol{R}_i + \xi)} & = &
- (\omega + \Frac{u}{(\ol{R}_i + \xi)}) \Frac{(A + 1)\,
(\ol{\sigma}_{\theta r} + \ol{\sigma}_{r
\theta})}{(\ol{\sigma}_{\theta r} +
A \ol{\sigma}_{r \theta})}, \label{eqn-nd_fr1_cyl} \\
\varepsilon \alpha \der{\omega}{\xi} & = & - (\omega +
\Frac{u}{(\ol{R}_i + \xi)}) \Frac{2(A + 1) \, m}{
(\ol{\sigma}_{\theta r} + A \ol{\sigma}_{r \theta})},
\label{eqn-nd_fr2_cyl}
\end{eqnarray}
which determine the variation of the velocity fields $u$ and
$\omega$. Here, $\ol{R}_i \equiv R_i/H$, and $\varepsilon \equiv
d_p/H$ is the ratio of the particle diameter to the Couette gap.
An important point to note is that we must seek a solution in
which the couple stress is negative, so the magnitude of $\omega$
(and the vorticity) decrease with distance from the inner
cylinder, as is physically reasonable.

    The mass balance is identically satisfied, and the balances of linear
and angular momentum assume the form
\ba
\pder{\ol{\sigma}_c(\nu)}
{\xi} & = & \Frac{C \, \nu \, u^2}{(1 + \xi/\ol{R}_i)},
\label{eqn-nd_rmom_cyl}\\
\pder{\ol{\sigma}_{r \theta}}{\xi} + \Frac{\ol{\sigma}_{r \theta}
+ \ol{\sigma}_{\theta r}}{(\ol{R}_i  + \xi)} & = & 0 ,
\label{eqn-nd_thetamom_cyl}\\
\pder{\ol{\sigma}_c(\nu)}{\zeta} & = & - \nu ,  \label{eqn-nd_zmom_cyl}\\
\varepsilon \left( \pder{m}{\xi} + \Frac{m}{(\ol{R}_i  + \xi)}
\right) & = & \ol{\sigma}_{r \theta} - \ol{\sigma}_{\theta r}
\label{eqn-nd_zangmom_cyl},
\ea
where $C = V_i^2/(R_i g)$ is the Froude number. Note that
\eqref{eqn-nd_thetamom_cyl} differs from form given in most books
on fluid mechanics, where the symmetry of the stress tensor is
implicitly assumed.

    We now show that the forms assumed for the velocity and
angular velocity fields in
\eqref{eqn-vel_cyl}-\eqref{eqn-omega_cyl} lead to an
inconsistency. Equations \eqref{eqn-nd_rmom_cyl} and
\eqref{eqn-nd_zmom_cyl} imply that $\nu = \nu(\xi,\zeta)$. Hence
the yield condition \eqref{eqn-nd_yc_cyl} and the momentum balance
\eqref{eqn-nd_thetamom_cyl} imply that $\ol{\sigma}_{r \theta}$,
$\ol{\sigma}_{\theta r}$, and therefore the right-hand sides of
\eqref{eqn-nd_fr1_cyl} and \eqref{eqn-nd_fr2_cyl}, depend on $\xi$
and $\zeta$. However, their left-hand sides have been assumed to
depend only on $\xi$. Thus it appears that we must relax some of
the assumptions made above by allowing variation of $u$ and
$\omega$ with $\zeta$, i.e.~in the vertical direction, and also to
allow a non-zero $\omega_r$. This however leads to a large set of
coupled partial differential equations whose solution appears to
be a formidable task. This is contrary to our purpose of
developing simple solutions that bring out the qualitative
features of our model. However, it is possible construct a simple
approximate solution, as described below.

    For slow flows, the centripetal acceleration is small compared
to the acceleration due to gravity, i.e.~$C \ll 1$. We may
therefore seek a solution as a regular perturbation in $C$,
i.e.~$m = m^{(0)} + C m^{(1)} + C^2 m^{(2)} + \ldots$. For $C \ll
1$, the exact solution does not differ significantly from the
leading order solution of $\mathrm{O}(C^0)$, except near the ends
of the Couette cell where additional boundary conditions have to
be satisfied. We therefore determine only the leading order
solution below, leaving the higher order corrections to a later
investigation.

    At $\mathrm{O}(C^0)$, it follows from \eqref{eqn-nd_rmom_cyl}
that $\ol{\sigma}_c$ is not a function of the radial position
$\xi$, and hence $\ol{\sigma}_c = - \nu \,\zeta$ (the normal
stress vanishes at the upper free surface). When this is
substituted in the equations \eqref{eqn-nd_fr1_cyl} and
\eqref{eqn-nd_fr2_cyl} to determine the velocity fields, there is
no longer an inconsistency, and hence the simple forms of the
velocity fields assumed in
\eqref{eqn-vel_cyl}-\eqref{eqn-omega_cyl} are valid; in other
words, when scaled by the hydrostatic head, the stress is
independent of $\zeta$.  We now replace the definitions of the
dimensionless stress and couple stress in
\eqref{eqn-dimless_vars_cyl} by the following:
\begin{equation}
\ol{\sigma}_{ij} = \Frac{\sigma_{ij}}{\rho_p \, g \, H \,
\ol{\sigma}_c}, \quad m = \Frac{M_{rz}}{\rho_p \, g \, H \,
\ol{\sigma}_c \, d_p \, L\sqrt{2(A+1)} }.
\label{newdimless}
\end{equation}
Equations \eqref{eqn-nd_thetamom_cyl} and \eqref{eqn-nd_zmom_cyl}
remain unchanged even after the above definitions.  We note that
our analysis is not applicable at the free surface, where the
stresses vanish.  To simplify the analysis, we assume as in
\S\ref{subsec_gravity} that the material is incompressible and
treat $\ol{\sigma}_c$ as an independent field.

    Consider the case where both walls have the same roughness.
Referring to figure \ref{fig-schem_cyl_couette}, $\ol{\sigma}_{r
\theta} < 0$ at the inner cylinder ($\xi = 0$).  If it is assumed
that $\ol{\sigma}_{\theta r}$ has the same sign as $\ol{\sigma}_{r
\theta}$, (\ref{eqn-nd_thetamom_cyl}) implies that $\ol{\sigma}_{r
\theta}$ decreases in magnitude as $\xi$ increases.  (Our results
are consistent with this assumption.)  As \eqref{eqn-nd_rmom_cyl} and
\eqref{newdimless} imply that $\ol{\sigma}_{rr}$ is a
constant, the ratio $|\ol{\sigma}_{r \theta}|/\ol{\sigma}_{rr}$
decreases with distance from the inner cylinder. Therefore, the
friction boundary condition applies at the inner cylinder, but
cannot be satisfied at the outer cylinder. This situation is
identical to that at the lower wall for the case of plane shear
under gravity.  We therefore follow the arguments outlined in
\S\ref{subsec_gravity} and adopt boundary conditions
\begin{equation}
    \ol{\sigma}_{r \theta} = \ol{\sigma}_{\theta r}, \;\;
u = 0 \label{eqn-nd_bc1_cyl}
\end{equation}
at the outer cylinder ($\xi=1$). At the inner cylinder ($\xi=0$),
we use the friction and slip boundary conditions
\begin{equation}
 \ol{\sigma}_{r \theta} = - \tan \delta, \;\; u =
-1 - \varepsilon K \omega, \label{eqn-nd_bc2_cyl}
\end{equation}
where we have the inner cylinder rotating in the anti-clockwise
direction (see figure \ref{fig-schem_cyl_couette}). In the limit
$\ol{R}_i \to \infty$, the first of \eqref{eqn-nd_bc2_cyl} holds
at the outer cylinder also. Hence the velocity boundary condition
$u = \varepsilon K \omega$ is applied at the outer wall.  This
corresponds to plane shear between walls of equal roughness in the
absence of gravity, discussed in \S\ref{subsec-zerog}.

\begin{figure}
  \begin{center}
%    \psfrag{inf}[t][]{$\infty$}
    \psfrag{xi}[br][][1][-90]{$\xi$}
    \psfrag{vel}[t][]{$u$}
    \psfrag{omega}[t][]{$\omega + 1/2\der{u}{\xi}$}
    \psfrag{sigma}[t][]{$\ol{\sigma}_{r \theta} - \ol{\sigma}_{\theta r}$}
    \psfrag{m}[t][]{$m$}
    \vspace{1in}
    \hspace*{0.1in} (a) \hspace{2.2in} (b) \\
    \includegraphics[width= 0.45\textwidth]{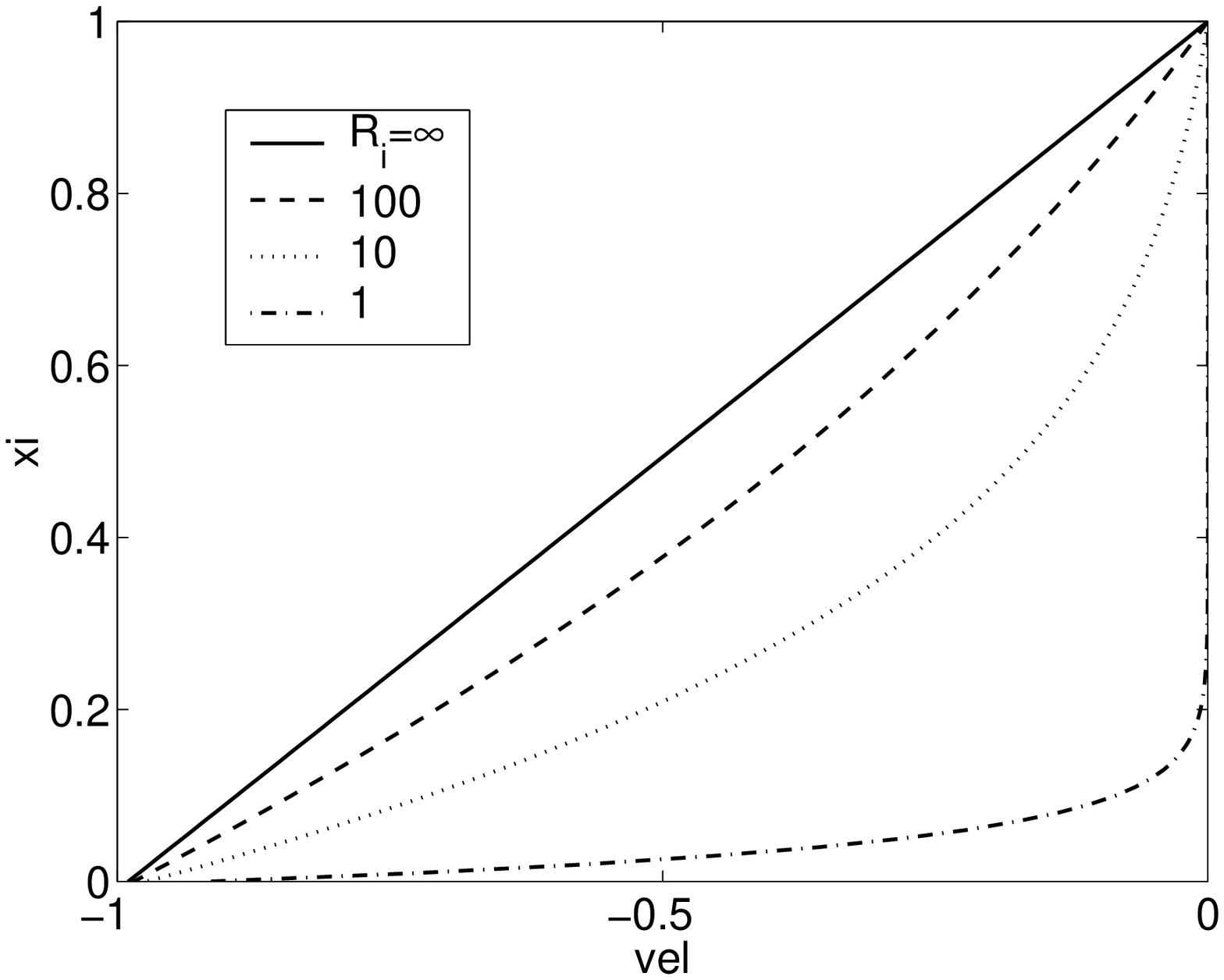}
    \hspace{0.2in} \includegraphics[width= 0.45\textwidth]{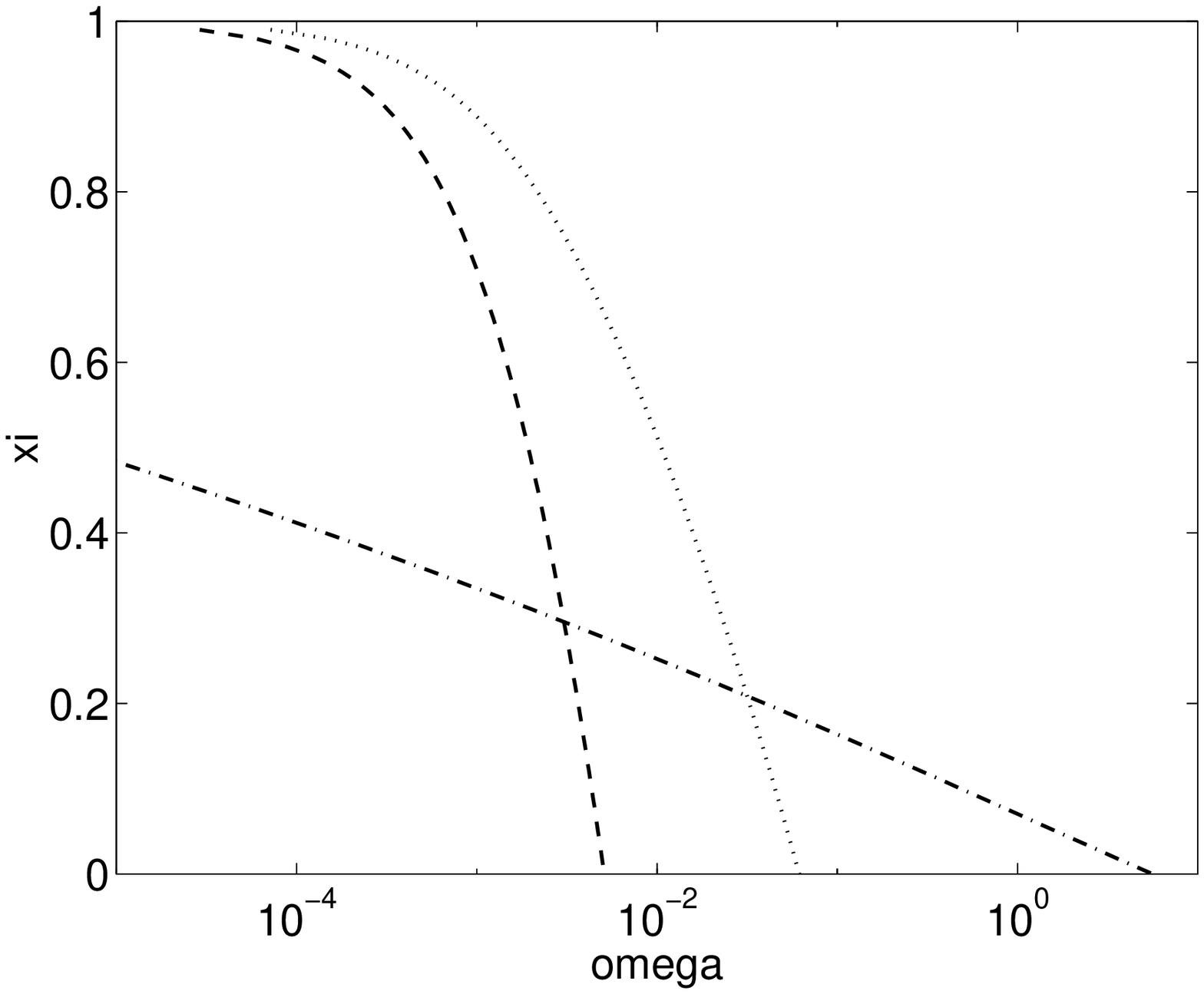} \\
    \vspace{0.5in}
    \hspace*{0.1in} (c) \hspace{2.2in} (d) \\
    \includegraphics[width= 0.45\textwidth]{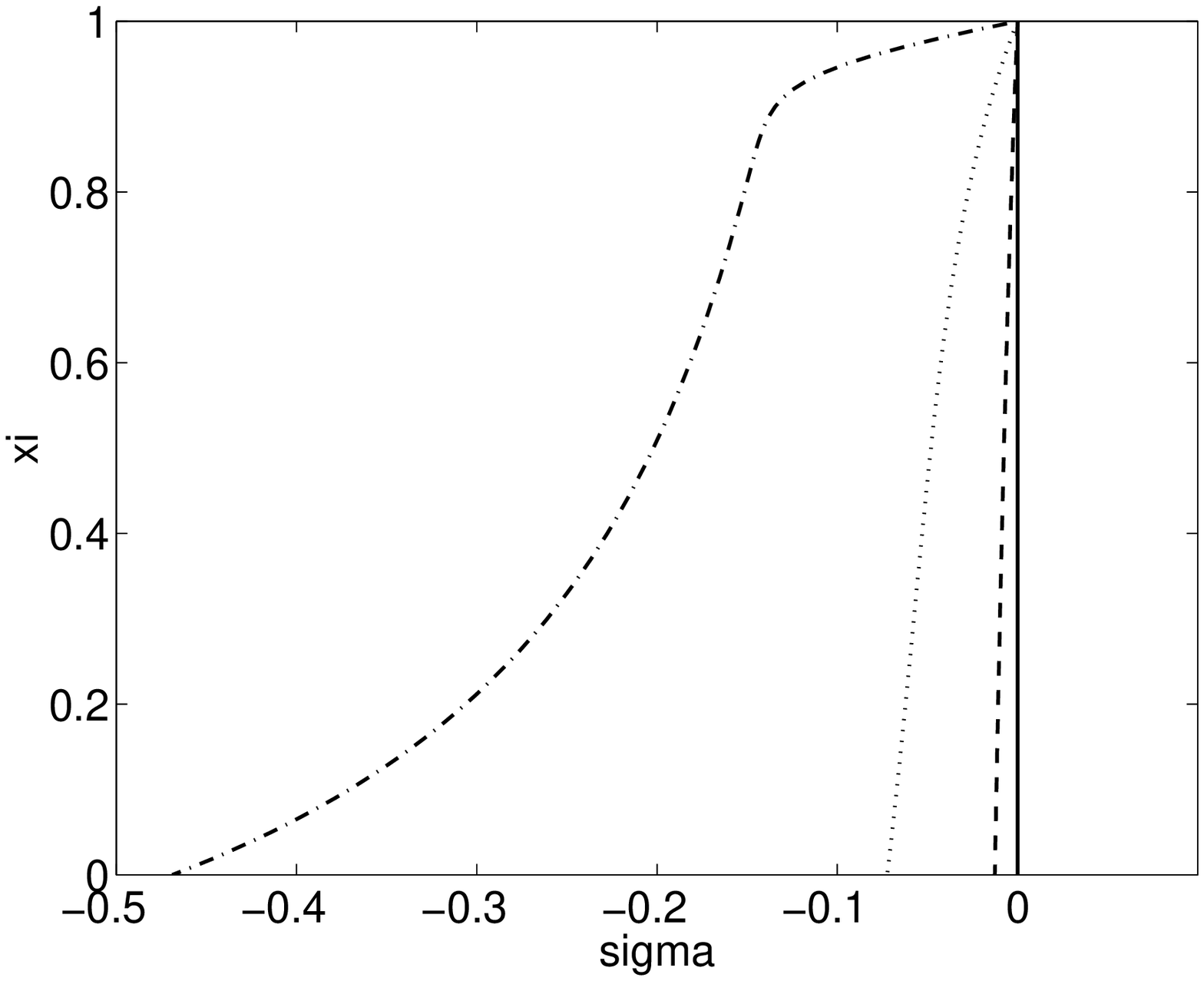}
    \hspace{0.2in} \includegraphics[width=0.45\textwidth]{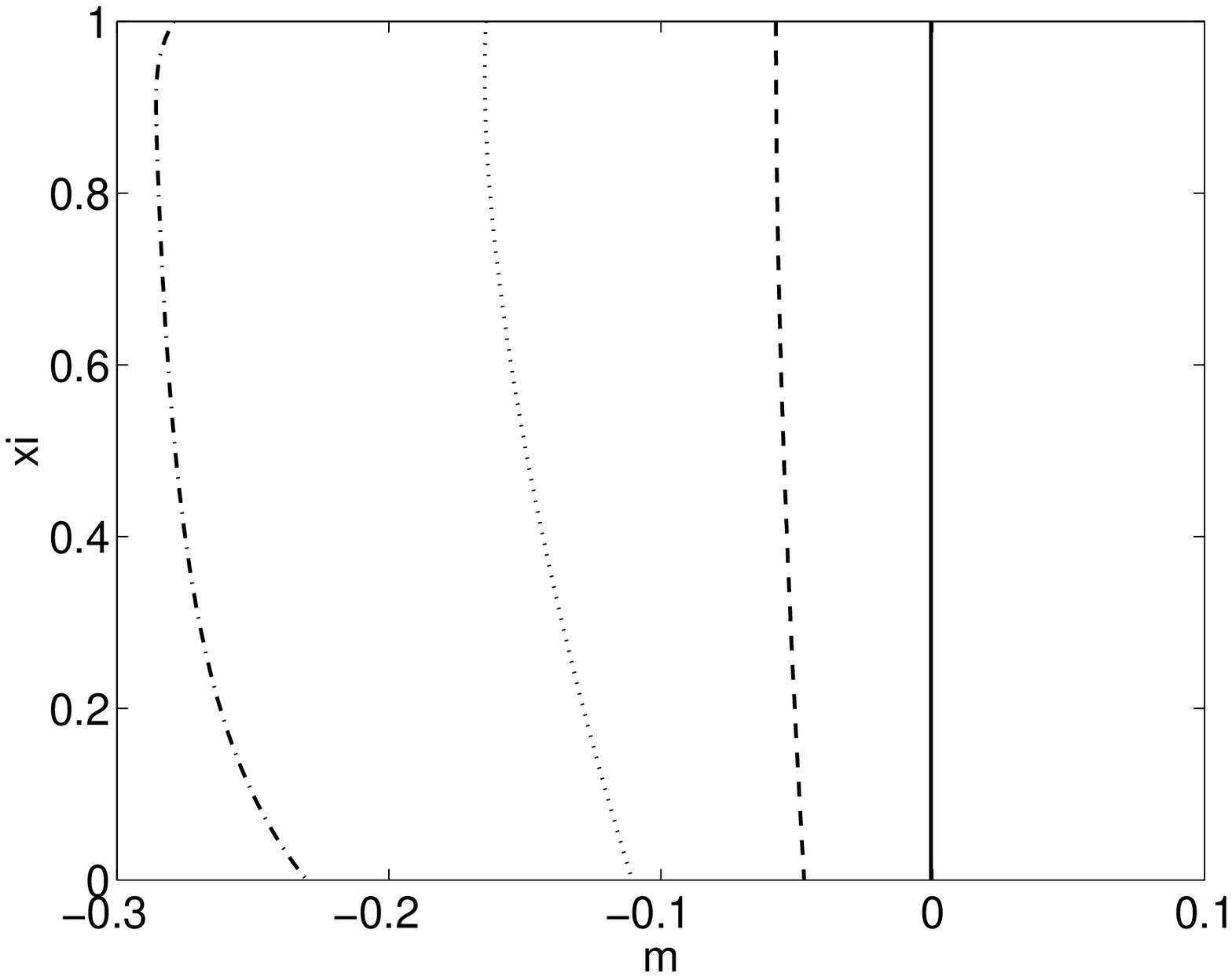}
    \caption{Profiles of the stress and kinematic fields for shear between
    fully rough walls in a cylindrical Couette cell for four
    values of $\ol{R}_i$, the ratio of the inner cylinder radius to the
    Couette gap.  Panel (b) shows the difference between $\omega$ and half
   the     vorticity, and panel (c) shows the asymmetry in the shear
stress.  The solid line is absent in panel (b) because $\omega +
1/2\, du/d\xi = 0$ for this case.  $\ol{R}_i = \infty$ (solid
line) corresponds to plane shear, for which $\omega$ is equal to
the vorticity and the Cauchy stress is symmetric.  Parameter
values are: $\phi=28.5^\circ$, $\varepsilon=0.04$, $L=10$, $A=1/3$
and $K=0.5$. \label{fig-cyl_couette}} \end{center}
\end{figure}

    The governing equations
\eqref{eqn-nd_yc_cyl}-\eqref{eqn-nd_zangmom_cyl}, with boundary
conditions  \eqref{eqn-nd_bc1_cyl}-\eqref{eqn-nd_bc2_cyl}, are
integrated using the procedure described in \S\ref{subsec-zerog}.
The  stress and velocity profiles for the case of
fully rough walls are presented in figure~\ref{fig-cyl_couette}
for four values of $\ol{R}_i$.  As expected, we recover the
uniformly shearing solution for plane shear in the limit $\ol{R}_i
\to \infty$.  The localization of shear near the inner cylinder
increases as $\ol{R}_i$ decreases.  Though the magnitude of the
couple stress is maximum at the outer cylinder, the asymmetry of
the Cauchy stress and the deviation of $\omega$ from half the
vorticity are maximum at the inner cylinder.

    The solution for large Couette gap is provided by the asymptotic
analysis in \S\ref{subsubsec-asymp_cyl}.  The leading order
solution is the same as in plane shear (with or without gravity)
when the walls are not fully rough.  If the walls are fully rough,
the solution is similar in form to that for plane shear under
gravity: the couple stress rises to $m_0$ within a boundary layer
of thickness $\sim \varepsilon^{2/3}$ near the inner cylinder, and
remains at $m_0$ outside the boundary layer.  The velocity fields
are determined by the couple stress within the boundary layer, and
are independent of the wall roughness.  The leading order
asymptotic solution for the velocity fields is compared with the
numerical solution for fully rough walls in
figure~\ref{fig-asymp_g_c}.

\subsection{Comparison with experimental data}
\label{subsec-comparison}

     Experimental measurements of the velocity profile in cylindrical
Couette flow of granular materials have recently been reported by
\citet{mueth_etal00} and \citet{bocquet_etal00}.
\citet{mueth_etal00} used mustard and poppy seeds and obtained the
velocity profiles using magnetic resonance imaging.
\citet{bocquet_etal00} used glass beads and measured the velocity
profile by video imaging of the upper surface of the granular
column; they also applied upward aeration across the granular
column, and observed that it has negligible effect on the velocity
profile.  From the data reported in their papers, we have
determined the parameter ${\cal R}$, defined in \S\ref{sec-intro},
to be in the range $6 \times 10^{-9} -\, 10^{-3}$  for the
experiments of \citet{mueth_etal00}, and $7 \times 10^{-9} - \, 2
\times 10^{-4}$ for that of \citet{bocquet_etal00}.  The Froude
number $C$, defined below \eqref{eqn-nd_zangmom_cyl}, for the two
studies was in the range $10^{-6} -\, 6 \times 10^{-2}$ and $2
\times 10^{-6} - \, 2 \times 10^{-2}$, respectively.  As ${\cal
R}$ and $C$ are both small for these studies, it is appropriate to
compare our model predictions with their data. Both find that the
shape of the velocity profile is independent of the rotation rate
of the inner cylinder, which, of course, is also a feature of our
model.

\begin{figure}
  \begin{center}
    \psfrag{xi}[t][]{$\xi$}
    \psfrag{vel}[br][][1][-90]{$-u$}
    \vspace{1in}
    \hspace*{0.1in} (a) \hspace{2.2in} (b) \\
    \includegraphics[width= 0.45\textwidth]{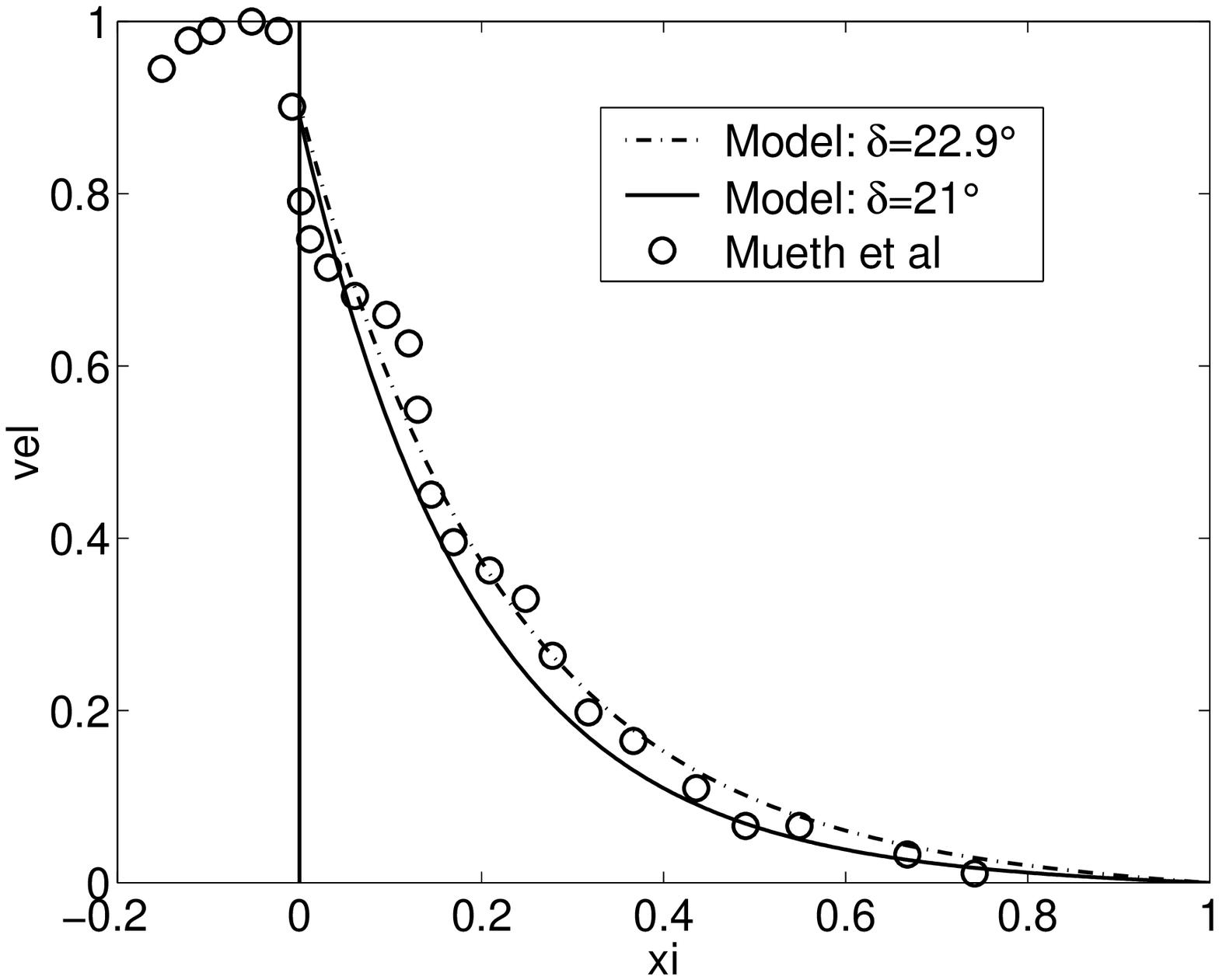}
    \hspace{0.2in} \includegraphics[width= 0.45\textwidth]{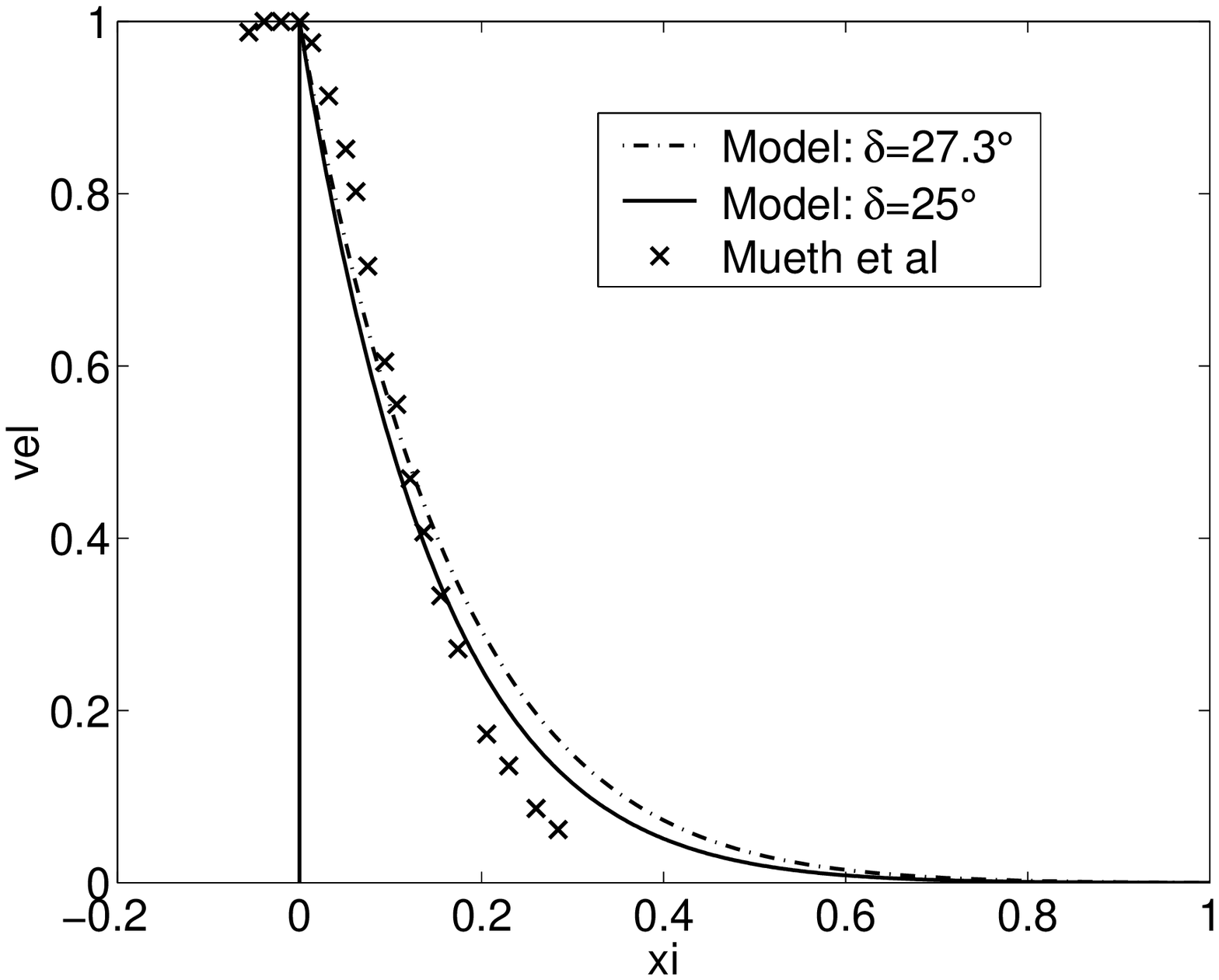}
    \caption{Comparison of model predictions (lines) with the data (symbols) of
\citet{mueth_etal00} for shear of (a) mustard seeds and (b) poppy
seeds in a cylindrical Couette cell.  In the experiments, a layer
of particles was glued to the cylinders.  The data points to the
left of $\xi=0$ in the figures are the velocity of the glued
layer.   The dot-dash line represents the model prediction for
fully rough walls ($\tan\delta = \sin\phi$, see
eqn.~\ref{eqn-bc_fullyrough}). See text for other
parameters.\label{fig-mueth}} \end{center}
\end{figure}

\begin{figure}
  \begin{center}
    \psfrag{xi}[t][]{$\xi$}
    \psfrag{vel}[br][][1][-90]{$-u$}
    \vspace{1in}
    \includegraphics[width= 0.45\textwidth]{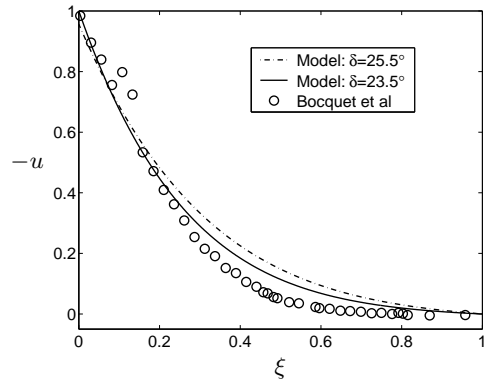}
    \caption{Comparison of model predictions (curves) with the data
(symbols) of \citet{bocquet_etal00} for the shear of glass beads
in a cylindrical Couette cell. In the experiments, a layer of
particles was glued to the cylinders.  The dot-dash line
represents the model prediction for fully rough walls
($\tan\delta= \sin\phi$, see eqn.~\ref{eqn-bc_fullyrough}).
See text for other parameters.\label{fig-bocquet}}
  \end{center}
\end{figure}

    The predictions of the velocity profile using our Cosserat model
are compared with the data of \citet{mueth_etal00} in
figure~\ref{fig-mueth} and with the data of \citet{bocquet_etal00}
in figure~\ref{fig-bocquet}.  In both studies, the inner and outer
cylinders were coated with a layer of particles; we have therefore
taken the walls to be at the outer edges of the glued layers for
the purpose of comparison. Neither of the studies has reported the
angle of internal friction of the material they used.  For mustard
seeds, \citet{tuzun_nedderman85} report it to be in the range
23-25$^{\circ}$, and we have used $\phi=25^\circ$ for our
calculations. Data on this property is lacking for the more
angular and rough poppy seeds, and we have taken the value of
$31^\circ$. For glass beads, we have used the value of
$\phi=28.5^\circ$ reported by \citet{nott91}.  The value of  $K$
was set by matching the wall slip velocity to the reported values,
yielding $K=0.65$ for mustard seeds, and zero for poppy seeds and
glass beads.  No attempt was made to determine the best-fit values
of $L$ and $A$, and the values used so far in the paper have been
retained, viz.\ $L=10$, $A=1/3$.  The values of $\varepsilon$ and
$R_i$, determined from reported values of the mean grain diameter
and the dimensions of the Couette cell are: $\varepsilon = 0.15$,
0.058, and 0.067 for mustard, poppy, and glass, respectively, and
$\ol{R}_i = 2.29$, 1.89, and 4.25 in the same order.

    The dot-dash lines in figures~\ref{fig-mueth} and \ref{fig-bocquet}
are model predictions assuming fully rough walls ($\tan\delta =
\sin\phi$), and the solid lines are predictions for $\delta$
roughly $2^\circ$ lower than $\tan^{-1}(\sin\phi)$. We find that the
latter fits the data better, in agreement with our argument in
\S\ref{subsec-zerog} that a fully rough wall may be difficult to
achieve practically.  It is clear that there is, in general, good
agreement between the model predictions and the data.  Though our
model underestimates the velocity for glass beads
(figure~\ref{fig-bocquet}), we must bear in mind that
\citet{bocquet_etal00} made their measurements at the upper free
surface, where the velocity is expected to be higher than in the
bulk. Moreover, we have only made plausible guesses for almost all
the material properties, and have not attempted to achieve a
better fit by adjusting the parameters.

    Our model predicts only a small increase in the solids
fraction with distance from the inner cylinder, while the data of
\citet{mueth_etal00} suggests a substantial variation.  In spite
of this difference, the predicted velocity profiles are in good
agreement with their data.  Thus it appears that the velocity
field is relatively insensitive to the dilation of the granular
medium.

\section{Parameter sensitivity of model predictions}
\label{subsec-par_sens}

    The properties of the material and the boundaries
in the Cosserat model are characterized by the angles of internal
friction $\phi$, the angles of wall friction $\delta_L$ and
$\delta$, the parameters $A \equiv a_2/a_1$ and $L$ occurring in
the yield condition \eqref{eqn-yc2}, and the parameter $K$ that
determines the extent of slip at a boundary in the boundary
condition \eqref{eqn-bc3}.  The significance of the angles of
friction is well understood; their physical meaning remains the
same as in classical plasticity models.  We now consider the
sensitivity of the model to the parameters $A$ and $K$, which are
associated with Cosserat effects in our model. To this end, we
consider the problem of plane shear under gravity, and explore the
effect of varying these parameters.

\begin{figure}
  \begin{center}
    \psfrag{xi}[br][][1][-90]{$\xi$}
    \psfrag{vel}[t][]{$u$}
    \psfrag{omega}[t][]{$\omega$}
    \psfrag{m}[t][]{$m$}
    \vspace{1in}
    \hspace*{0.1in} (a) \hspace{2.2in} (b) \\
    \includegraphics[width= 0.45\textwidth]{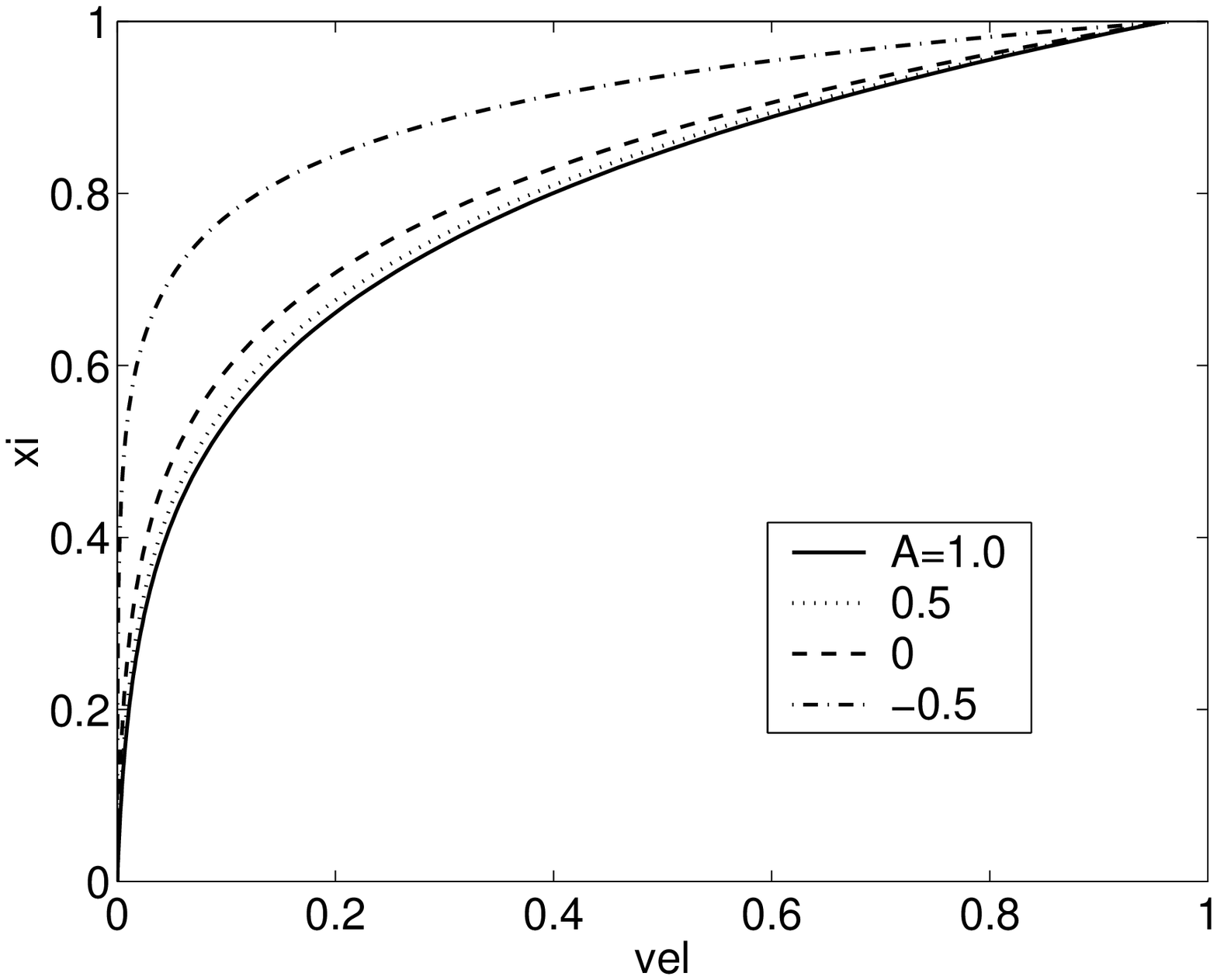}
    \hspace{0.2in} \includegraphics[width= 0.45\textwidth]{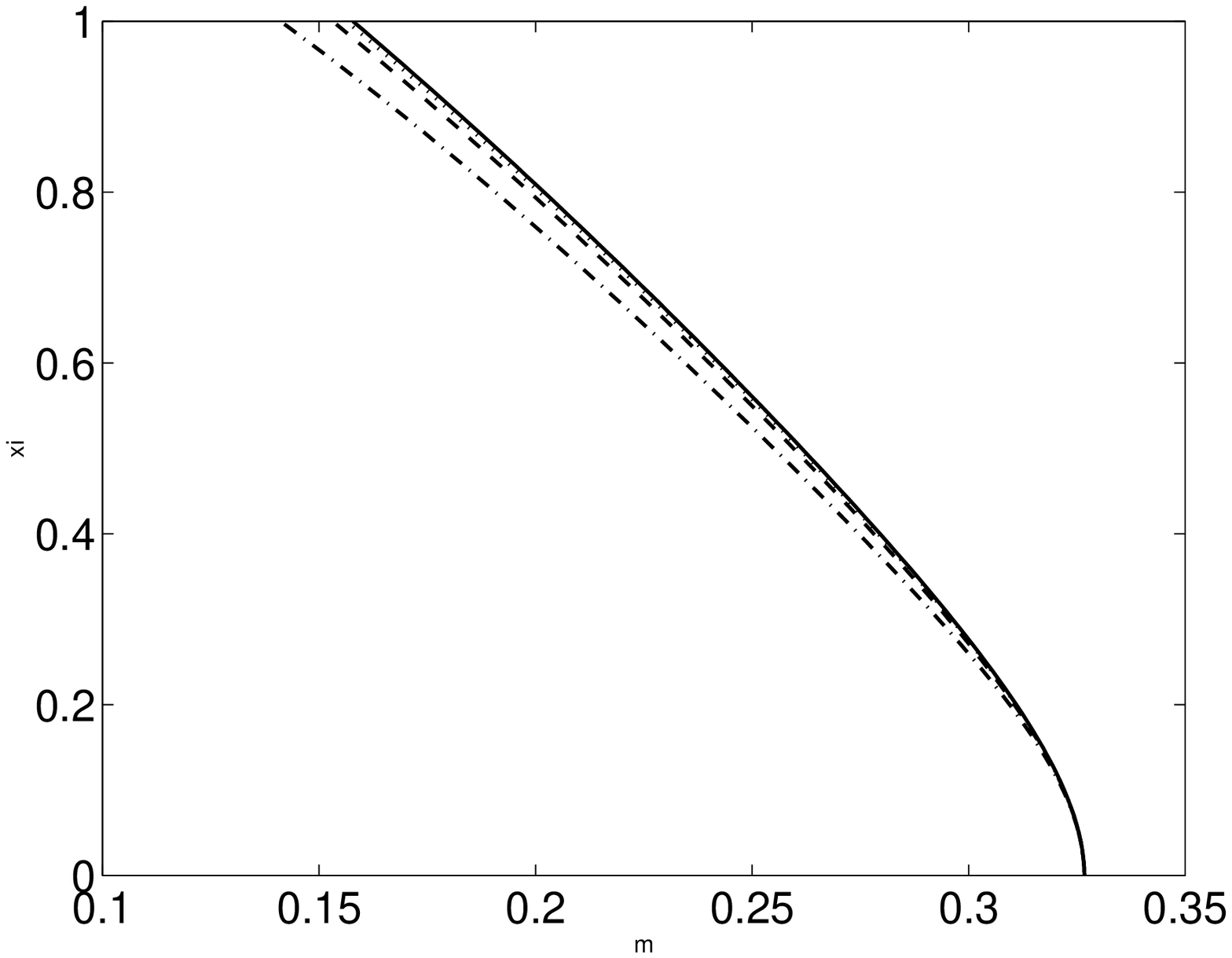} \\
    \caption{Sensitivity of model predictions to the parameter $A$ in
    the yield condition \eqref{eqn-nd_yc}.  The solutions are for plane shear
    between fully rough walls under gravity with $B=1$.   As discussed in
\S\ref{subsec-YC}, $|A|$ must be $\le 1$, and for the set of
parameters used here, solutions do not exist for $A < -0.57$. The
values of other parameters are as in the caption of
figure~\ref{fig-zero_g}.\label{fig-A_sensitivity}} \end{center}
\end{figure}

    As discussed in \S\ref{subsec-YC}, $A$ can assume values only
in the range (-1,1). However, solutions may not exist for all
negative values of $A$ within this range, as illustrated in
figure~\ref{fig-A_sensitivity}.  For the set of values of the
other parameters used here, the shear stress $\sigma_{xy}$ assumes
complex values when $A < -0.57$, and hence a physically acceptable
solution does not exist. Solutions exist for $-0.57 < A < 1$, and
figure~\ref{fig-A_sensitivity} shows that in this range of values,
there is not much variation in the velocity profile and even less
of the couple stress.

\begin{figure}
  \begin{center}
    \psfrag{xi}[br][][1][-90]{$\xi$}
    \psfrag{vel}[t][]{$u$}
    \psfrag{omega}[t][]{$\omega$}
    \vspace{1in}
    \hspace*{0.1in} (a) \hspace{2.2in} (b) \\
    \includegraphics[width= 0.45\textwidth]{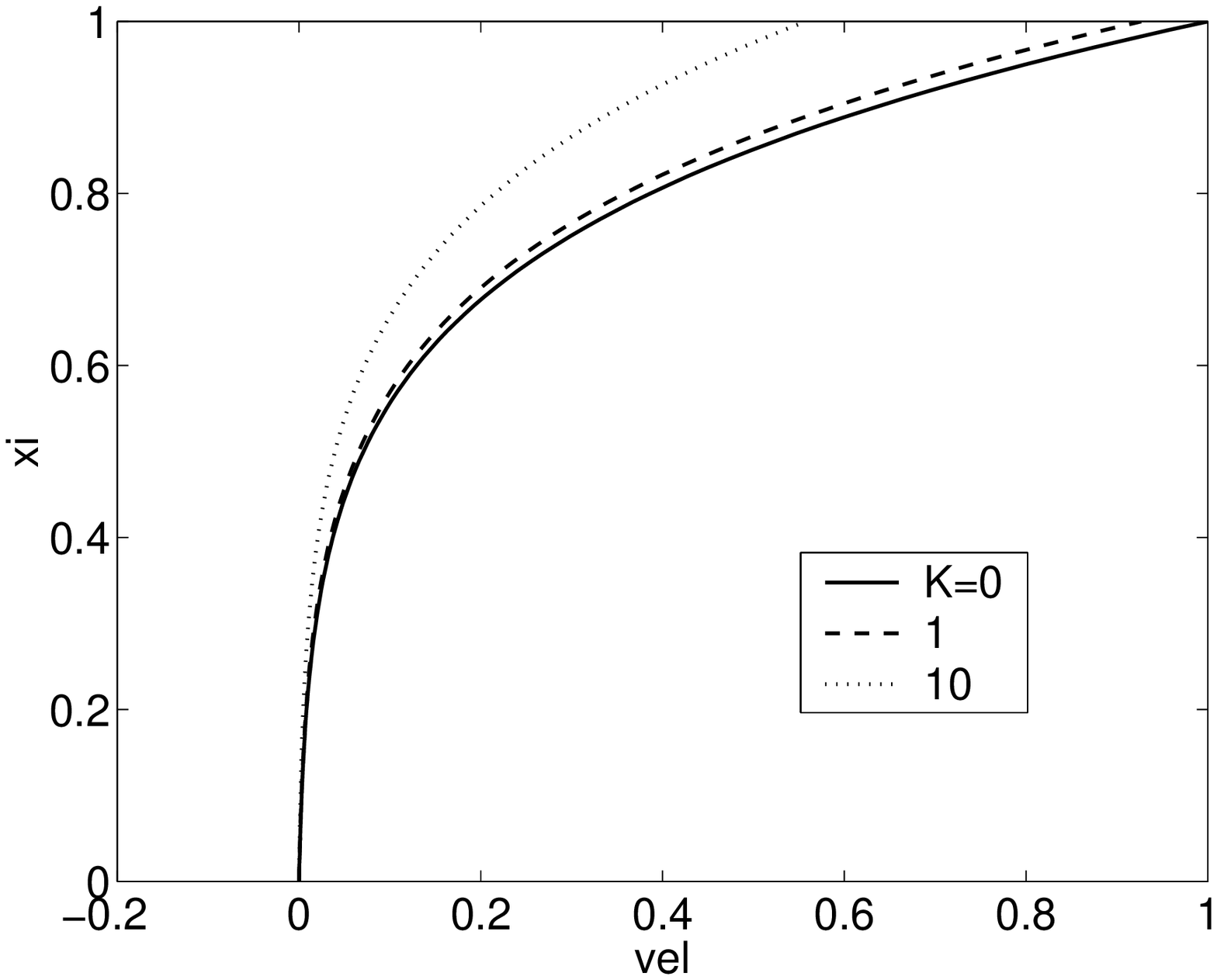}
    \hspace{0.2in} \includegraphics[width= 0.45\textwidth]{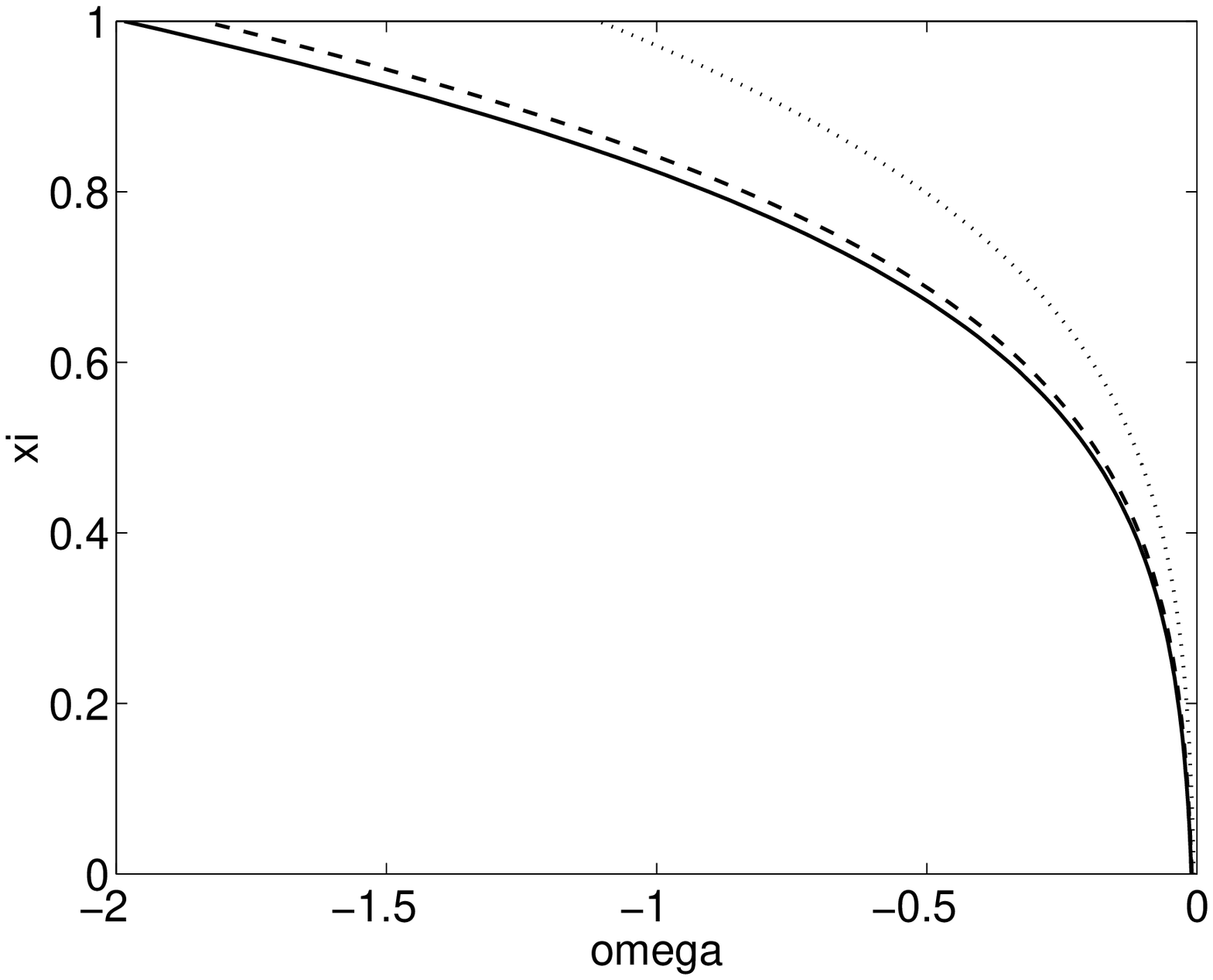} \\
    \caption{Sensitivity of model predictions to the parameter $K$ in
    the kinematic boundary condition \eqref{eqn-bc3}.  The solutions are for
    shear between fully rough walls under gravity with $B=1$.  As $K$
increases, the slip velocity at the upper wall increases and the
magnitude of the angular velocity at the wall decreases.  The
stress and couple stress profiles are not affected by the value of
$K$, and are given by the dotted lines in
figure~\ref{fig-finite_g}.  The values of parameters other than
$K$ are as in the caption of figure~\ref{fig-zero_g}.
\label{fig-K_sensitivity}} \end{center}
\end{figure}

    The parameter $K$ determines the extent of velocity slip at
the boundaries, but does not affect the stress and couple stress
fields. As shown in figure~\ref{fig-K_sensitivity}, there is no
slip at the upper wall when $K=0$, and the velocity slip increases with
$K$.  The important point is that the  shapes of the
velocity profiles are unaltered by varying $K$, a result which is
also shown clearly by the asymptotic solutions in
\S\ref{appendix}.

\section{Shear layer thickness}
\label{sec-sl_thickness}

    Though our model assumes that the granular material yields
everywhere within the Couette gap, the results in the preceding
two sections show the velocity decaying rapidly with distance from
the upper wall for plane shear (with the exception of gravity-free
shear between fully rough walls), and from the inner cylinder for
cylindrical Couette flow. In the experiments of
\citet{mueth_etal00} and \citet{bocquet_etal00}, grain motion is
not detectable beyond a distance of a few particle diameters from
the wall. It is therefore useful to define the dimensionless shear
layer thickness $\Delta$ as the distance, in terms of particle
diameters, from the upper wall (or the inner cylinder) at which
the velocity decays to a small fraction $f$ of the wall velocity,
\begin{subeqnarray}
    (u(1 -  \varepsilon \Delta) - u_0) = f \, (u(1) - u_0) \;\;\;
    \mbox{for plane shear,}\\
    (u(\varepsilon \Delta) - u_0) = f \, (u(0) - u_0) \;\;\; \mbox{for
    cylindrical Couette flow.}
\label{eqn-delta_defn}
\end{subeqnarray}
The reference velocity $u_0$ is equal to zero, except in the case
of gravity-free plane shear between identical walls, where it is
1/2 (the velocity at the center).  We set $f=0.05$, as in earlier
studies, and determine the shear layer thickness as a function of
the Couette gap, or equivalently, $\varepsilon$. While $\Delta$
must in general be determined numerically, an analytical
expression can be obtained in the limit of large Couette gap
($\varepsilon \to 0$), as discussed below.

    In most practical instances of granular flow, the size of the
vessel is much larger than the grain size. It is therefore useful
to determine the dependence of the shear layer thickness $\Delta$
on the system size in the limit $\varepsilon \to 0$.  In our
earlier work on flow in vertical channels~\citep{mohan_etal99}, we
showed that in the limit of small $\varepsilon$, $\Delta$ is
independent of $\varepsilon$, except in the singular case of a
fully rough wall ($\tan\delta = \sin\phi$), when it grows as
$\epsilon^{-1/3}$. Here, we apply the same approach to determine
the asymptotic behaviour of $\Delta$ for plane and cylindrical
Couette flow.

    For each of the problems considered, the leading order asymptotic
solutions for the stress and velocity fields are given in the
appendix.  These suffice to determine the shear layer thickness.
The details of the asymptotic analysis are given in
\S\ref{appendix}, and the solutions are used here to determine the
shear layer thickness.

    When the walls are not fully rough, i.e.~$\tan\delta < \sin\phi$,
the leading order linear velocity fields for plane shear and
cylindrical Couette flow may be written in the common form
(\ref{eqn-asymp_vels}d) (see eqns.~\ref{eqn-asymp_soln_vels_g0}b
and \ref{eqn-asymp_soln_vels_g0_aw}b), where  the reference
velocity $u_0$ is as defined below \eqref{eqn-delta_defn}.  The
shear layer thickness can now be determined using
(\ref{eqn-asymp_vels}d) and \eqref{eqn-delta_defn},
\begin{equation}
    \Delta = - \ln(f)/k = \Frac{2.996 \, L \tan\delta }{2 (\sin^2\phi
    - \tan^2\delta)^{1/2}}. \label{eqn-delta_soln}
\end{equation}
Thus, we find $\Delta$ to be independent of the Couette gap when
the latter is large compared to particle size (i.e.~$\varepsilon
\ll 1$).  A feature observed in some experiments is that the shear
layer thickness decreases as the wall becomes smoother
\citep{nedderman_laohakul80}, and this too is captured
by~\eqref{eqn-delta_soln}.

    We note that the shear layer thickness increases as
$\tan\delta$ approaches $\sin\phi$.  In the precise limit of fully
rough walls, the solution was already given in
\S\ref{subsec-zerog} for plane shear in the absence of gravity,
showing that shear rate is equal throughout the gap, and therefore
$\Delta = (1 - u_0)(1 - f) \varepsilon^{-1}$.  Thus, the parameter
$L$ does not determine the thickness of the shear layer when the
walls are fully rough, in contrast to the case of non-fully rough
walls.  In the former case, the couple stress vanishes, and hence
the governing equations do not involve $L$.

For shear under gravity and cylindrical Couette flow, the material
does not shear uniformly for fully rough walls, and the asymptotic
velocity profile (\ref{eqn-asymp_vels}d) is not valid in region of
thickness $\sim \varepsilon^{2/3}$ near the upper wall and inner
cylinder, respectively. A uniformly valid solution is found by
rescaling $\xi$ and $m$ in this ``inner'' region, as shown in
\S\ref{subsubsec-asymp_g} and \S\ref{subsubsec-asymp_cyl}, and the
leading order velocity profile then assumes the form
\begin{equation}
    u = 1 - I(\hat{\xi})/I(\infty).
\end{equation}
where $I(\hat{\xi}) \equiv (\airyai(0))^{-2}
\int_0^{\hat{\xi}}{(\airyai(z))^2 dz}$.  Here $\airyai(x)$ is the
Airy function, and the rescaled independent variable is
\begin{equation}
\bea{l}
    \hat{\xi}= (1 - \xi) \, \varepsilon^{-2/3} \left(\Frac{L^2}{2 B \nu}
    \right)^{-1/3}\;\;\; \mbox{for plane shear under gravity, and}\\
    \hat{\xi} = \xi \, \varepsilon^{-2/3} \left(\Frac{L^2
    \ol{R}_i}{4} \right)^{-1/3} \;\;\; \mbox{for cylindrical Couette flow.}
\eea
\end{equation}
The value of $\hat{\xi}$ at which \eqref{eqn-delta_defn} holds is
found to be 1.275, resulting in the following expression for the
shear layer thickness:
\begin{subeqnarray}
    \Delta & = & 1.275 \, \varepsilon^{-1/3} \left(\Frac{L^2}{2 B
    \nu}\right)^{1/3} \;\;\; \mbox{for plane shear under gravity},\\
    \Delta & = & 1.275 \, \varepsilon^{-1/3} \left(\Frac{L^2
    \ol{R}_i}{4}\right)^{1/3} \;\;\; \mbox{for cylindrical Couette flow}.
\label{eqn-delta_soln_fr}
\end{subeqnarray}
Thus, $\Delta$ grows as the one-third power of the ratio of the
Couette gap to grain size when the walls are fully rough.  In
contrast, $\Delta$ is independent of the Couette gap
(cf~\ref{eqn-delta_soln}) when the walls are not fully rough. It
also grows as $(B \nu)^{-1/3}$, in plane shear under gravity, as
the gravitational body force is reduced, but the above asymptotic
analysis is not valid when $B \nu \sim \varepsilon$.  In the limit
$B = 0$, which corresponds to plane shear without gravity, the
solution in \S\ref{subsec-zerog} shows a linear velocity profile,
and therefore $\Delta = (1 - u_0)(1 - f) \varepsilon^{-1}$.
Similarly $\Delta$ grows as $(\ol{R}_i)^{1/3}$, in cylindrical
Couette flow, as the radius of the inner cylinder increases, and
approaches $(1 - u_0)(1 - f) \varepsilon^{-1}$ in the limit
$\ol{R}_i \to \infty$.

\begin{figure}
  \begin{center}
    \psfrag{delta}[br][][1][-90]{$\Delta$}
    \psfrag{eps}[t][]{$\varepsilon$}
    \psfrag{Delta}[br][][1][-90]{$\Delta$}
    \vspace{1in}
    \hspace*{0.1in} (a) \hspace{2.2in} (b) \\
    \includegraphics[width= 0.45\textwidth]{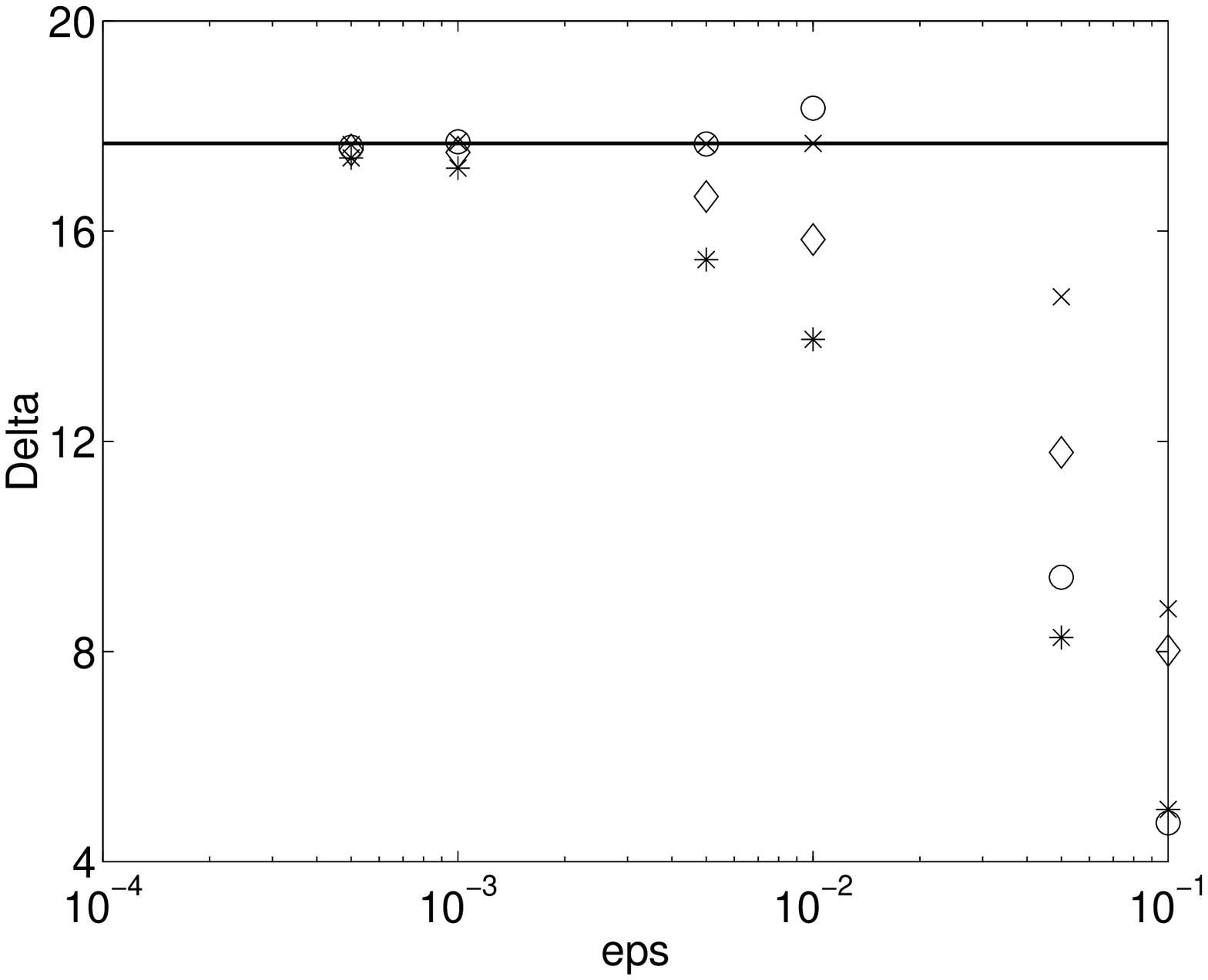}
    \hspace{0.2in} \includegraphics[width= 0.45\textwidth]{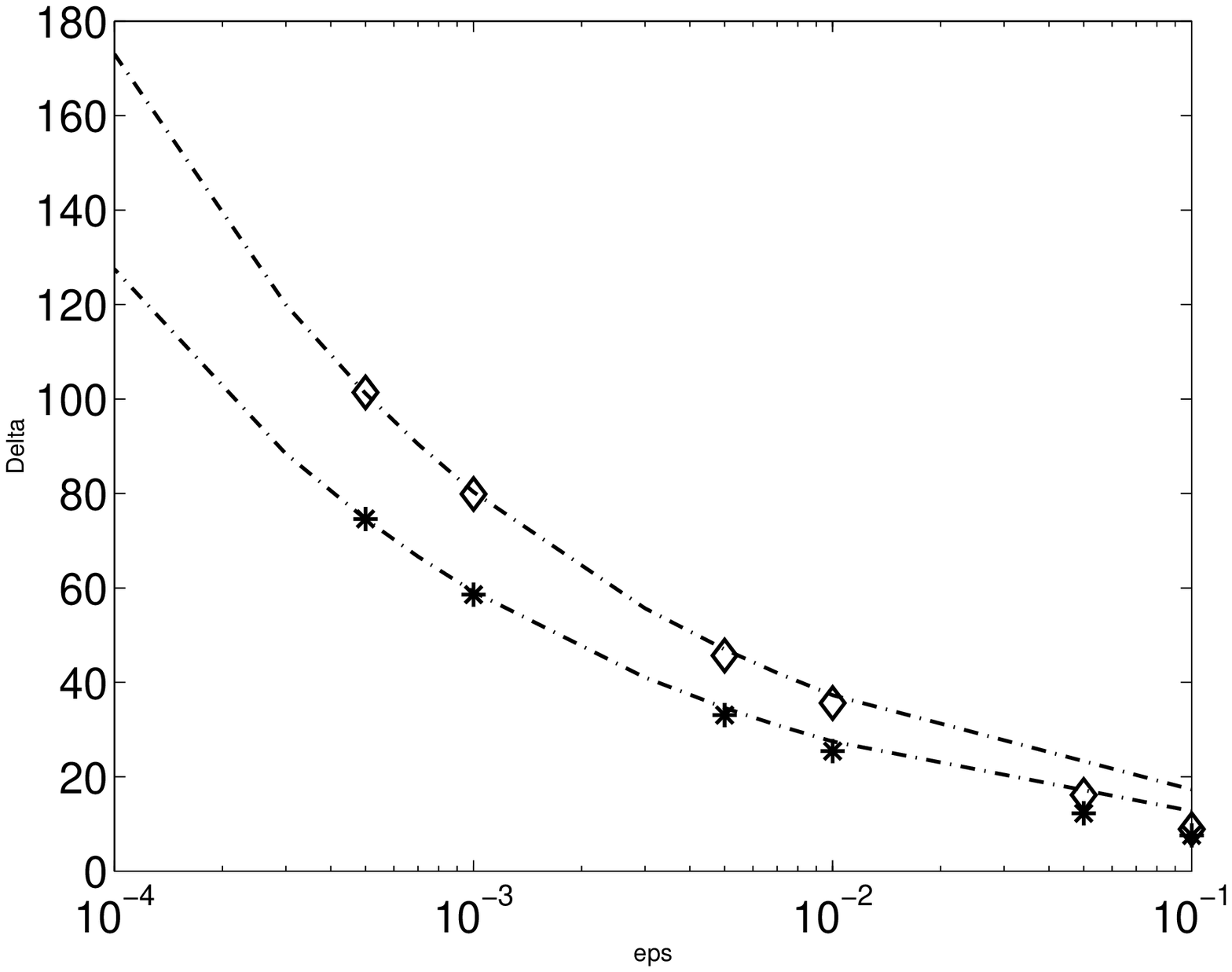} \\
    \caption{The shear layer thickness $\Delta$ as a function of
     $\varepsilon$ for plane shear and cylindrical Couette flow.  The
     results shown in panel (a) are for a non-fully rough upper wall (or
     inner cylinder), while panel (b) gives results for fully rough walls.
     In both panels, the curves represent  asymptotic solutions for small
     $\varepsilon$, given in \S\ref{sec-sl_thickness}, and the symbols
     represent numerical solutions.  The circles and crosses are for plane
     shear in the absence of gravity between identical walls and with a
     rougher lower wall, respectively; the asterisks are for plane shear
     under gravity and the diamonds are for cylindrical Couette flow.
     Parameter values for panel (a) are: $\phi=28.5^\circ$, $\delta =
     20^\circ$, $L=10$, $A=1/3$ and $K=0.5$, $B=1$ (for
    plane shear under gravity) and $\ol{R}_i=1$ (for cylindrical Couette
    flow).  Parameter values for panel (b) are the same with the exception
    that $\delta = \tan^{-1}\sin\phi = 25.5^{\circ}$.\label{fig-slt}}
\end{center}
\end{figure}

    The variation of $\Delta$ with $\varepsilon$ is shown in
figure \ref{fig-slt} for plane and cylindrical Couette flow; panel
(a) shows the results for non-fully rough walls and panel (b) for
fully rough walls.  The symbols are the numerical results obtained
by the procedure described in
\S\ref{sec-results_planecouette}-\ref{sec-cyl_couette}, and the
lines are the asymptotic solutions.  For non-fully rough walls,
the shear layer thickness for all the four problems converges to
the common asymptote of $\Delta \approx 17.67$ as $\varepsilon \to
0$, as shown in figure \ref{fig-slt}a.  For shear between fully
rough walls, the numerical solutions for $\Delta$ asymptote to the
forms given by \eqref{eqn-delta_soln_fr} as $\varepsilon \to 0$.

\section{Other models}
\label{sec-other_models}

    Models for the shear of dense granular materials have also been
proposed recently by \citet{savage98} and \citet{bocquet_etal00}.
Here we briefly discuss some aspects of their models.

\subsection{The model of \citet{savage98}}
\label{savage}

        In a recent paper, \citet{savage98} proposed a model that
attempts to merge the critical state theory for quasistatic flows
with the results from kinetic theory for rapid flows, with the
stated primary aim of predicting the behaviour of transitional and
rapid flows.  The model starts with a yield condition and an
associated flow rule (as in this work), but Savage argues that the
rate of deformation tensor \te{D} at any location fluctuates in
time, with a standard deviation $\epsilon$.  The mean stress
tensor $\langle \te{\sigma} \rangle$ is  determined by computing
its average with respect to a Gaussian distribution of deformation
rates.  In the limit $\vert \langle D_{ij} \rangle \vert \ll
\epsilon$, where $D_{ij}$ is a component of $\te{D}$, a Newtonian
constitutive relation is obtained for $\langle \te{\sigma}
\rangle$, with the shear viscosity given by $\sigma_c \, A/
\epsilon$, where $\sigma_c$ is the frictional mean stress at a
critical state and $A$ is a material constant. In order to obtain
a theory which resembles  kinetic theories for rapid flow, it is
assumed that $\sigma_c$ may be replaced by $\sigma_{c1}(\nu) +
\sigma_{c2}(\nu, T)$.  Here $\sigma_{c1}$ is the frictional mean
stress, $\sigma_{c2}$ is the mean stress obtained from kinetic
theory, and $T$ is the grain temperature or equivalently, the
kinetic energy of velocity fluctuations.  To proceed further, it
is assumed that $\epsilon$ is proportional to $\sqrt{T}$, and the
proportionality constant is determined by matching the form
obtained for large values of $T$ with the corresponding kinetic
theory result.

    The idea of accounting for fluctuations in the deformation rate
appears to have merit, but we find the following aspects of the
model unconvincing. (i) The physical origin of large fluctuations
in the velocity gradients relative to the mean velocity gradients
is not clear. Data bearing on this issue do not appear to be
available in the literature. (ii) Even though the constitutive
relations are derived by assuming that $| \langle D_{ij}\rangle |
\ll \epsilon$, this ratio turns out to be O(1) in the examples
discussed in \citet{savage98}.  Thus the theory is applied beyond
its range of validity.  As noted by \citet{savage98}, some of the
kinetic constitutive equations also suffer from this defect when
applied to plane shear. (iii) As the velocity gradients are
decomposed into mean values and fluctuations, it appears that the
inertial terms, which involve products of velocity fluctuations,
may contribute non-zero terms to the averaged momentum balances.
Such terms have been omitted in the analysis. (iv) The solution of
bounded flow problems requires the specification of the grain
temperature at the boundaries, which is usually unknown {\it a
priori}\/; realistically one would like to determine the
temperature as part of the solution.

\subsection{The model of \citet{bocquet_etal00}}
\label{subsec-bocquet}

    This is a minor variant of the high density kinetic theory of
\citet{haff83}, though \citet{bocquet_etal00} have arrived at the
high density limit using the kinetic theory of \citet{jenkins83}.
The only change the authors have made is in proposing a modified
expression for the shear viscosity; this  gives a better fit for
their data of the grain temperature as a function of the shear
rate.  Making some simplifying assumptions about the form of the
temperature profile, they determine the velocity profile with
three adjustable parameters. It is shown that the profile provides
a good fit to their data for shear in a cylindrical Couette cell
(see figure~\ref{fig-bocquet}).  However, their solution is
derived for plane shear, whereas the data they compare with (and
achieve a good fit) is for cylindrical Couette flow.

    More importantly, it seems likely that the underlying assumptions
of kinetic theory, such a molecular chaos and instantaneous binary
collisions will break down in the limit of small deformation rate
and high solids fraction. Therefore, as in the case of Savage's
model discussed above, it appears that the theory has been used
beyond its range of validity.  Even if one were to just view the
high density kinetic theory as a phenomenological model, its
applicability to slow granular flow is suspect as it does not
yield a rate-independent stress for small deformation rates.

\section{Summary and conclusions}
\label{sec-summary}

    We have shown that our frictional Cosserat model for slow granular
flows captures the formation of thin shear layers in viscometric
flows. The principal features of our continuum model are: (1) the
presence of a couple stress field, which is a result of tangential
frictional forces between grains; (2) an angular velocity field
which is not necessarily determined by the local vorticity; (3)
solution of the balance of angular momentum, in addition to the
balances of mass and linear momentum; (4) the extension of the
yield condition and flow rule used in classical plasticity to
incorporate the couple stress and the angular velocity. By
including these features, we incorporate a microscopic length
scale in our model, which determines the thickness of the shear
layer.

    For plane shear in the absence of gravity, we have considered two
cases: in the first, the two walls are of equal roughness, and in
the second, the upper wall is smoother. In the former, for which
the shear rate is symmetric about the mid plane, we find that
Cosserat effects (a finite couple stress, asymmetry of the Cauchy
stress tensor and deviation of the angular velocity from half the
vorticity) are maximum near the solid boundaries, and decay with
distance from the boundaries. In the second case, the couple
stress is finite everywhere in the gap, but other Cosserat effects
are  absent. In both cases, the velocity fields decay rapidly with
distance from the upper wall if the Couette gap is large compared
to the grain size. If the roughness of the granular medium is
exactly equal to that of the wall, it shears uniformly in the
entire gap.

    For shear between walls of equal roughness in a gravitational
field (and in cylindrical Couette flow), the couple stress is
finite throughout the Couette gap, but other Cosserat effects are
present only in the shear layer near the upper wall (inner
cylinder). Here too the velocity and angular velocities decay
rapidly with distance from the upper wall (inner cylinder). While
experimental measurements of the couple stress or asymmetry in the
Cauchy stress have not been reported, our predictions are in good
agreement with the velocity measurements reported recently by
\citet{mueth_etal00} and \citet{bocquet_etal00}.

    If the wall roughness is less than that of the granular medium
(i.e.~$\tan\delta < \sin\phi$), the shear layer thickness $\Delta$
increases with the Couette gap, but reaches an asymptotic value
independent of the Couette gap in the limit $H/d_p \to \infty$
($H$ and $d_p$ are the Couette gap and grain diameter,
respectively). Further, $\Delta$ decreases when the angle of wall
friction $\delta$ is reduced, and vanishes when the wall is
perfectly smooth, in qualitative agreement with available
experimental data. In the singular case of fully rough walls
($\tan\delta = \sin\phi$), the material shears uniformly over the
entire Couette gap for plane shear in the absence of gravity; for
plane shear under gravity or cylindrical Couette flow, $\Delta$
increases with the Couette gap as $(H/d_p)^{1/3}$, and does not
depend on $\phi$.

    Many of the above predictions regarding the behaviour of $\Delta$
have so far not been confirmed experimentally. We believe that
these are important issues to be probed in future experimental
investigations, as they will serve to distinguish between existing
models for slow granular flows. Another prediction we make, also
requiring experimental verification, is that the shear layer is
located near the lower wall (outer cylinder) if it is sufficiently
smoother than the upper wall (inner cylinder), and the material
elsewhere suffers little deformation.

    While the predictions of our model are for the particular forms
of the constitutive relations (viz.~the yield condition and the
flow rule) and the boundary conditions we have chosen, we believe
that the main qualitative features of our results have a general
validity. For instance, the yield condition \eqref{eqn-yc1} is a
modification of the extended von Mises yield condition, a relation
between the first and second invariants of $\te{\sigma}$; by
incorporating the couple stress in this relation, we introduce a
microscopic length scale naturally into the constitutive
relations.  Any other yield condition, such as that proposed by
\citet{lade_duncan75} (extended to allow critical states) which
involves also the third invariant, could also be modified in the
same spirit, and the qualitative effect would be the same: the
microscopic length scale will set the length scale for the
thickness of the shear layer.

    In cylindrical Couette flow, our model predicts a small increase
in the solids fraction (depending on the magnitude of the Froude
number $C$) as we move towards the outer cylinder.  This increase,
however, is substantially smaller than what was observed in the
experiments of \citet{mueth_etal00}. This defect may perhaps be
corrected by incorporating elastic effects in the model, which is
a direction worthy of further study. For example,
\citet{tejchman_gudehus01} have found that the use of a Cosserat
model based on hypoplasticity permits the density to vary across
the shear layer.  Nevertheless, we believe that our model  is a
natural extension of classical plasticity models for slow granular
flows.

While the models of \citet{bocquet_etal00} and \citet{savage98}
also attempt to explain thin shear layers, they do not yield a
rate independent stress, which has shown to be an important
characteristic of slow flows.  In any case, further experiments
are required to assess the performance of our model and others in
various flow problems, before attempting to decide which model, if
any, is more realistic.

\appendix

\section{Asymptotic solution for small $\varepsilon$}
\label{appendix}

    To determine the asymptotic behaviour for small $\varepsilon$, we
seek a perturbation solution of the form
\begin{equation}
m = m^{(0)} + \varepsilon m^{(1)} + \varepsilon^2 m^{(2)} + \ldots
\label{eqn-perturb_soln}
\end{equation}
for the couple stress, and similarly for the other fields. For
each of the problems considered, we derive only the leading order
solutions of $\mathrm{O}(\varepsilon^0)$ for the stress and
velocity fields, as they suffice to determine the shear layer
thickness.  However, we ensure that the asymptotic expansion is
uniformly valid by checking that the ratio $m^{(1)}/m^{(0)}$ is
$\mathrm{O}(1)$ everywhere in the domain $0 < \xi < 1$.

\subsubsection{Plane shear in the absence of gravity}
\label{subsubsec-asymp_g0}

        On substituting \eqref{eqn-perturb_soln} in  \eqref{eqn-eta_g0},
we find that the leading order solution for the stresses are
\begin{subeqnarray}
    \ol{\sigma}_{xy}^{(0)} & = & \ol{\sigma}_{yx} = -\tan\delta,\\
    m^{(0)} & = & m_0 \equiv \sqrt{b^2 - a^2},
\label{eqn-asymp_soln_eta_g0}
\end{subeqnarray}
and the solutions at all higher orders in $\varepsilon$ vanish.
The constants $a$ and $b$ are defined in \eqref{eqn-a_b_def}. This
asymptotic solution is valid only in the outer region $\xi - 1/2
\gg \varepsilon$; in the inner region $(\xi - 1/2) \sim
\varepsilon$, $m$ deviates from $m_0$ to satisfy the first of
\eqref{eqn-nd_bc6}, as evident from the exact solution
\eqref{eqn-eta_soln_g0}.  For fully rough walls, $m_0 = 0$, and
hence \eqref{eqn-asymp_soln_eta_g0} is a uniformly valid solution
as it satisfies the boundary condition $m(1/2) = 0$.

    The $\omega$ and $u$ fields in the outer region are obtained from
\eqref{eqn-nd_fr1}-\eqref{eqn-nd_fr2} and
\eqref{eqn-asymp_soln_eta_g0}, yielding
\begin{subeqnarray}
    \omega & = & \Frac{-1/2}{\varepsilon (K + 2/k)} \exp \left(-k(1 -
    \xi)/\varepsilon \right),\\
    u & = & \frac{1}{2}\left(1  + \Frac{2/k}{(K + 2/k)}  \exp \left( -k(1 -
    \xi)/\varepsilon \right)\right),
\label{eqn-asymp_soln_vels_g0}
\end{subeqnarray}
where $k$ is defined in \eqref{eqn-k_def}. Thus, the linear and
angular velocities decay rapidly with distance from the walls. The
variation of $m$ within the inner region near $\xi = 1/2$ has
little influence on the velocity and fields, as they are
negligibly small within this region.

    When the two walls are of different roughness,
the exact solution is given in \S\ref{subsec-zerog}. The couple
stress is constant, $m = m_0$, regardless of $\varepsilon$, and
the solutions for $\omega$ and $u$ given in
\eqref{eqn-soln_aw_vels} reduce in the asymptotic limit to
\begin{subeqnarray}
    \omega & = & \Frac{-1}{\varepsilon (K + 2/k)} \exp \,
    \left(-k(1 - \xi)/\varepsilon \right)\\
    u & = & \Frac{2/k}{(K + 2/k)} \exp \, \left( -k(1 - \xi)
    /\varepsilon \right)
\label{eqn-asymp_soln_vels_g0_aw}
\end{subeqnarray}
It should be  noted that if the couple stress at the lower wall is
changed by replacing \eqref{eqn-nd_bc4} with an alternative
boundary condition (see \S\ref{sec-BC}), the asymptotic solution
for $m$ in the outer region $\xi \gg \varepsilon$ will still be $m
= m_0$, but the solution in the inner region $\xi \sim
\varepsilon$ will be different.  However, the velocity fields are
affected only by the couple stress in the outer region, as they
decay rapidly with distance from the upper wall, and therefore the
boundary condition for $m$ at $\xi = 0$ does not influence the
velocity profile in the limit $\varepsilon \to 0$.

\subsubsection{Plane shear under gravity}
\label{subsubsec-asymp_g}

        An equation for the couple stress, similar to
\eqref{eqn-eta_g0} for zero gravity, can be obtained by combining
the yield condition and the angular momentum balance, to get
\begin{equation}
\alpha \, \varepsilon \der{m}{\xi} = -a + \left( b^2 + c^2 \left[
(1 + B \nu(1 - \xi))^2 - 1 \right] - m^2 \right)^{1/2},
\label{eqn-eta_g}
\end{equation}
where $c^2 = \textfrac{\sin^2\phi}{(2A+2)}$, and $a$ and $b$ are
defined in \eqref{eqn-a_b_def}.

    Substituting \eqref{eqn-perturb_soln} in \eqref{eqn-eta_g},
the solution at the first two orders in $\varepsilon$ is
\begin{subeqnarray}
    m^{(0)} & = & \left(m_0^2  + c^2 \left[ (1 + B \nu(1 - \xi))^2
    - 1 \right] \, \right)^{1/2},\\
    m^{(1)} & = & \Frac{a\, c^2 \alpha }{{m^{(0)}}^2} \left[B^2\nu^2
    (1 - \xi) + B\nu \right], \label{eqn-asymp_eta_g}
\end{subeqnarray}
with $m_0$ defined in (\ref{eqn-asymp_soln_eta_g0}b).  Unlike in
\S\ref{subsubsec-asymp_g0}, the expression for $m^{(0)}$ above is
valid in the entire domain $0 < \xi < 1$, as it satisfies the
boundary condition \eqref{eqn-nd_bc4} at the lower wall. To
determine if the expansion is uniformly valid, we inspect the
ratio $\varepsilon \, m^{(1)}/m^{(0)}$. When $\tan\delta <
\sin\phi$, the ratio is $\mathrm{O}(\varepsilon)$. Thus the above
asymptotic expansion is uniformly valid when the walls are not
fully rough. The leading order solutions for the velocity and
angular velocity, obtained by integrating \eqref{eqn-nd_fr1} and
\eqref{eqn-nd_fr2} are,
\begin{subeqnarray}
    \omega^{(0)} & = & \omega_1 \, \exp \left( -\Frac{k}{\varepsilon}
    \int_{\xi}^{1} \left(1 + (\Frac{c}{m_0})^2
    \left[ (1 + B \nu(1 - \xi'))^2 - 1 \right]\right)^{1/2} d\xi' \right),\\
    u^{(0)} & =  & -2 \int_0^{\xi} {\omega(\xi') d\xi'},
\end{subeqnarray}
where $\omega_1 = \omega(1)$ and $k$ is defined in
\eqref{eqn-k_def}. To simplify the above expressions, we
anticipate the rapid decay of the velocity within a region of
thickness $\mathrm{O}(\varepsilon)$ from the upper wall,
substitute $\ol{\xi} = (1 - \xi)/\varepsilon$ and expand the above
expressions in $\varepsilon$. Retaining only the leading order
terms for consistency, we get expressions for the velocity fields
which are identical to \eqref{eqn-asymp_soln_vels_g0_aw}, the
latter corresponding to shear in the absence of gravity.

    For the singular case of  fully rough walls
($\tan\delta = \sin\phi$), $m^{(0)}(1) = 0$.  Hence the asymptotic
expansion \eqref{eqn-asymp_eta_g} is not valid in a region near
the upper wall where $(1 - \xi) \sim \varepsilon^{2/3}$, because
the ratio $\varepsilon \, m^{(1)}/m^{(0)}$ is $\mathrm{O}(1)$
within this region.  Noting that the scaling for the latter is
exactly what we obtained for flow through a vertical channel
\citep{mohan_etal99}, we follow the same procedure to determine a
uniformly valid asymptotic expansion in the entire domain $0 < \xi
< 1$.

    To determine the solution within the inner region near the
upper wall, we follow the prescription of
\citet[p.~104]{van_dyke64} and introduce the following rescaled
variables
\begin{equation}
    \hat{\xi} = (1 - \xi) \, \varepsilon^{-2/3} \left(\Frac{L^2
   }{2 B \nu} \right)^{-1/3},\;\;\; \hat{m} =
    m\, \varepsilon^{-1/3} (4 a \, \alpha B \nu c^2)^{-1/3},
\end{equation}
and seek a solution a solution of the form $\hat{m} =
\hat{m}^{(0)} + f(\varepsilon) \hat{m}^{(1)} + \ldots$. Upon
substituting the above in \eqref{eqn-eta_g}, we get at leading
order
\begin{equation}
    -\der{\hat{m}^{(0)}}{\hat{\xi}} + (\hat{m}^{(0)})^2 - \hat{\xi} = 0.
\label{eqn-asymp_eta_inner}
\end{equation}
This is a Riccati equation, whose solution can be written in terms
of the Airy functions \citep[p.~20, 569]{bender_orszag},
\begin{equation}
    \hat{m}^{(0)} = -\der{}{\hat{\xi}}
    \ln \left( c_1 \airyai(\hat{\xi}) + c_2 \airybi(\hat{\xi}) \right).
\end{equation}
where $\airyai(\hat{\xi})$ and $\airybi(\hat{\xi})$ are the two
independent Airy functions and $c_1$ and $c_2$ are constants of
integration. The constants are determined by matching the inner
and outer solutions in an overlap region, as suggested by
\citet[p.~105]{van_dyke64}.  Upon writing the outer solution (
\ref{eqn-asymp_eta_g}a) in terms of the inner variable $\hat{\xi}$
and expanding for small $\varepsilon$, we determine the leading
order term to be $(4 a \, \alpha B \nu c^2)^{1/3}
\varepsilon^{1/3} \, \hat{\xi}^{1/2}$. This must match with the
leading order outer expansion of the inner solution, which is
\begin{equation}
    -(4 a \, \alpha B \nu c^2)^{1/3} \, \varepsilon^{1/3}
\,\hat{\xi}^{1/2}\;\; \mbox{if} \;\; c_1 = 0,\;\; \mbox{and}\;\;
(4 a \, \alpha B \nu c^2)^{1/3} \, \varepsilon^{1/3} \,
\hat{\xi}^{1/2}\; \mbox{if} \; c_2 = 0.
\end{equation}
Hence $c_2 = 0$ and the solution at leading order in the inner
region is
\begin{equation}
    m^{(0)} = -\varepsilon^{1/3} \Frac{(4 a \, \alpha B \nu
    c^2)^{1/3}}{\airyai(\hat{\xi})} \der{}{\hat{\xi}}
    (\airyai(\hat{\xi})),
\end{equation}
from which we may derive the uniformly valid (additive) composite
solution in the entire domain,
\ba
    m^{(0)} & = & c \left([1 + B \nu(1 - \xi)]^2 - 1 \right)^{1/2}
     - \label{eqn-eta_g_unif}\\
    & & \varepsilon^{1/3} \Frac{(4 a \, \alpha B \nu c^2)^{1/3}}
    {\airyai(\hat{\xi})} \der{}{\hat{\xi}}
    (\airyai(\hat{\xi})) - (2 B \nu c^2(1 - \xi))^{1/2}. \nonumber
\ea
In the same manner, the uniformly valid solutions for the velocity
fields at the leading order are found to be
\begin{subeqnarray}
    \omega & = & \omega_1 \left( \Frac{\airyai(\hat{\xi})}{\airyai(0)}
    \right)^{2},\\
    u - u_1 & = & \varepsilon^{2/3} \left(\Frac{L^2}{2 B \nu}
    \right)^{1/3} 2 \omega_1 I(\hat{\xi}),
\label{eqn-asymp_soln_vels_g_fr}
\end{subeqnarray}
where $\omega_1 = \omega(1)$, $u_1 = u(1)$ and
\begin{equation}
    I(\hat{\xi}) \equiv (\airyai(0))^{-2} \int_0^{\hat{\xi}}{(\airyai(z))^2 dz}.
\label{eqn-defn_I}
\end{equation}
Enforcing the boundary condition u(0) = 0 and the velocity
boundary condition \eqref{eqn-nd_bc3u} at the upper wall, the
angular velocity and velocity at the wall can be determined,
giving to leading order
\begin{subeqnarray}
% define subequation numbers
\gdef\thesubequation{\theequation \textit{a,b}}
    \omega_1 = \Frac{-\varepsilon^{-2/3}}{2} \left(\Frac{L^2}
    {2 B \nu} \right)^{-1/3} \Frac{1}{I(\infty)},
    \quad \quad    u_1 = 1.
\label{eqn-asymp_soln_vels_g__fr_consts}
\end{subeqnarray}
% reinstate the original definition of \thesubequation
\returnthesubequation
 The uniformly valid velocity profile to leading order therefore is
\begin{equation}
    u = 1 - I(\hat{\xi})/I(\infty).
\label{eqn-vel_g_unif}
\end{equation}

\subsubsection{Cylindrical Couette flow}
\label{subsubsec-asymp_cyl}

        Subtracting \eqref{eqn-nd_zangmom_cyl} from \eqref{eqn-nd_thetamom_cyl} and
integrating, we get the following relation between the shear
stress $\ol{\sigma}_{r \theta}$ and the couple stress,
\begin{equation}
    \ol{\sigma}_{r \theta} = \Frac{- \tan\delta}{(1 +
    \xi/\ol{R}_i)^2} + 2 (A+1) \varepsilon \, \alpha \left( \Frac{m}{(\ol{R}_i
    + \xi)} - \Frac{m(0) \ol{R}_i}{(\ol{R}_i + \xi)^2} \right).
\label{eqn-nd_shsts_cyl}
\end{equation}
 At $\mathrm{O}(\varepsilon^0)$, \eqref{eqn-nd_zangmom_cyl} and
\eqref{eqn-nd_shsts_cyl} result in $\ol{\sigma}_{\theta r} =
\ol{\sigma}_{r \theta} = -\textfrac{\tan\delta}{(1 +
\xi/\ol{R}_i)^2}$, which when substituted in the yield condition
\eqref{eqn-nd_yc_cyl} gives the leading order couple stress,
\begin{equation}
        m^{(0)} = - \left(m_0^2 + d^2
        \left[ 1 - \Frac{1}{(1 + \xi/\ol{R}_i)^4} \right] \right)^{1/2}
\end{equation}
where $d^2 = \textfrac{\tan^2\delta}{2(A+1)}$. At
$\mathrm{O}(\varepsilon)$, \eqref{eqn-nd_yc_cyl} and
\eqref{eqn-nd_zangmom_cyl}, together with \eqref{eqn-nd_shsts_cyl}
yield,
\begin{equation}
        m^{(1)} = \Frac{a \, \alpha}{(m^{(0)})^2 \, \ol{R}_i \, (1 +
        \xi/\ol{R}_i)^3} \left( m_0^2 + d^2 \left[ 1 - \Frac{3}
        {(1 + \xi/\ol{R}_i)^4} \right] \right)
        - \Frac{2 a \alpha \, m^{(0)}(0)} {m^{(0)} \ol{R}_i \,
        (1 + \xi/\ol{R}_i)^4}
\end{equation}
Here too $\varepsilon m^{(1)}/m^{(0)} \sim
\mathrm{O}(\varepsilon)$, except when $\tan\delta = \sin\phi$.
Therefore the above expansion is uniformly valid when the wall is
not fully rough. On integrating
\eqref{eqn-nd_fr1_cyl}--\eqref{eqn-nd_fr2_cyl} and expanding the
results for the region $\xi \sim \varepsilon$ near the inner
cylinder, we see that the leading order velocity fields are again
identical to \eqref{eqn-asymp_soln_vels_g0_aw}.

        As in \S\ref{subsubsec-asymp_g}, the above asymptotic expansion
for $m$ is not uniformly valid when the inner cylinder is fully
rough, as the ratio $\varepsilon \, m^{(1)}/m^{(0)}$ is
$\mathrm{O}(1)$ in the inner region $\xi \sim \varepsilon^{2/3}$.
A solution in the inner region may be found by re-scaling the
variables,
\begin{equation}
    \hat{\xi} = \xi \, \varepsilon^{-2/3} \left(\Frac{L^2 \ol{R}_i}{4}
    \right)^{-1/3},\;\;\; \hat{m} =
    \Frac{m}{d}\, \varepsilon^{-1/3} \left(\Frac{4 L}{\ol{R}_i}
\right)^{-1/3},
\end{equation}
 yielding the following equation for $\hat{m}^{(0)}$:
\begin{equation}
    \der{\hat{m}^{(0)}}{\hat{\xi}} + (\hat{m}^{(0)})^2 - \hat{\xi} = 0.
\end{equation}
Following the procedure given in \S\ref{subsubsec-asymp_g}, the
uniformly valid expression for the couple stress is
\begin{equation}
        m^{(0)} = -\, d \left(1 - \Frac{1}{(1 + \xi/\ol{R}_i)^4}\right)^{1/2}
    + \varepsilon^{1/3} \Frac{d \, (4 L/\ol{R}_i)^{1/3}}{\airyai(\hat{\xi})}
        \der{}{\hat{\xi}} (\airyai(\hat{\xi})) + d (4 \xi/\ol{R}_i)^{1/2},
\end{equation}
and for velocity fields is
\begin{subeqnarray}
    \omega & = & \Frac{-\varepsilon^{-2/3}}{2} \left(\Frac{L^2
\ol{R}_i}{4} \right)^{-1/3} \Frac{1}{I(\infty)} \left(
\Frac{\airyai(\hat{\xi})}{\airyai(0)} \right)^{2},\\ u & = & -1 +
I(\hat{\xi})/I(\infty), \label{eqn-asymp_soln_vels_cyl_fr}
\end{subeqnarray}
where $I(\hat{\xi})$ is defined by \eqref{eqn-defn_I}.

\newpage
\bibliography{cosserat}

\newpage

\begin{figure}
  \begin{center}
    \psfrag{xi}[br][][1][-90]{$\xi$}
    \psfrag{vel}[t][]{$u$}
    \psfrag{omega}[t][]{$\omega$}
    \vspace{1in}
    \hspace*{0.1in} (a) \hspace{2.2in} (b) \\
    \includegraphics[width= 0.45\textwidth]{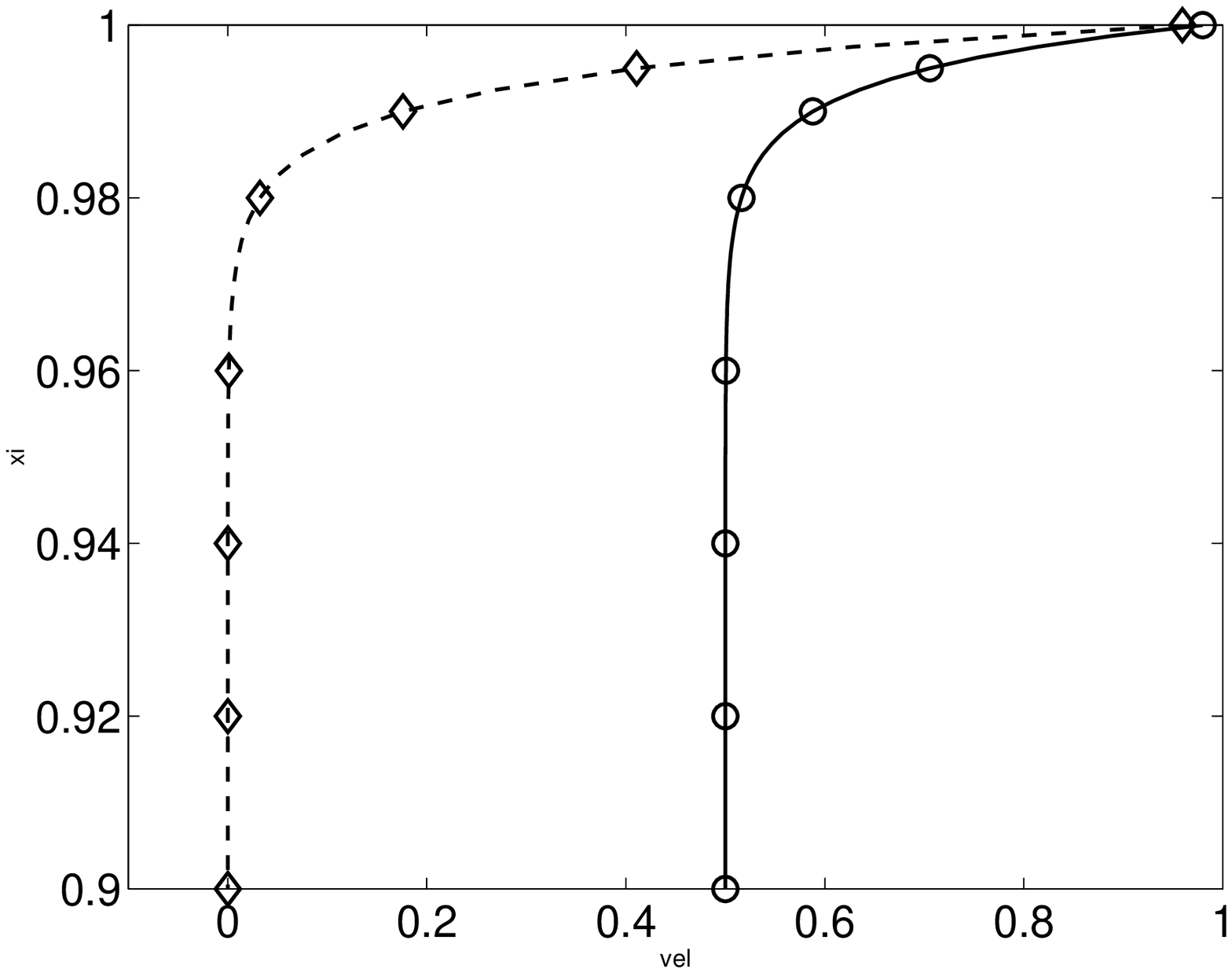}
    \hspace{0.2in} \includegraphics[width= 0.45\textwidth]{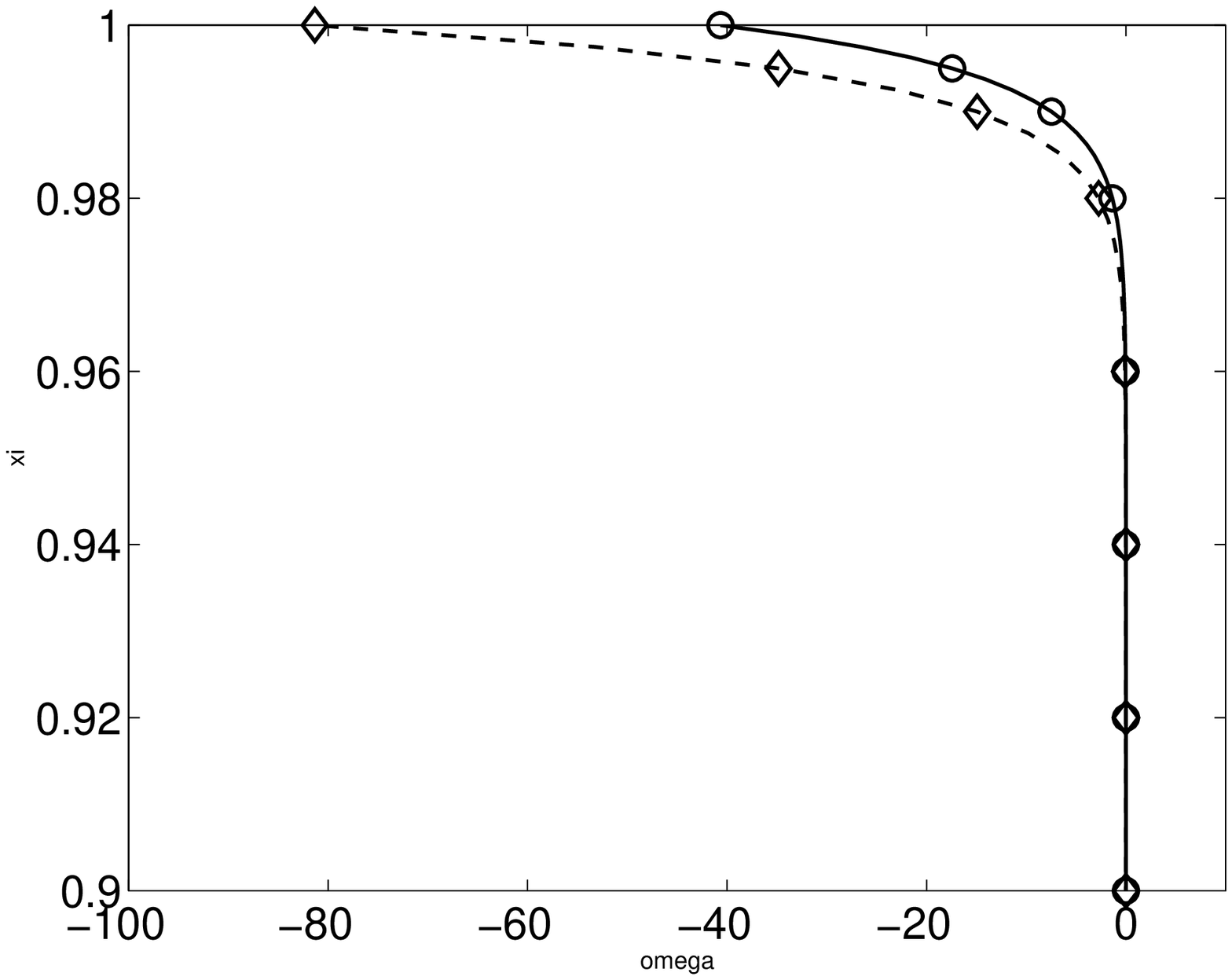} \\
    \caption{Comparison of the asymptotic solutions (for small $\varepsilon$)
    for the (a) linear velocity and (b) angular velocity fields with
    their respective numerical solutions, for plane shear in the absence of
    gravity.  The solid line and  circles represent numerical and
    asymptotic solutions, respectively, for shear between identical walls.  The
    dashed line and diamonds represent numerical and asymptotic solutions,
    respectively, for the case of a rougher lower wall.  Parameter values are:
    $\phi=28.5^\circ$, $\delta=20^\circ$, $\varepsilon=10^{-3}$,
    $L=10$, $A=1/3$ and $K=0.5$.\label{fig-asymp_g0}}
  \end{center}
\end{figure}

\begin{figure}
  \begin{center}
    \psfrag{xi}[br][][1][-90]{$\xi$}
    \psfrag{vel}[t][]{$u$}
    \psfrag{m}[t][]{$m$}
    \vspace{1in}
    \hspace*{0.1in} (a) \hspace{2.2in} (b) \\
    \includegraphics[width= 0.45\textwidth]{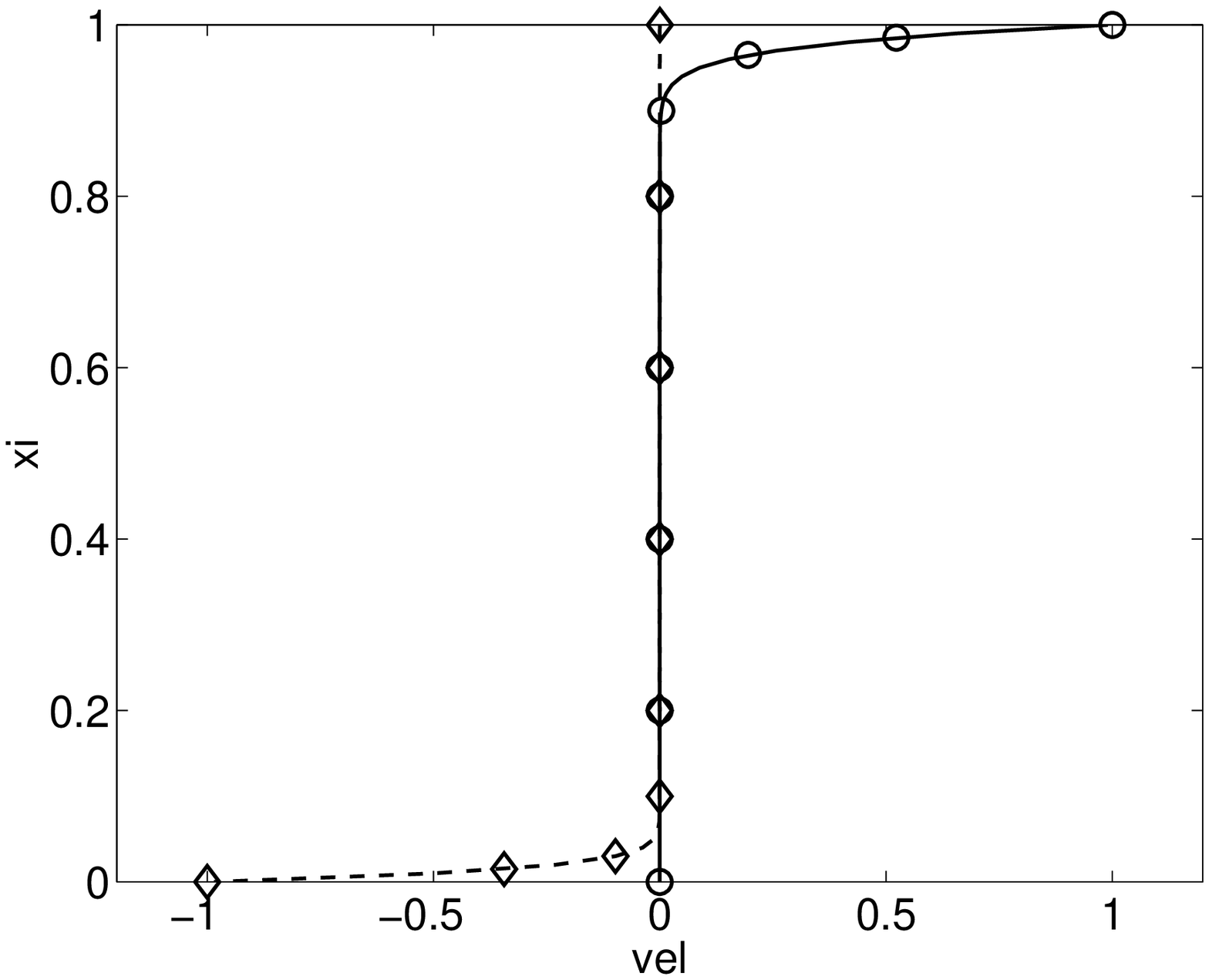}
    \hspace{0.2in} \includegraphics[width= 0.45\textwidth]{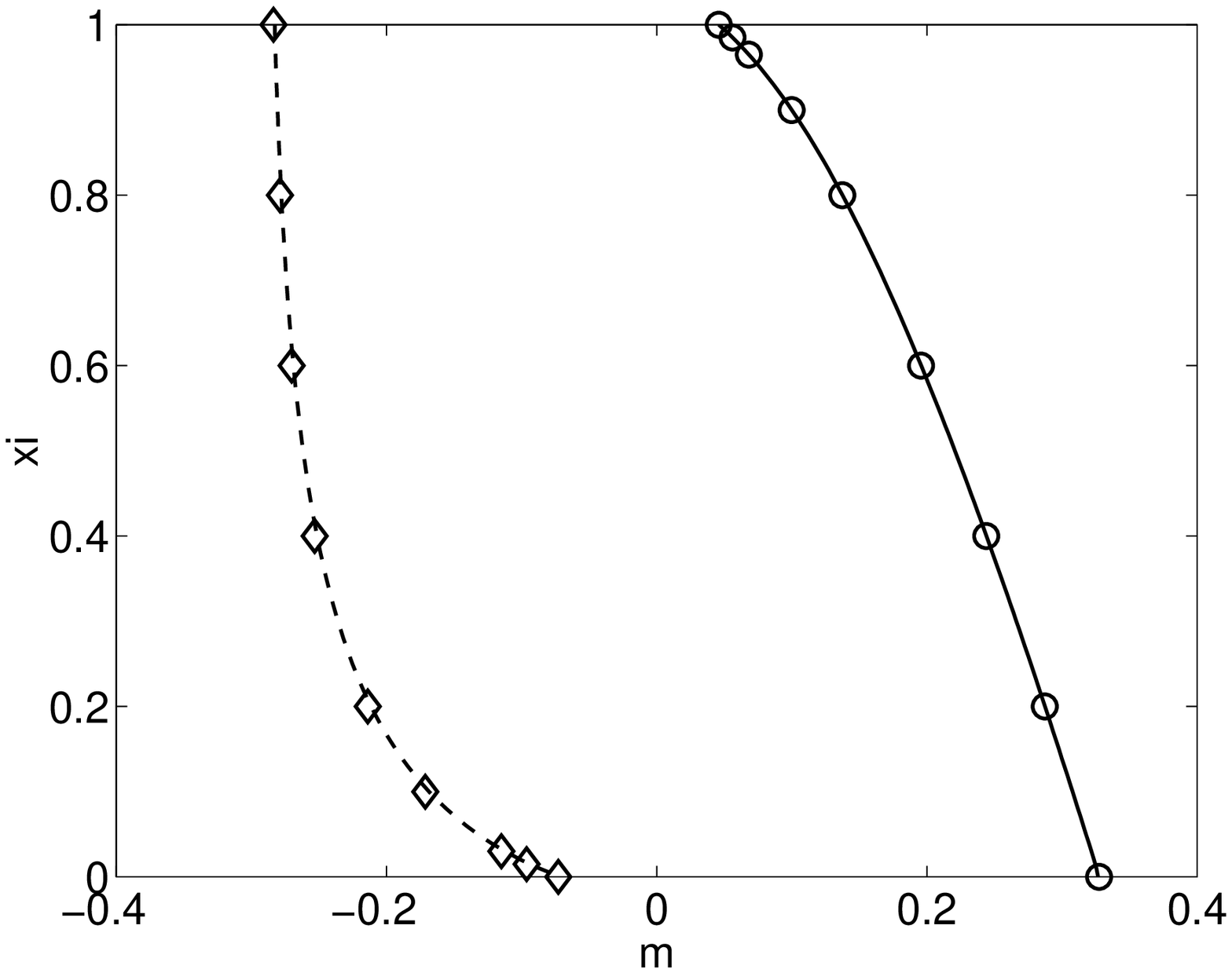} \\
    \caption{The asymptotic solutions for small $\varepsilon$ compared with
    the numerical solutions for plane shear under gravity (solid line and
     circle), and cylindrical Couette flow (dashed line and diamond).
    The lines represent numerical solutions, and the symbols asymptotic solutions.
    The walls are fully rough for both cases.  Parameter values are:
    $\phi=28.5^\circ$, $\varepsilon=10^{-3}$, $L=10$, $A=1/3$, and
    $K=0.5$.  In addition, $B=1$ for plane shear and $\ol{R}_i=1$ for cylindrical
    Couette flow (see \S\ref{sec-cyl_couette}).\label{fig-asymp_g_c}}
  \end{center}
\end{figure}

\end{document}